\documentclass[traditabstract]{aa}
\usepackage{graphicx}
\usepackage{natbib}
\usepackage{lscape}
\usepackage{afterpage}

\def\sun{\ifmmode\odot\else$\odot$\fi}
\def\micron{\hbox{\,$\mu$m}}

\newcommand{\HII}{H\,{\scriptsize II}}
\newcommand{\Feka}{Fe\,K$\alpha$}
\newcommand{\xmm}{\textit{XMM-Newton}}
\newcommand{\spitzer}{\textit{Spitzer}}
\newcommand\nodata{ ~$\cdots$~ }

\bibpunct{(}{)}{;}{a}{}{,}

\begin{document}

\title{The X-ray emission of local luminous infrared galaxies}

\author{Miguel Pereira-Santaella\inst{\ref{inst1}} \and Almudena Alonso-Herrero\inst{\ref{inst1}} \and Mar\'ia Santos-Lleo\inst{\ref{inst2}} \and Luis Colina\inst{\ref{inst1}} \and Elena Jim\'enez-Bail\'on\inst{\ref{inst3}} \and Anna L. Longinotti\inst{\ref{inst4}} \and George H. Rieke\inst{\ref{inst5}} \and  Martin Ward\inst{\ref{inst6}} \and Pilar Esquej\inst{\ref{inst1}}}

\institute{Departamento de Astrof\'isica, Centro de Astrobiolog\'ia, CSIC/INTA, Carretera de Torrej\'on a Ajalvir, km 4, 28850, Torrej\'on de Ardoz, Madrid, Spain \email{mpereira@cab.inta-csic.es}\label{inst1} \and XMM-Newton Science Operation Centre, European Space Agency, 28691, Villanueva de la Ca\~nada, Madrid, Spain\label{inst2} \and Instituto de Astronom\'ia, Universidad Nacional Aut\'onoma de M\'exico, Apartado Postal 70-264, 04510, M\'exico DF, M\'exico\label{inst3} \and  MIT Kavli Institute for Astrophysics and Space Research, 77 Massachusetts Avenue, NE80-6011, Cambridge, MA, 02139, USA\label{inst4} \and Steward Observatory, University of Arizona, 933 North Cherry Avenue, Tucson, AZ 85721, USA\label{inst5} \and Department of Physics, Durham University, South Road, Durham, DH1 3LE, UK\label{inst6}}

\abstract{
We study the X-ray emission of a representative sample of 27 local luminous infrared galaxies (LIRGs). The median IR luminosity of our sample is $\log L_{\rm IR}\slash L_{\odot}$ $=$ 11.2, thus the low-luminosity end of the LIRG class is well represented. We used new \xmm\ data as well as \textit{Chandra} and \xmm\ archive data.
The soft X-ray (0.5-2\,keV) emission of most of the galaxies ($>$80\,\%), including LIRGs hosting a Seyfert 2 nucleus, is dominated by star-formation related processes. These LIRGs follow the star-formation rate (SFR) versus soft X-ray luminosity correlation observed in local starbursts. We find that $\sim$15\,\% of the non-Seyfert LIRGs (3 out of 20) have an  excess hard X-ray emission relative to that expected from star-formation that might indicate the presence of an obscured AGN. The rest of the non-Seyfert LIRGs follow the SFR versus hard X-ray (2-10\,keV) luminosity correlation of local starbursts.
The non-detection of the 6.4\,keV \Feka\ emission line in the non-Seyfert LIRGs allows us to put an upper limit to the bolometric luminosity of an obscured AGN, $L_{\rm bol}$ $<$10$^{43}$\,erg\,s$^{-1}$. That is, in these galaxies, if they hosted a low luminosity AGN, its contribution to total luminosity would be less than 10\,\%.
Finally we estimate that the AGN contribution to the total luminosity for our sample of local LIRGs is between 7\,\% and 10\,\%.
}


\titlerunning{The X-ray emission of local LIRGs}
\authorrunning{Pereira-Santaella et al.}

\maketitle

\section{Introduction}\label{s:intro}

Luminous infrared galaxies (LIRGs) are galaxies with infrared (IR) luminosities ($L_{\rm IR}$\,$=$\,$L_{8-1000\mu m}$) from 10$^{11}$ to 10$^{12}$\,$L_{\rm \odot}$.  They are powered by star-formation and/or an active galactic nucleus (AGN; see \citealt{Sanders96} for a review).
Together with ultraluminous infrared galaxies (ULIRGs; $L_{\rm IR} > 10^{12}$ $L_{\rm \odot}$), they are the major contributors to the star-formation rate (SFR) density at $z$\,$\sim$\,1--2 \citep{PerezGonzalez2005, LeFloch2005, Caputi2007}.

The star-formation in local LIRGs is distributed over few kpc scales \citep{AAH06s, Hattori04, RodriguezZaurin2011}. This is similar to local starbursts and $z$\,$\sim$\,2 infrared bright galaxies \citep{Daddi2007, Rigby2008, Farrah08, Rujopakarn2010}, but at odds with local ULIRGs where most of the activity is taking place in very compact regions (the central kpc). Similarly, the fraction of AGN dominated local ULIRGs increases with increasing IR luminosity. About 40\,\% of the ULIRGs are classified as Seyfert \citep{Veilleux1995,Kim1998}. Thus the study of local LIRGs is motivated as they might be scaled-down versions of more distant IR bright galaxies.

The X-ray emission of starburst galaxies is mainly produced by high-mass X-ray binaries (HMXB), supernova remnants (SNR), O stars and hot gas heated by the energy originated in supernova explosions \citep{Persic2002, Fabbiano2006}.
The hard X-ray (2--10\,keV) emission is dominated by HMXB, although the contribution of hot gas ($kT$ $>$ 3\,keV) heated by supernovae may dominate the hard X-ray emission for the most intense starbursts \citep{Strickland2009, Iwasawa2011, Colina2011}.
On the other hand, the soft part of the X-ray emission (0.5--2\,keV) emission is mostly produced by gas at $kT$ $\sim$ 0.3--0.7\,keV.

It has been shown that there is a good correlation between the hard X-ray luminosity and the SFR for local starbursts (e.g., \citealt{Ranalli2003, Grimm2003, Persic2004}). However, the contribution to the hard X-ray luminosity from low-mass X-ray binaries (LMXB), which is not related to the current SFR, is not always negligible. For instance \citet{Colbert2004} and \citet{Lehmer2010} estimated that the LMXB contribution is important for galaxies with low SFR\slash $M_\star$.
It should be noted that the X-ray emission of a star-formation burst is delayed with respect to other SFR tracer. Thus, an evolution with time is expected in the X-ray emission of the star-forming galaxies \citep{Mas2008, Rosa2009}. This evolution might explain part of the scatter in the X-ray luminosity vs. SFR correlations.

According to their IR luminosity, the SFR of LIRGs ranges from $\sim$20 to 200\,$M_{\rm \odot}$\,yr$^{-1}$ \citep{Kennicutt1998}. Therefore strong X-ray emission ($\sim$10$^{41}$\,erg\,s$^{-1}$) associated to star-formation is expected from these galaxies. The AGN contribution to the X-ray emission of LIRGs is expected to be low. Pure Seyfert AGN emission is detected in $\sim$15\,\% of the LIRGs using optical spectroscopy \citep{Kim1998}, however a dust embedded AGN could be present in some of the them.
Thanks to X-ray observations of LIRGs we are be able to determine whether an obscured AGN is present, or, in the case of non-detection, set an upper limit to the AGN contribution.

Previous studies of the X-ray emission produced by star-formation have been focused on nearby starbursts (e.g., \citealt{Ptak1999, JimenezBailon2003, Ranalli2003, Grimm2003, Persic2004, Colbert2004}) or ULIRGs (e.g., \citealt{Rieke1988, PerezOlea1996, Ptak2003, Franceschini2003, Teng2005, Teng2010}). 

Although there is a number of papers on individual LIRGs (e.g., \citealt{Lira2002, Blustin2003, Jenkins2004, Jenkins2005, Levenson2004, Levenson2005, Miniutti2007}), there are few studies of the X-ray properties of LIRGs as a class (e.g., \citealt{Risaliti2000, Lehmer2010, Iwasawa2011}).
\citet{Risaliti2000} studied a sample of 78 objects, biased towards Seyfert LIRGs (90\,\%). They concluded that many of the sources might be completely Compton thick ($N_{\rm H}$ $>$ 10$^{25}$). Using an unbiased subsample of LIRGs they found that $\sim$60\,\% of the LIRGs host AGN although they are weak or heavily obscured.
More recently, \citet{Iwasawa2011} carried out a study of the most luminous local IR galaxies (11.7 $<$ $\log L_{\rm IR}/L_\odot$ $<$ 12.5). About 50\,\% of them are likely to contain an AGN, increasing the fraction of AGN sources with increasing $L_{\rm IR}$. They found that their non-AGN galaxies have lower hard X-ray luminosities than that expected from the local starbursts hard X-ray vs. SFR correlation. They suggested that the hard X-ray emission of these (U)LIRGs is dominated by hot gas and not by HMXB as in local starbursts.

In this paper we present a study of a sample of 27 local LIRGs (median $\log L_{\rm IR}/L_\odot$ $=$ 11.2) observed with \xmm\ and \textit{Chandra}. The sample is described in Sect. \ref{s:sample}. In Sect. \ref{s:observations} we describe the X-ray data reduction. In Sects. \ref{s:spatial} and \ref{s:spec_analysis} we present the spatial and spectral analysis of the X-ray data. The properties of the X-ray emission produced by star-formation and AGN related processes are discussed in Sects. \ref{s:sfr_xrayir} and \ref{s:agn_activity} respectively. Sect. \ref{s:conclusions} summarizes the main conclusions.

Throughout this paper we assume a flat cosmology with $H_0 = 70$ km
s$^{-1}$Mpc$^{-1}$, $\Omega_{\rm M} = 0.3$ and $\Omega_{\rm \Lambda} = 0.7$.

\section{The sample of LIRGs}\label{s:sample}

\subsection{Definition of the sample}\label{ss:sample_def}

Our sample of LIRGs contains 27 galaxies with \xmm\ or \textit{Chandra} data drawn from the volume limited sample of local LIRGs (40\,Mpc $<$ d $<$ 75\,Mpc) of \citet{AAH06s}.
The \citet{AAH06s} sample was selected from the the \textit{IRAS} Revised Bright Galaxy Sample \citep[RBGS;][]{SandersRBGS}\footnote{The \textit{IRAS} RBGS is a complete flux limited sample including all the extragalactic objects with a 60\micron\ flux density greater than 5.25\,Jy and Galactic latitude \textbar b\textbar $>$ 5\,\degr.} to have 2750 $<$ $v_{\rm hel}$\,(km s$^{-1}$) $<$ 5200 and 11.05 $<$ log $L_{\rm IR}$\slash$L_{\rm \odot}$ $<$ 11.88. Such criteria were imposed to allow for narrow-band observations of the Pa$\alpha$ emission line with the NICMOS instrument on the \textit{HST}.
Further details about the parent sample are given in \citet{AAH06s}.
Additionally we extended the \citet{AAH06s} sample to include all the galaxies in the \textit{IRAS} RBGS that fulfill their selection criteria but were not included in their sample (mostly optically classified Seyfert galaxies, see \citealt{AAH2011}).

\begin{figure}[h]
\center
\resizebox{\hsize}{!}{\includegraphics{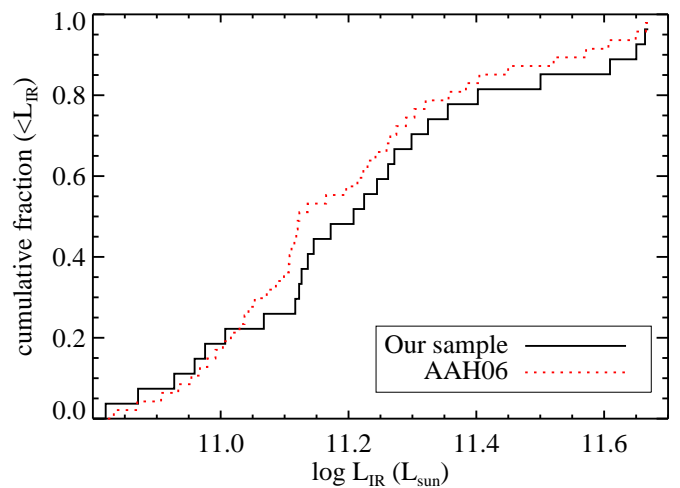}}
\caption{Comparison of the cumulative galaxy distributions as a function of the $L_{\rm IR}$ of our sample of LIRGs and the extended parent sample (AAH06; see Sect. \ref{ss:sample_def}).}
\label{fig_sample}
\end{figure}

In Fig. \ref{fig_sample} we compare the $L_{\rm IR}$ distribution (adapted to the cosmology used throughout this paper) of our sample with that of the extended \citet{AAH06s} sample.
The Kolmogorov-Smirnov two-sample test shows that it is not possible to reject (p $>$ 0.49) that both samples come from the same distribution.
According to their nuclear activity classification\footnote{In Appendix \ref{apx:optical_class} we present the optical activity classification of 7 LIRGs with no classification published. This new classification is based on the optical spectra available in the six-degree Field Galaxy Survey (6dFGS) database.}, 44$\pm$10\,\% and 46$\pm$13\,\% are \HII-type in the parent sample and in our sample respectively. On the other hand, Seyfert galaxies represent 22$\pm$7\,\% of the parent sample and 27$\pm$10\,\% of our sample. That is, our X-ray sample is not biased towards active galaxies.
Thus these 27 galaxies constitute a representative sample of the local LIRGs in terms of both IR luminosity and nuclear activity. The median $\log L_{\rm IR}\slash L_{\rm \odot}$ of the sample is 11.2, thus low luminosity LIRGs are well represented.
The selected galaxies are listed in Table \ref{tbl_sample}.

\begin{table*}[ht]
\caption{The sample of local LIRGs}
\label{tbl_sample}
\centering
\begin{tiny}
\begin{tabular}{llccccccc}
\hline \hline
Galaxy Name &  \textit{IRAS} Name & $v_{\rm hel}$\tablefootmark{a} & $D_{\rm L}$ & Nuclear & Ref.\tablefootmark{c} & log $L_{\rm IR}$\tablefootmark{d} & Ref.\tablefootmark{e} & X-ray data\\
& & (km\,s$^{-1}$) & (Mpc) & Spect. class\tablefootmark{b} & & ($L_{\rm \odot}$) \\
\hline
NGC23 & IRAS~F00073+2538 & 4478 & 64.7 & composite & 1 & 11.1 & 9 & \citet{Lehmer2010}\\
NGC1614 & IRAS~F04315--0840 &4778&69.1& composite & 11 &11.7& 9 & \xmm\ (archive) \\
NGC2369 & IRAS~F07160--6215 &3196&46.0& composite & 8 &11.1& 9 & \xmm\ (our data)\\
NGC3110 & IRAS~F10015--0614 &5014&72.6& \HII\ & 11 &11.3& 9 & ''\\
NGC3256 & IRAS~F10257--4339 &2790&40.1& \HII\ & 5 &11.7& 9 &\xmm\ (archive)\\
NGC3690\tablefootmark{*} & IRAS~F11257+5850 &3057&44.0& Sy2 & 4 &11.4& 2 &''\\
IC694\tablefootmark{*} & '' &3098&44.6& LINER & 4 &11.6& 2 &''\\
ESO320-G030 & IRAS~F11506--3851 &3038&43.7& \HII & 8 &11.2& 9 & \xmm\ (our data)\\
IC860 & IRAS~F13126+2453 & 3859 & 55.7 & \nodata & 1 & 11.1 & 9 & \citet{Lehmer2010}\\
MCG$-$03-34-064 & IRAS~F13197--1627 &5009&72.5& Sy2 & 11 &11.1& 10 &\xmm\ (archive)\\
NGC5135 & IRAS~F13229--2934 &4074&58.8& Sy2 & 11 &11.3& 9 & \citet{Levenson2004}\\
NGC5653 & IRAS~F14280+3126 & 3513 & 50.7 &  \HII\ & 11 & 11.0 & 9 & \citet{Lehmer2010}\\
NGC5734 & IRAS~F14423--2039 &3998&57.7& composite & 8 &11.0& 10 & \xmm\ (our data)\\
NGC5743 & IRAS~F14423--2042 &4121&59.5& \HII & 8 &10.9& 10 & ''\\
IC4518W\tablefootmark{*} & IRAS~F14544--4255 &4720&68.2& Sy2 & 3 &11.2& 8 &\xmm\ (archive)\\
Zw049.057 & IRAS~F15107+0724 & 3858 & 55.7 & composite & 7 & 11.2 & 9 & \citet{Lehmer2010}\\
IC4686\tablefootmark{*} & IRAS~F18093--5744 &4948&71.6& \HII\ & 11 &11.0& 8 & \xmm\ (our data)\\
IC4687\tablefootmark{*} &  '' &5105&73.9& \HII\ & 11 &11.3& 8 & ''\\
IC4734 & IRAS~F18341--5732 &4623&66.8& \HII\ & 3 &11.3& 9 & ''\\
MCG+04-48-002 & IRAS~20264+2533 &4199&60.6& \HII\ & 6 &11.0& 10 &\xmm\ (archive)\\
NGC7130 & IRAS~F21453--3511 &4837&70.0& Sy2 & 11 &11.4& 9 & \citet{Levenson2005} \\
IC5179 & IRAS~F22132--3705 &3363&48.5& \HII\ & 11 &11.2& 9 & \xmm\ (our data)\\
NGC7469 & IRAS~F23007+0836 &4840&70.0& Sy1 & 1 &11.6& 9 &\xmm\ (archive)\\
NGC7679 & IRAS~23262+0314 &5162&74.7& Sy1 & 11 &11.1& 9 &''\\
NGC7769 & IRAS~F23485+1952 &4158&60.0& composite & 7 &10.9& 10 &'' \\
NGC7770\tablefootmark{*} & IRAS~F23488+1949 &4128&59.6& \HII & 11 &10.8& 8 &''\\
NGC7771\tablefootmark{*} & '' &4302&62.1& \HII\ & 1 &11.3& 8 &'' \\
\hline
\end{tabular}
\end{tiny}
\tablefoot{
\tablefoottext{a}{Heliocentric velocity from \textit{Spitzer} spectra \citep{Pereira2010c}.}
\tablefoottext{b}{Classification of the nuclear activity from optical spectroscopy. Galaxies classified as composite are likely to be a combination of AGN activity and star-formation.}
\tablefoottext{c}{Reference for the optical spectroscopic data.}
\tablefoottext{d}{Logarithm of the IR luminosity, $L$(8--1000\micron), calculated as defined in \citet{Sanders96}.}
\tablefoottext{e}{Reference for the IR luminosity (adapted to the cosmology used throughout this paper).}
\tablefoottext{*}{The logarithm of the integrated $L_{\rm IR}$ in solar units of these systems are: NGC~3690\,+\,IC~694, 11.8; IC~4518W\,+\,IC~4518E, 11.2; IC~4686\,+\,IC~4687, 11.5; and NGC~7770\,+\,NGC7771, 11.4.}
\tablebib{(1) \citet{AAH09PMAS}; (2) \citet{Charmandaris2002}; (3) \citet{Corbett2003}; 
(4) \citet{GarciaMarin06}; (5) \citet{Lipari2000}; (6) \citet{Masetti2006}; (7) \citet{Parra2010}; (8) This work; (9) \citet{SandersRBGS}; (10) \citet{Surace2004}; (11) \citet{Yuan2010}.}
}
\end{table*}

\subsection{Star-formation rate and stellar mass}\label{ss:sfr_sm}

The IR luminosity in bright-IR galaxies is produced by dust heated by massive young stars. The dust absorbs a large fraction ($>$90\,\% for LIRGs, \citealt{Buat2007}) of the UV light from these stars that is re-emitted as thermal radiation in the mid- and far-IR. Thus the IR luminosity is a good tracer of the SFR for these galaxies (see \citealt{Kennicutt1998} for review).
We note, however, that the far-IR luminosity may also include the emission of a cooler dust component heated by the interstellar radiation field not be related with the current SFR.
To minimize the contribution of the cooler dust we used the \spitzer\slash MIPS 24\micron\ luminosity instead of the $L_{\rm IR}$, which includes longer wavelengths, to estimate the SFR. Moreover, the better spatial resolution of the \spitzer\slash MIPS 24\micron\ data, compared with \textit{IRAS}, allowed us to separate the individual galaxy emission of interacting systems.
We used the SFR calibration of \citet{Rieke2009}:

\begin{eqnarray}
{\rm SFR_{\rm IR}}\,(M_\odot\;{\rm yr^{-1}}) = 7.8\times 10^{-10}L_{\rm 24\,\mu m}\,(L_{\odot}) \nonumber \\
\times \lbrace 7.76\times 10^{-11} L_{\rm 24\,\mu m}\,(L_{\odot}) \rbrace ^{0.048}.
\end{eqnarray}

For this calibration \citet{Rieke2009} assumed a \citet{Kroupa2001} initial mass function (IMF) over the stellar mass range from 0.08 to 100 $M_{\rm \odot}$.
A correction to account for the leaked UV light from young stars is included in the calibration.
\citet{AAH2011} estimated the AGN contribution at 24\micron\ in these LIRGs. They decomposed their \spitzer\slash IRS mid-IR spectra into AGN and starburst components using clumpy torus models and star-forming galaxy templates. We used these estimations to subtract the AGN luminosity at 24\micron\ before calculating the obscured SFR.
The AGN fractions and the calculated SFR$_{\rm IR}$ are listed in Table \ref{tbl_sample_data}.

Some galaxies were observed with the \xmm\ optical monitor (OM) in the UVW2 band (2120\,\AA, see Sect. \ref{s:observations}). We corrected the observed UV flux from Galactic extinction using the \citet{Fitzpatrick1999} extinction curve and the Galactic color excess, $E$($B$--$V$), from the NASA Extragalactic Database (NED).
For these galaxies we compared the SFR derived from the UV continuum with that from the IR luminosity. 
Assuming a flat continuum in $L_{\rm \nu}$ in the UV range (1500--2800\,\AA) and scaling to a Kroupa IMF the SFR$_{\rm UV}$ can be expressed as:

\begin{equation}
{\rm SFR_{UV}}\,(M_\odot\;{\rm yr^{-1}}) = 9.2\times 10^{-29}L_{\rm \nu}\,{\rm (erg\;s^{-1}\;Hz^{-1})}
\end{equation}

where $L_{\rm \nu}$ is the UV continuum luminosity \citep{Kennicutt1998}.

The SFR$_{\rm IR}$\slash SFR$_{\rm UV}$ ratio ranges from $\sim$2--115 for our LIRGs. This corresponds to an UV leakage from 1 to 30\,\% of the total luminosity if the IR and the UV emissions trace the same stellar populations. This is in good agreement with the ratio expected for LIRGs \citep{Buat2007}.

To estimate the stellar mass of the galaxies we used the integrated $J$-, $H$-, and $K$-band near-IR magnitudes from the Two Micron All Sky Survey (2MASS) large galaxy atlas \citep{Jarrett2003} and the 2MASS extended source catalog \citep{Jarret2000}. The near-IR emission is well suited to calculate the stellar mass since the contribution from young stars is usually negligible and the scatter in the mass-to-light ratio is relatively small ($\sim$0.4\,dex). 
Following \citet{Bell2001} we used the $K$-band luminosity together with the $J-H$ color to obtain the stellar mass. We adjusted the normalization for the Kroupa IMF:

\begin{equation}
\log\,\frac{M_{\star}}{M_{\odot}} = \log\,\frac{L_{\rm K}}{L_{\rm \odot, K}} + 1.44 (J - H) - 1.17
\end{equation}

The SFRs, IR\slash UV ratios and stellar masses for our sample are listed in Table \ref{tbl_sample_data}.

\begin{table*}[ht]
\caption{Galaxy Properties}
\label{tbl_sample_data}
\centering
\begin{tabular}{lcccccccccc}
\hline\hline
Galaxy Name & $E$($B$--$V$)\tablefootmark{a} & $F_{\rm UV}$\tablefootmark{b} & $F_{\rm UV}^{\rm corr}$\,\tablefootmark{c} & $J-H$\tablefootmark{d} 
&  $F_{\rm K}$\tablefootmark{e}  & $F_{\rm 24\mu m}$\tablefootmark{f}  & $\log M_{\star}$\tablefootmark{g} & $C_{\rm 24\mu m}^{\rm AGN}$\tablefootmark{h} & SFR$_{\rm IR}$\tablefootmark{i} & $\frac{\rm SFR_{\rm IR}}{\rm SFR_{\rm UV}}$\tablefootmark{j} \\ 
& (mag) & (mJy) & (mJy) & (mag) & (mJy) & (Jy) & ($M_{\rm \odot}$) & & ($M_{\rm \odot}$\,yr$^{-1}$) \\
\hline
NGC23 & 0.04 &  \nodata & \nodata & 0.70 & 175 & 0.89 & 11.2 & 0.05 & 11 & \nodata \\
NGC1614 & 0.15 & 0.93 & 3.5 & 0.77 & 101 & 5.64 & 11.1 & \nodata & 92 & 49 \\
NGC2369 & 0.11 & 0.15 & 0.4 & 0.78 & 253 & 1.83 & 11.2 & 0.08 & 12 & 115 \\
NGC3110 & 0.04 & 2.58 & 3.5 & 0.73 & 96 & 0.98 & 11.1 & $<$0.04 & 15 & 8 \\
NGC3256 & 0.12 & 5.12 & 14.8 & 0.74 & 360 & 12.86 & 11.1 & $<$0.04 & 69 & 26 \\
NGC3690 & 0.02 &  \nodata & \nodata & 0.76 & 116 & 8.19 & 10.8 & 0.28 & 38 & \nodata \\
IC694 & 0.02 &  \nodata & \nodata & 0.79 & 96 & 8.85 & 10.7 & $<$0.02 & 59 & \nodata \\
ESO320-G030 & 0.13 & 0.33 & 1.0 & 0.78 & 129 & 1.72 & 10.9 & $<$0.04 & 10 & 46 \\
IC860 & 0.01 &  \nodata & \nodata & 0.75 & 36 & 0.85 & 10.4 & $<$0.05 & 8 & \nodata \\
MCG$-$03-34-064 & 0.08 &  \nodata & \nodata & 0.73 & 77 & 2.45 & 11.0 & 0.85 & 7 & \nodata \\
NGC5135 & 0.06 &  \nodata & \nodata & 0.76 & 177 & 2.02 & 11.2 & 0.14 & 19 & \nodata \\
NGC5653 & 0.01 &  \nodata & \nodata & 0.70 & 132 & 1.11 & 10.9 & 0.04 & 9 & \nodata \\
NGC5734 & 0.10 & 0.41 & 0.9 & 0.78 & 140 & 0.59 & 11.1 & $<$0.04 & 6 & 17 \\
NGC5743 & 0.10 & 1.79 & 4.1 & 0.73 & 75 & 0.43 & 10.8 & 0.15 & 4 & 2 \\
IC4518W & 0.16 &  \nodata & \nodata & 0.78 & 44 & 1.00 & 10.8 & 0.67 & 5 & \nodata \\
Zw049.057 & 0.04 &  \nodata & \nodata & 0.80 & 22 & 0.52 & 10.3 & $<$0.05 & 5 & \nodata \\
IC4686 & 0.10 & 0.76 & 1.8 & 0.70 & 15 & 0.86 & 10.2 & \nodata & 13 & 13 \\
IC4687 & 0.10 & 0.65 & 1.6 & 0.75 & 61 & 1.66 & 10.9 & 0.05 & 29 & 30 \\
IC4734 & 0.09 & 0.39 & 0.9 & 0.78 & 77 & 1.03 & 11.0 & $<$0.05 & 15 & 33 \\
MCG+04-48-002 & 0.45 &  \nodata & \nodata & 0.92 & 71 & 0.69 & 11.1 & 0.41 & 5 & \nodata \\
NGC7130 & 0.03 &  \nodata & \nodata & 0.69 & 125 & 1.88 & 11.1 & 0.15 & 25 & \nodata \\
IC5179 & 0.02 & 5.12 & 6.1 & 0.71 & 222 & 1.90 & 11.1 & $<$0.03 & 14 & 9 \\
NGC7469 & 0.07 &  \nodata & \nodata & 0.81 & 166 & 4.80 & 11.4 & 0.40 & 48 & \nodata \\
NGC7679 & 0.07 &  \nodata & \nodata & 0.63 & 60 & 0.85 & 10.8 & 0.23 & 11 & \nodata \\
NGC7769 & 0.07 &  \nodata & \nodata & 0.70 & 162 & 0.50 & 11.1 & 0.10 & 5 & \nodata \\
NGC7770 & 0.07 &  \nodata & \nodata & 0.71 & 28 & 0.40 & 10.3 & 0.27 & 3 & \nodata \\
NGC7771 & 0.07 &  \nodata & \nodata & 0.76 & 287 & 1.29 & 11.5 & 0.04 & 15 & \nodata \\
\hline
\end{tabular}
\tablefoot{
\tablefoottext{a}{Galactic color excess $E$($B$--$V$) from NED.}
\tablefoottext{b}{Observed \xmm\slash OM UVW2 (2120\,\AA) flux.}
\tablefoottext{c}{\xmm\slash OM UVW2 (2120\,\AA) flux corrected for Galactic extinction using the \citet{Fitzpatrick1999} extinction curve.}
\tablefoottext{d}{$J-H$ color calculated from the 2MASS magnitudes.}
\tablefoottext{e}{Integrated $K$-band flux from 2MASS.}
\tablefoottext{f}{\spitzer\slash MIPS 24\micron\ flux from \citet{Pereira2011b}.}
\tablefoottext{g}{Logarithm of the stellar mass obtained from the K-band luminosity and the $J-H$ color.}
\tablefoottext{h}{AGN fractional contribution to the total 24\micron\ emission from \citet{AAH2011}.}
\tablefoottext{i}{Star-formation rate based on the 24\micron\ luminosity. The AGN contribution to the 24\micron\ luminosity is subtracted.}
\tablefoottext{j}{Ratio of the star-formation rates estimated from the IR and UV luminosities.}
}
\end{table*}

\section{X-ray observations}\label{s:observations}

\subsection{\xmm\ observations and data reduction}\label{ss:xmm_reduction}

We obtained new \textit{XMM-Newton} data for 9 galaxies (proposals 55046 and 60160). We also found in the \xmm\ archive X-ray data for 12 more galaxies.
Our proposal was focused on galaxies classified as \HII\ galaxies based on their optical spectra whereas most of the galaxies from the archive are active galaxies (Seyfert and LINER activity).
The observation IDs and effective exposure times are shown in Table \ref{tbl_obslog}.
The analysis of the \textit{Chandra} X-ray data for the other 6 galaxies in our sample is taken from the literature (see Sect. \ref{ss:chandra_lit}).

We reduced the observation data files (ODF) using SAS version 10.0.2. First we used the SAS \textit{epproc} and \textit{emproc} tasks to generate the calibrated events files from the raw European Photon Imaging Camera (EPIC) pn and MOS data respectively.
A circular aperture (\hbox{d$\sim$15\,\arcsec} depending on the source extent) was used to extract the spectra of the galaxies. We estimated the backgrounds from a region close to the source in the same CCD and free of any contaminating source. The background regions were $\sim$4--5 times larger than the aperture used for the galaxies.
Then we created the background and background$+$source light-curves that we used to filter out high-background periods.
The background count rate threshold was chosen to just filter out those high-background periods that would not increase the signal-to-noise (S\slash N) ratio of the source (see Appendix A of \citealt{Piconcelli2004}).
For the pn data we considered single and double pixel events (PATTERN $\leq$ 4). We rejected events close to the CCD borders or to bad pixels (\#XMMEA\_EP).
For the MOS we also considered triple and quadruple pixel events (PATTERN $\leq$ 12) and we used the recommended expression (\#XMMEA\_EM) to filter the events. The energy redistribution matrices were generated with \texttt{rmfgen} and \texttt{arfgen}.
We rebinned the combined MOS spectrum (MOS1 and MOS2 spectra) and the pn spectrum in order to have at least 20 counts in each spectral bin using \texttt{grppha}.
Likewise we obtained X-ray images of the galaxies using the pn calibrated event files.

Due to the low number of counts in the Reflection Grating Spectrometer (RGS) data we could only extract the RGS spectra for few galaxies: NGC~3256, MCG$-$03-34-064, and NGC~7469. The spectra of MCG$-$03-34-064 and NGC~7469 are analyzed in detail by \citet{Miniutti2007} and \citet{Blustin2003}, respectively.

Simultaneously with the X-ray observations we obtained optical and UV images of the galaxies using the \xmm\slash OM with all the available filters (V 5430\,\AA, B 4500\,\AA, U 3440\,\AA, UVW1 2910\,\AA, UVM2 2310\,\AA, and UVW2 2120\,\AA). We used the SAS script \texttt{omichain} for the data reduction. This script processes the OM ODF files and produces calibrated images taking into account the telescope tracking information and the flat fielding corrections.
For some filters there was more than one exposure that we combined to increase the S/N ratio.
Then we used aperture photometry to measure the fluxes. We estimated the background from the image with special care to avoid artifacts in the images such as the smoke rings, etc. (see the \xmm\slash OM Calibration Status document). We corrected the count rate for the detector sensitivity degradation and coincidence loss.  The count rates were converted into Jy using the conversion factors given in \xmm\slash OM Calibration Status document.

\begin{table}[h]
\centering
\caption{Log of the \xmm\ observations}
\label{tbl_obslog}
\begin{tabular}{llcccccc}
\hline\hline
Galaxy Name & Obs. ID. & Exposure\tablefootmark{a} \\
  & & (ks) \\
\hline
NGC1614 & 0150480201 & 21.8\\
NGC2369 & 0550460101 & 24.3\\
NGC3110 & 0550460201 & 15.9 \\
NGC3256 & 0300430101 & 125.6\\
NGC3690/IC694 & 0112810101 & 17.1\\
ESO320-G030 & 0550460301 & 23.9\\
MCG$-$03-34-064 & 0206580101 & 42.7 \\
'' & 0506340101\tablefootmark{b} & \nodata \\
NGC5734/5743 & 0601600101 & 27.0 \\
IC4518W & 0406410101 & 22.8\\
IC4686/4687 & 0550460601 & 26.5\\
IC4734 & 0550460701 & 18.6\\
MCG+04-48-002 & 0312192301 & 11.4\\
IC5179 & 0550460801 & 22.0\\
NGC7469 & 0112170301 & 23.0\\
NGC7679 & 0301150501 & 17.9 \\
NGC7769/7770/7771 & 0093190301 & 30.0 \\
\hline
\end{tabular}
\tablefoot{
\tablefoottext{a}{Exposure time after flare removal.}
\tablefoottext{b}{Only used for the \xmm/OM UVW2 image of MCG$-$03-34-064.}
}
\end{table}

\subsection{\textit{Chandra} data from the literature}\label{ss:chandra_lit}
We found in the literature \citep{Levenson2004, Levenson2005, Lehmer2010} \textit{Chandra} X-ray data for another 6 galaxies (two Seyfert 2 galaxies, two composite, one \HII, and one without classification; see Table \ref{tbl_sample}). For these galaxies we used the published galaxy integrated  X-ray fluxes (Table \ref{tbl_model_lit}).

\begin{table}[h]
\centering
\caption{Galaxies taken from the literature.}
\label{tbl_model_lit}
\begin{tabular}{lccccccccccccccc}
\hline\hline
Galaxy Name & $L_{\rm 0.5-2\,keV}$ & $L_{\rm 2-10\,keV}$ & Ref. \\
 & \multicolumn{2}{c}{($10^{40}$\,erg\,s$^{-1}$)} \\
\hline
NGC23 & 6.7 & 4.2 &  1 \\
IC860 & 0.3 & 1.1 &  1 \\
MCG$-$03-34-064\tablefootmark{\star} &  26.6 & 111 &  2 \\
NGC5135 & 17.9 & 18.9 & 3 \\
NGC5653 & 2.8 & 1.5 &  1 \\
Zw049.057 & 0.2 & 1.5 &  1 \\
NGC7130 & 15.4 & 15.4 &  4 \\
NGC7469\tablefootmark{\star} & 1630 & 1690 &  5 \\
\hline
\end{tabular}
\tablefoot{0.5--2\,keV and 2--10\,keV absorption corrected luminosities of the galaxies taken from the literature adapted to the cosmology used throughout this paper.
\tablefoottext{\star}{For the galaxies with \xmm\ observations we fitted the data using the best-fit model given in the corresponding reference.}
\tablebib{(1) \citealt{Lehmer2010}; (2) \citealt{Miniutti2007}; 
(3) \citealt{Levenson2004}; (4) \citealt{Levenson2005}; (5) \citealt{Blustin2003}.}
}
\end{table}

\section{Spatial analysis of the \xmm\ data}\label{s:spatial}

\subsection{Morphologies}
We obtained X-ray images of these LIRGs as described in Sect. \ref{ss:xmm_reduction}. 
Fig. \ref{fig_ximage} shows the soft (0.5--2\,keV) and hard (2--7\,keV) X-ray images for the LIRGs together with the \xmm\slash OM UV (2120\,\AA) and near-IR \spitzer\slash IRAC (3.6\,$\mu$m) images for comparison.

In our sample of LIRGs we find different X-ray emission morphologies.
Most of them are dominated by the nuclear emission and appear as point like (or slightly resolved) sources at the \xmm\ angular resolution, 4--6\,arcsec. At the distances of these LIRGs this corresponds to 0.9--2\,kpc.
Six galaxies (NGC~3110, NGC~3256, NGC~5734, NGC~5743, IC~5179, and NGC~7769), $\sim$20\,\% of the sample, show extended soft X-ray emission. This indicates that at least some of the sources responsible for the origin of the X-ray emission (X-ray binaries, SNR, diffuse hot plasma, etc.) are extended over several kpc ($>$1\,kpc). Higher angular resolution images with \textit{Chandra} of LIRGs confirm that the X-ray emission comes from both: multiple point sources and diffuse emission distributed over the galaxies \citep{Zezas2003,Levenson2004, Levenson2005,Lehmer2010}.

The S\slash N ratio in the hard \xmm\ X-ray band (2--10\,keV) of the \HII\ galaxies is too low to accurately measure the size of the X-ray emitting region. The only exception is NGC~3256 that appears approximately as extended as its soft X-ray emission. The higher spatial resolution \textit{Chandra} X-ray images of NGC~3256 reveal that both the soft and hard X-ray emissions are resolved into multiple point sources, besides the two nuclei, and diffuse emission \citep{Lira2002}.

The hard X-ray emission of the Seyfert galaxies is dominated by the AGN, thus they appear as point sources in this energy range.

\begin{figure*}
\center
\includegraphics[width=17cm]{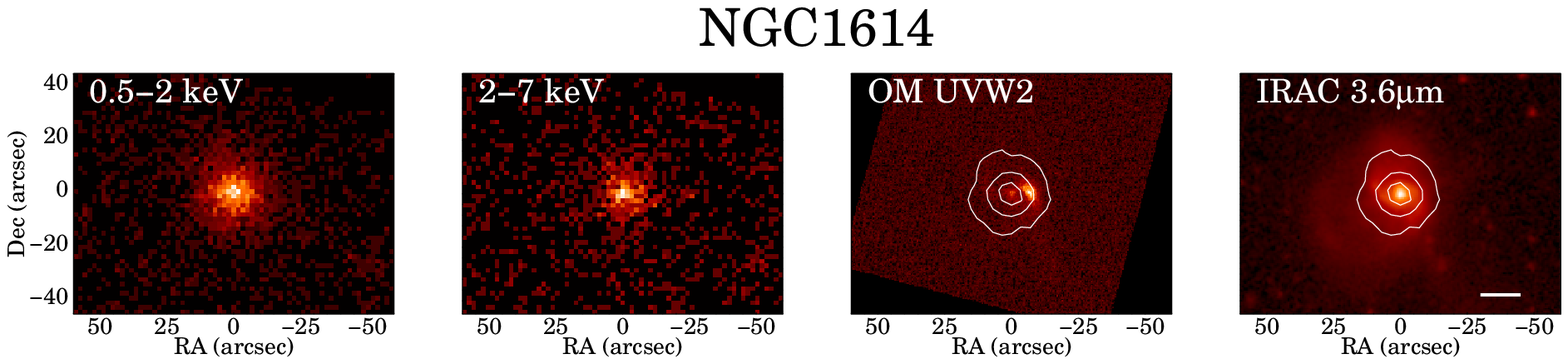}
\includegraphics[width=17cm]{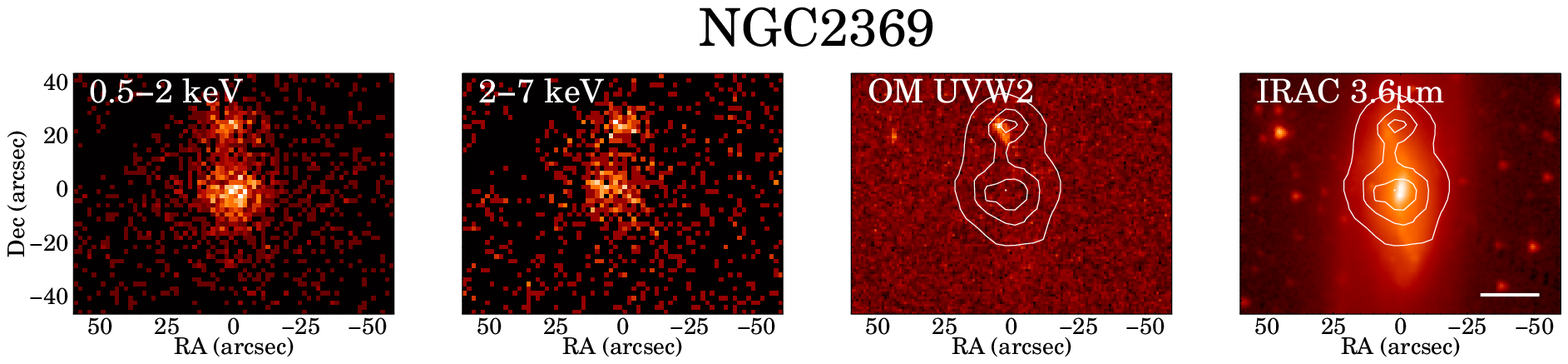}
\includegraphics[width=17cm]{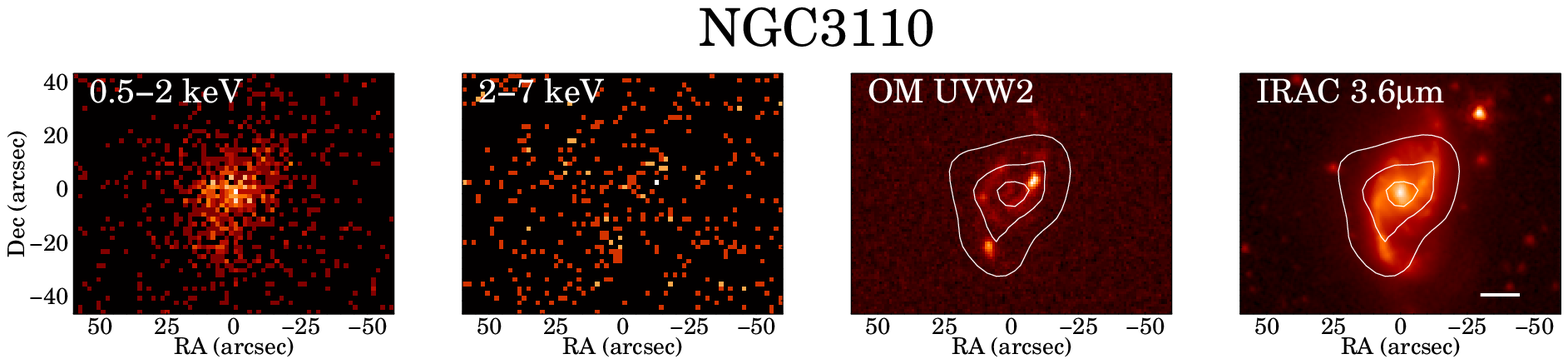}
\includegraphics[width=17cm]{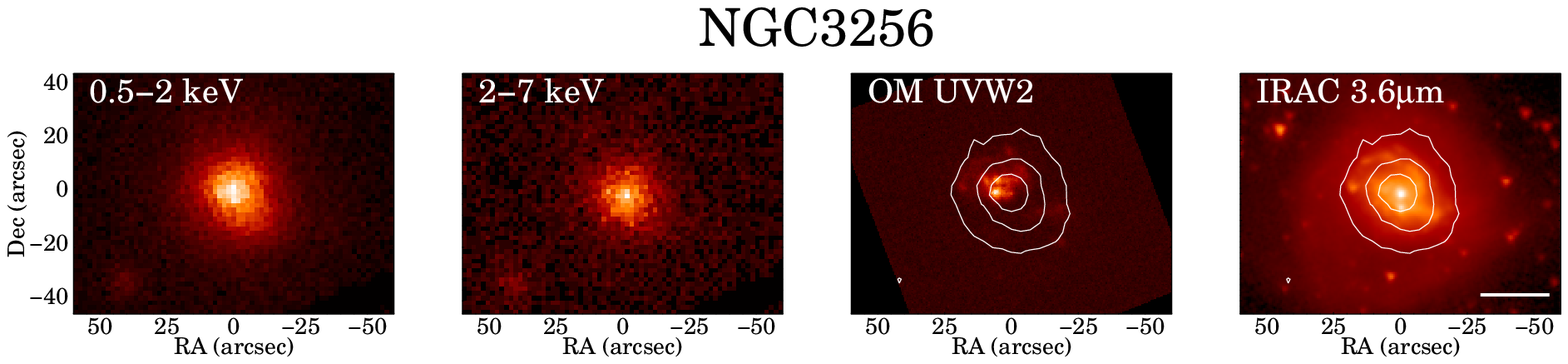}
\includegraphics[width=17cm]{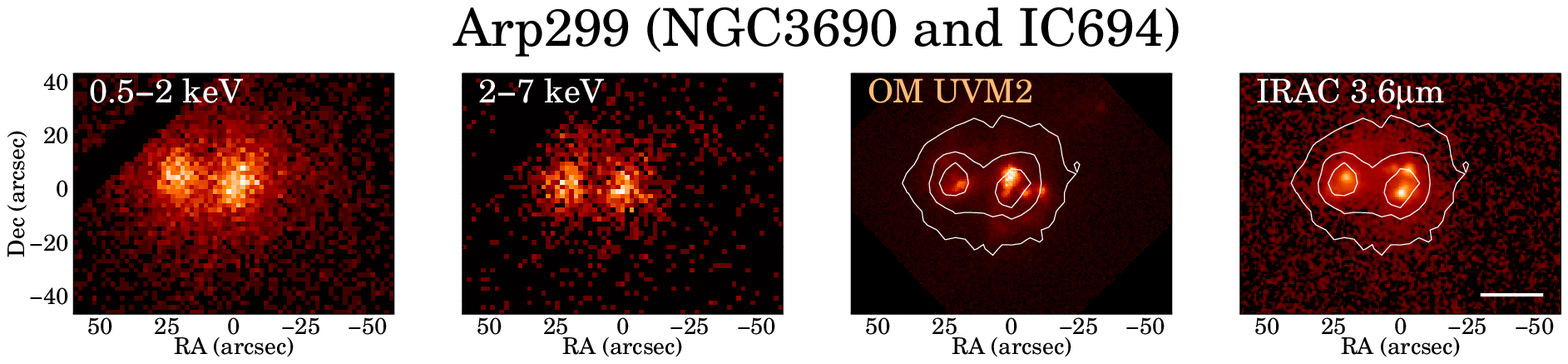}
\caption{\xmm/EPIC pn 0.5-2\,keV and 2-7\,keV images (first and second panels), \xmm/OM UVW2 (2120\,\AA) images for the galaxies observed with this filter (third panel). The third panel of Arp299 and MCG+04-48-002 corresponds to the \xmm/OM UVM2 (2310\,\AA) filter. \spitzer/IRAC 3.6\,\micron\ images (forth panel). For reference we represent in the third and forth panels the smoothed 0.5-7\,keV contours.
The white line in the right panels represents 5\,kpc at the distance of the galaxy. All images are shown in a square root scale. North is up and east is to the left.}
\label{fig_ximage}
\end{figure*}

\begin{figure*}
\addtocounter{figure}{-1}
\center
\includegraphics[width=17cm]{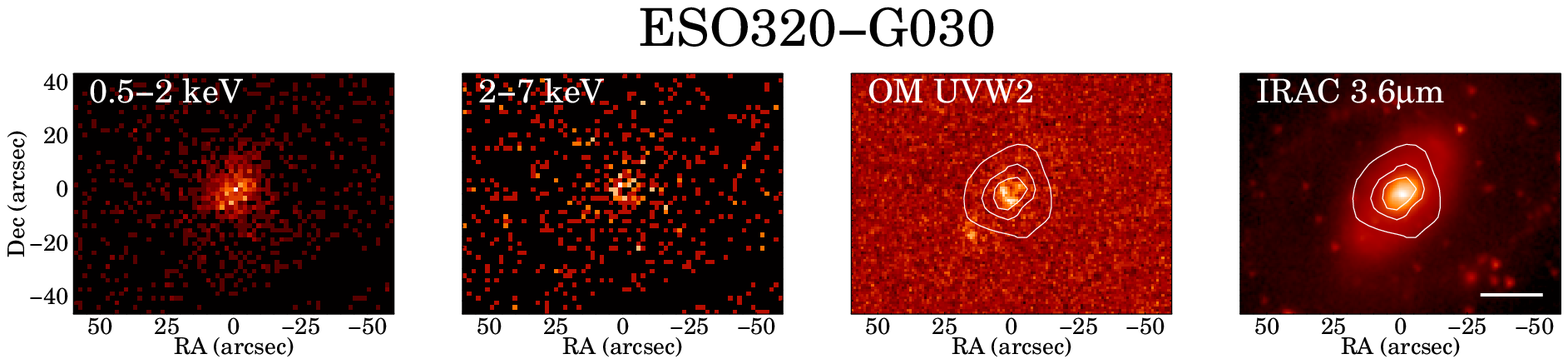}
\includegraphics[width=17cm]{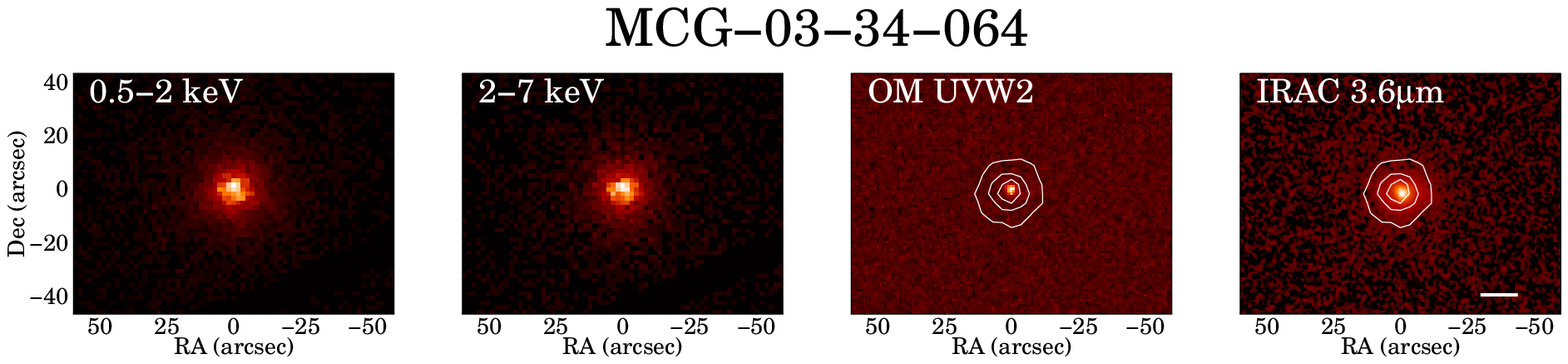}
\includegraphics[width=17cm]{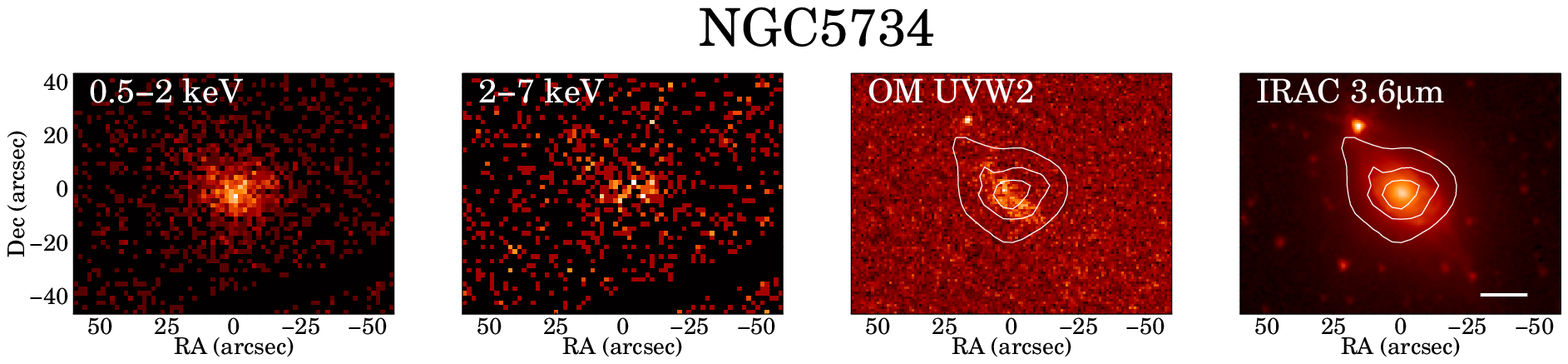}
\includegraphics[width=17cm]{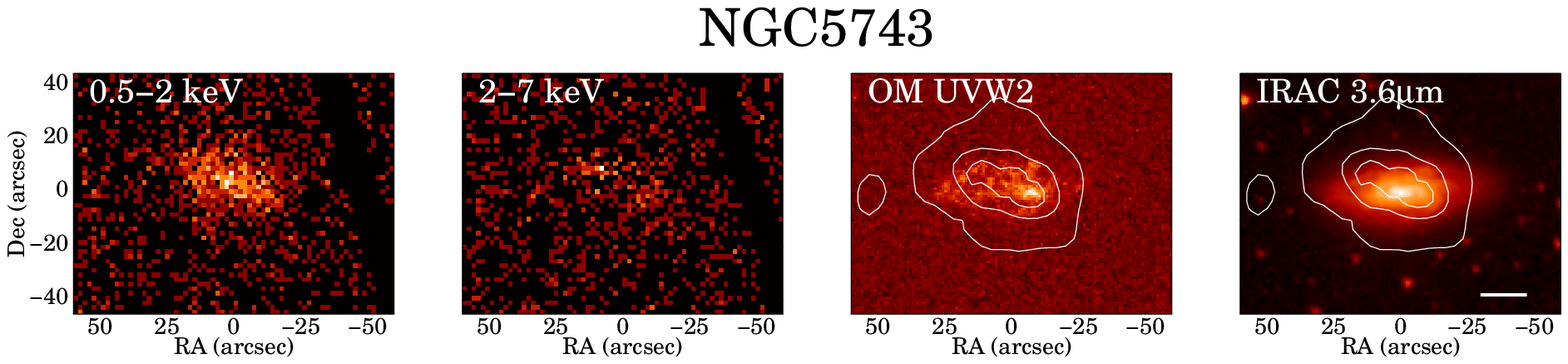}
\includegraphics[width=17cm]{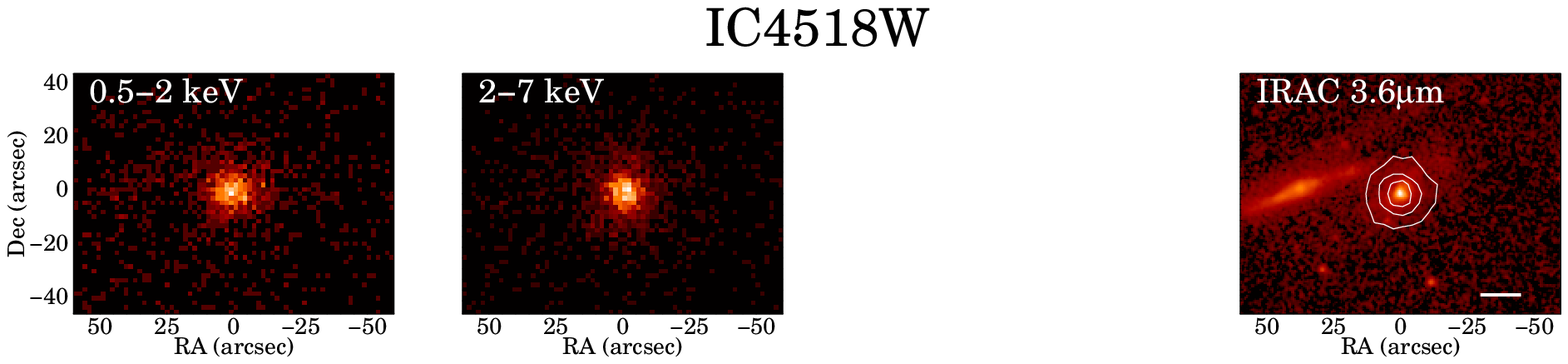}
\caption{Cont.}
\end{figure*}

\begin{figure*}
\addtocounter{figure}{-1}
\center
\includegraphics[width=17cm]{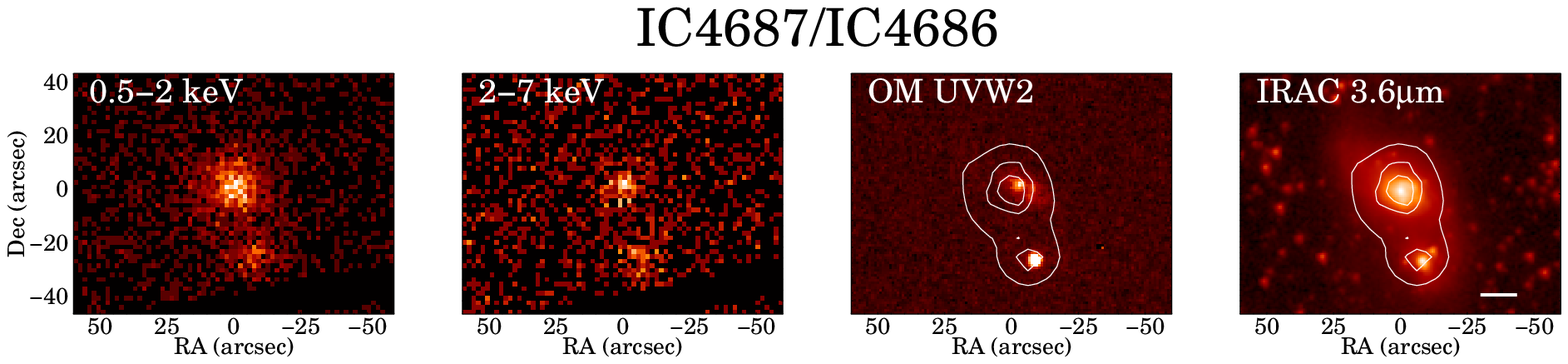}
\includegraphics[width=17cm]{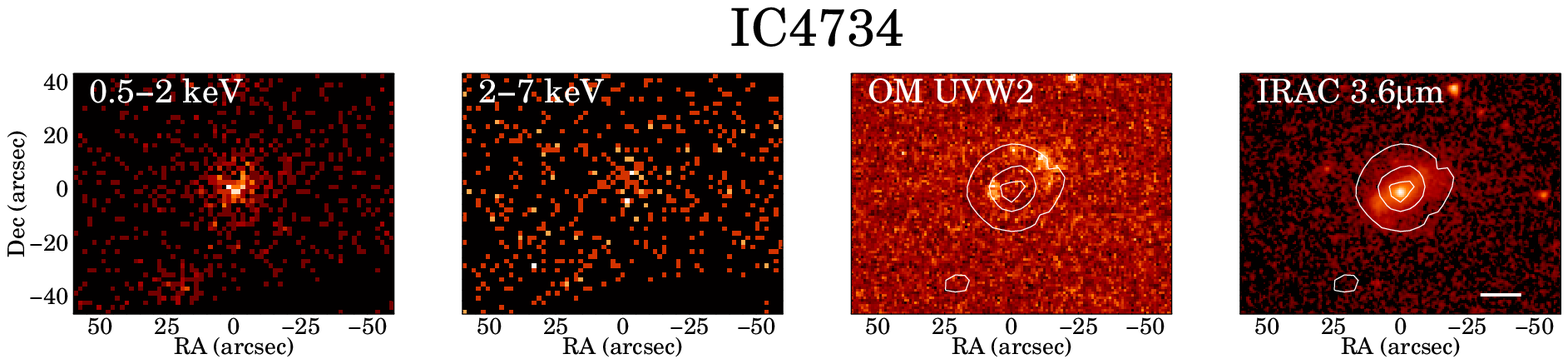}
\includegraphics[width=17cm]{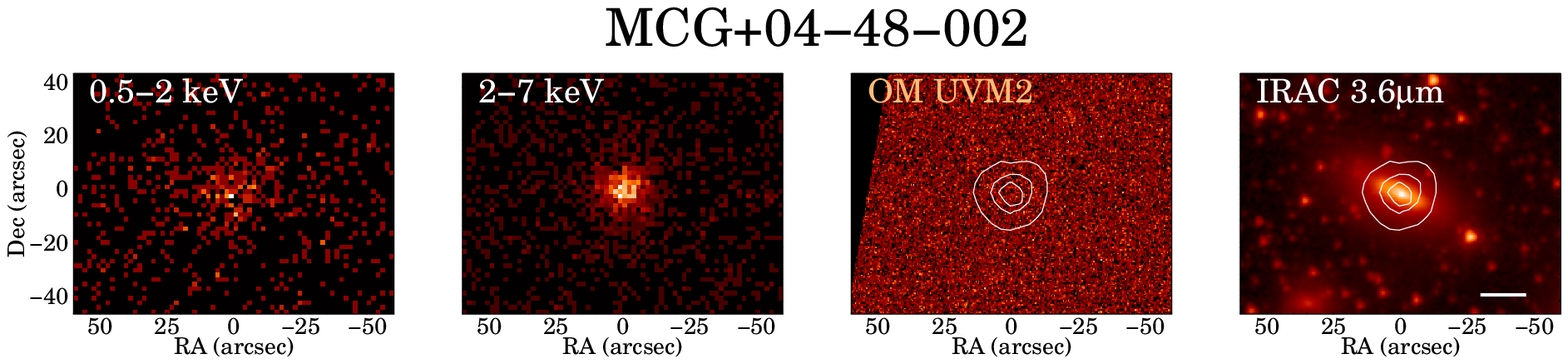}
\includegraphics[width=17cm]{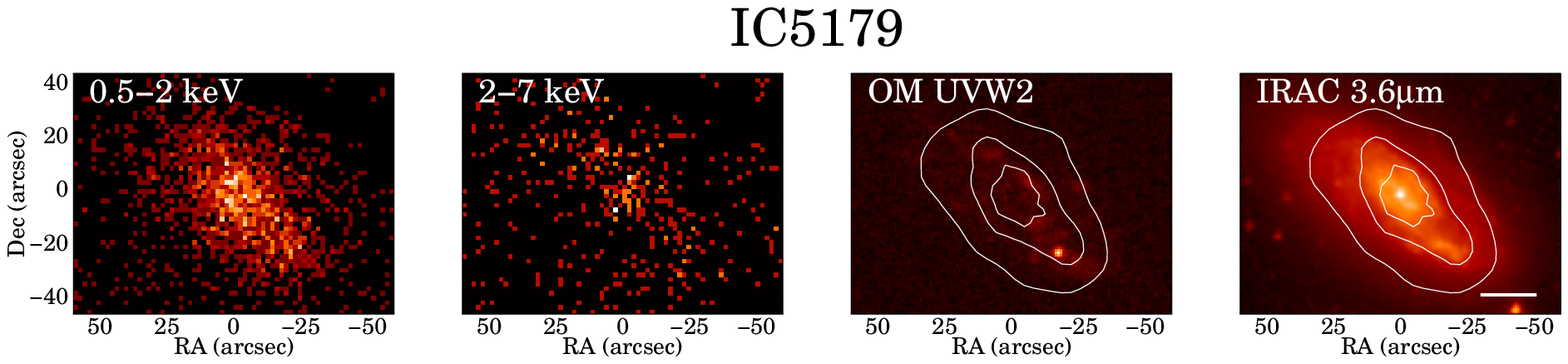}
\includegraphics[width=17cm]{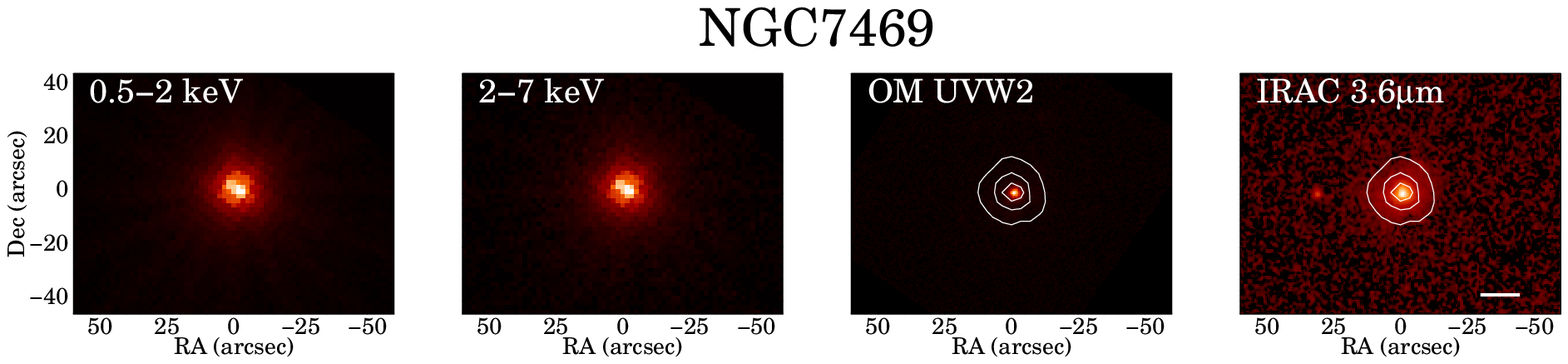}
\caption{Cont.}
\end{figure*}

\begin{figure*}
\addtocounter{figure}{-1}
\center
\includegraphics[width=17cm]{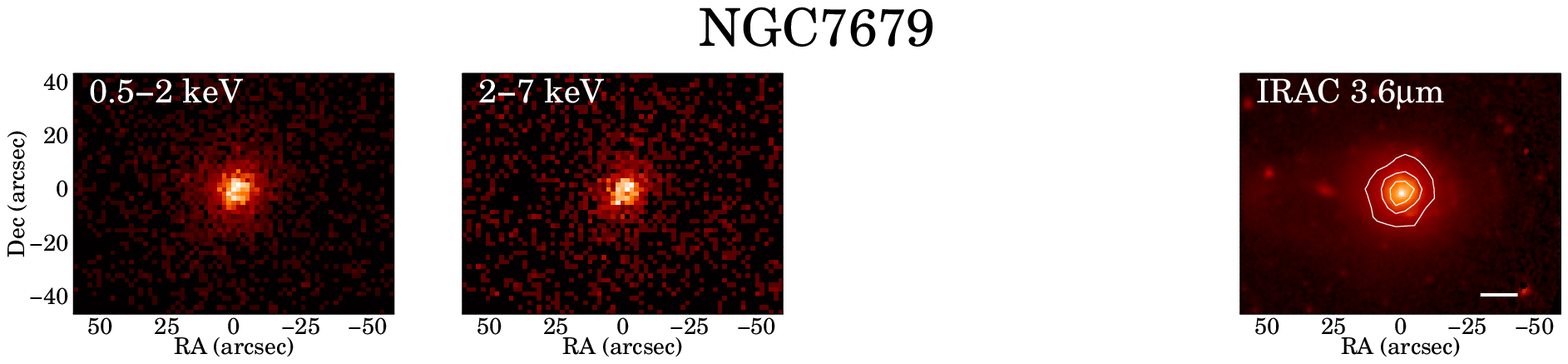}
\includegraphics[width=17cm]{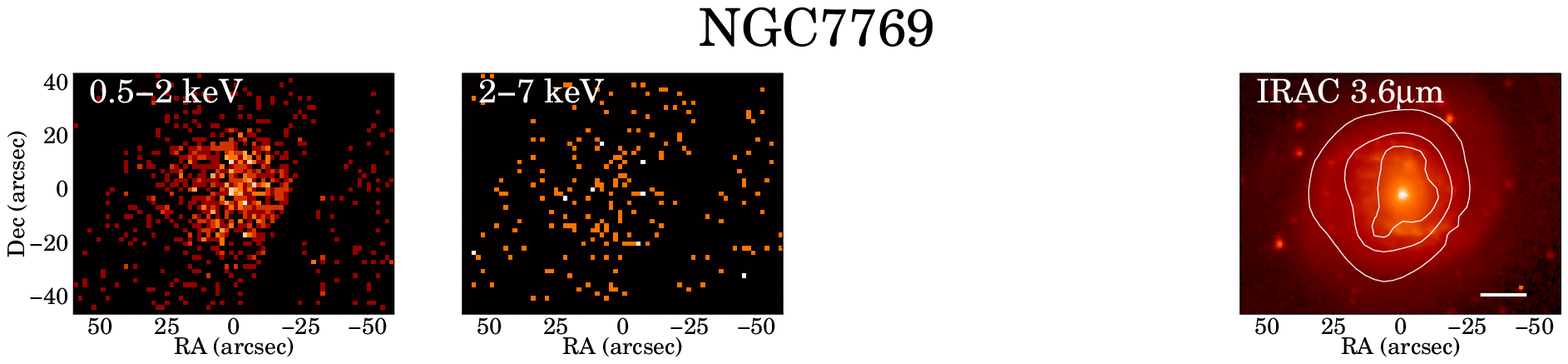}
\includegraphics[width=17cm]{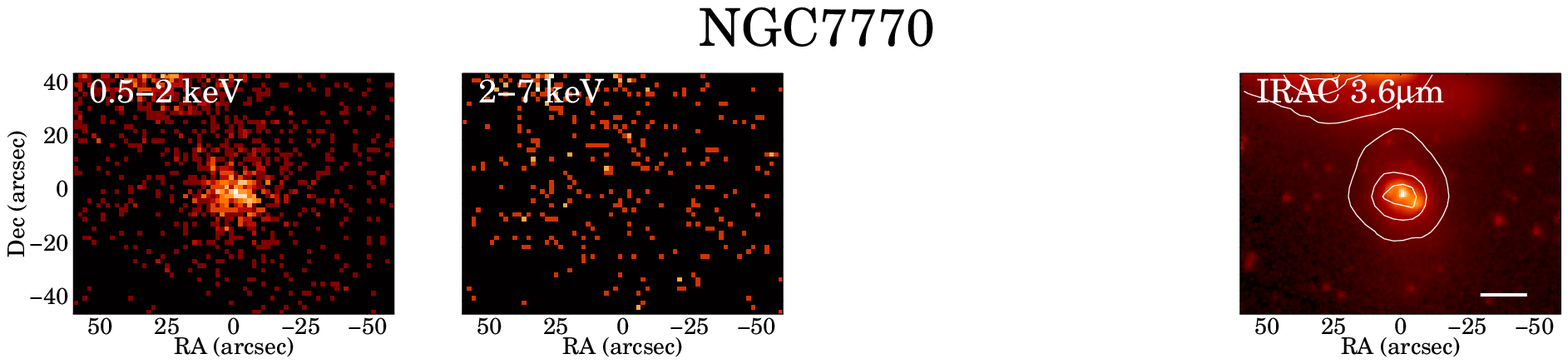}
\includegraphics[width=17cm]{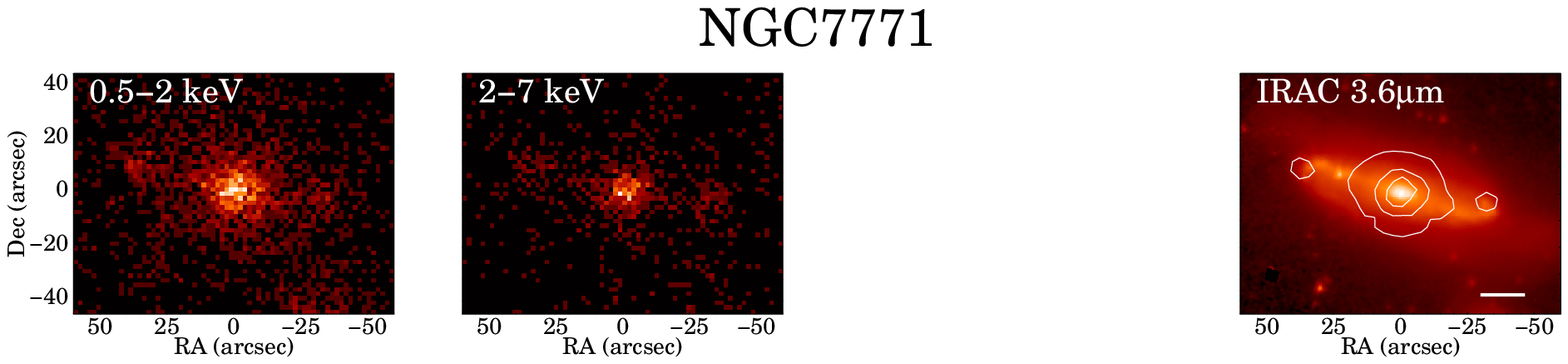}
\caption{Cont.}
\end{figure*}

\subsection{Extranuclear sources}\label{ss:ulx}

In two LIRGs observed with \xmm\ (NGC~2369 and NGC~7771) we find bright extranuclear X-ray sources that might be ultraluminous X-ray sources (ULXs). Note that we can only isolate such X-ray sources if they are located more than 0.9--2\,kpc away from the nucleus due to the spatial resolution of the images. That is, ULXs may exist in the rest of the sample within the central 0.9--2\,kpc.

The two sources located at either side of the nucleus of NGC~7771 (NGC~7771 X-1 and NGC~7771 X-2) were studied by \citet{Jenkins2005}.
The spectra of both sources are well fitted with an absorbed power-law ($\Gamma$ $=$ 1.6 and 1.7)
plus a soft component (thermal plasma at 0.3\,keV and a blackbody disk at 0.2\,keV). 
The unabsorbed luminosities ($L^{int}_{\rm 0.5-8\,keV}$) of these sources are $1.7 \pm 1.0 \times 10^{40}$\,erg s$^{-1}$ and $1.4 \pm 0.8 \times 10^{40}$\,erg s$^{-1}$.

\begin{figure}[h]
\center
\resizebox{\hsize}{!}{\includegraphics{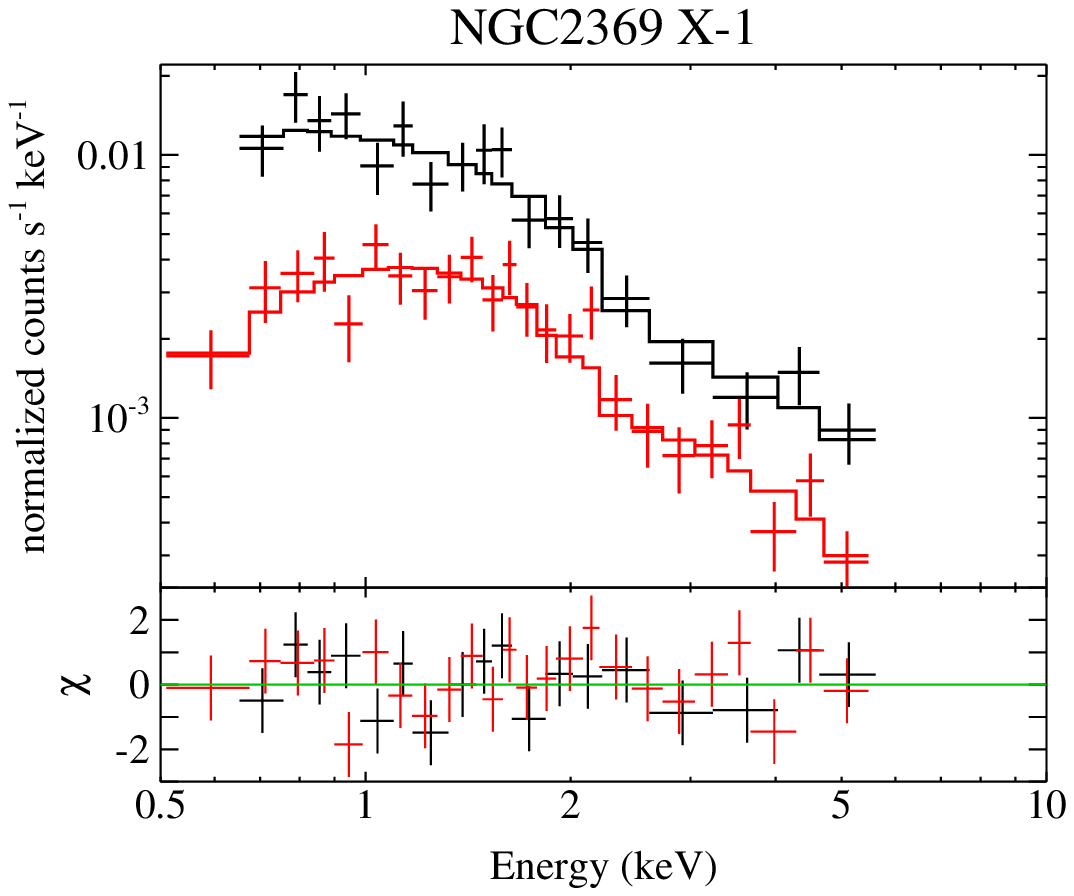}}
\caption{EPIC pn (black) and combined EPIC MOS (red) 0.5--10\,keV spectrum of NGC~2369 X-1 together with the spectral model consisting of an absorbed power-law (see Sect. \ref{ss:ulx}).}
\label{fig_ulx}
\end{figure}

The ULX candidate in NGC~2369 is located 25\,arcsec ($\sim$5\,kpc) to the north of the nucleus (NGC~2369 X-1 at R.A. (J2000) $=$ 7:16:38.5 Decl. (J2000) $=$ --62:20:16).
An absorbed power-law model reproduces well the observed spectrum of this source ($\chi^2\slash {\rm dof} = 31\slash 42$, Fig. \ref{fig_ulx}) and implies an intrinsic X-ray luminosity $L^{int}_{\rm 0.5-8\,keV} = 3.2 \pm 0.6 \times 10^{40}$\,erg s$^{-1}$. This is one order of magnitude higher than the ULX luminosity threshold ($L^{int}_{\rm 0.5-8\,keV} > 10^{39}$\,erg s$^{-1}$) and comparable to the luminosities measured for other ULXs. The parameters of the model ($\Gamma= 1.6 \pm 0.2$ and $N_{\rm H}= 1.3 \pm 0.5 \times 10^{21}\,{\rm cm}^{-2}$) are also similar to those obtained for other ULXs \citep{Swartz2004}.
The angular resolution of the \xmm\ data does not allow us to determine if this emission comes from a single source (i.e., ULX or a background AGN) or, conversely, if it is the integrated emission from multiple sources. NGC~2369 X-1 is coincident with an extended UV emitting region that seems to be located in the spiral arms of NGC~2369 (Fig. \ref{fig_ximage}). Therefore the background AGN possibility is unlikely. The SFR of this region derived from their UV and IR luminosities\footnote{The UV (2120\,\AA) and IR fluxes of the NGC~2369 X-1 region are: $F^{\rm corr}_{\rm UV}=0.4$\,mJy and $F_{\rm 24\mu m}=0.14$\,Jy.} is 0.9\,$M_{\odot}$\,yr$^{-1}$ (see Sect. \ref{ss:sfr_sm}). Thus the expected hard X-ray luminosity from star-formation ($\sim$2.3$\times 10^{39}$\,erg s$^{-1}$, see Sect. \ref{s:sfr_xrayir}) is $\sim$10 times lower than the observed luminosity. Moreover no excess soft X-ray emission from hot ($\sim$0.7\,keV) gas, which is common in star-forming regions, is detected in its spectra (Fig. \ref{fig_ulx}). All of these pieces of evidence suggest that the X-ray emission of NGC~2369 X-1 is dominated by a single bright source.

\section{Spectral analysis of the \xmm\ data}\label{s:spec_analysis}

\begin{figure*}
\center
\includegraphics[width=0.3\hsize]{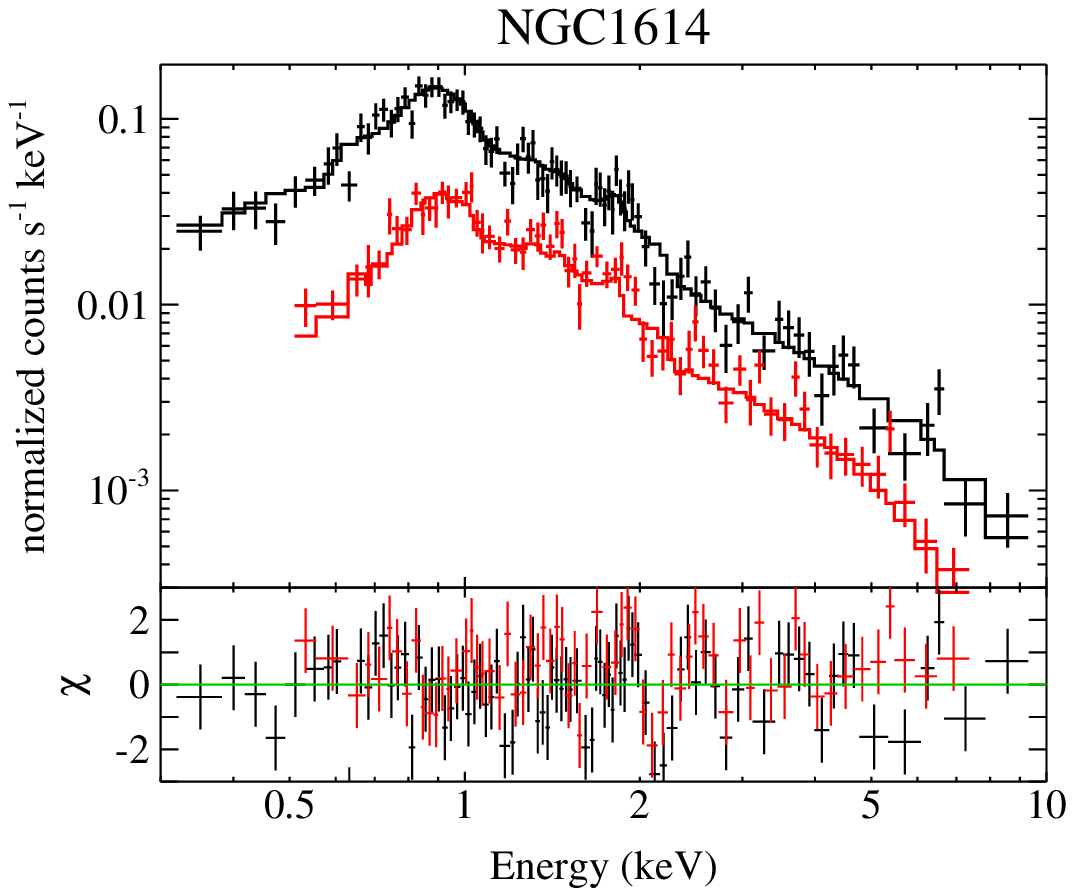}
\includegraphics[width=0.3\hsize]{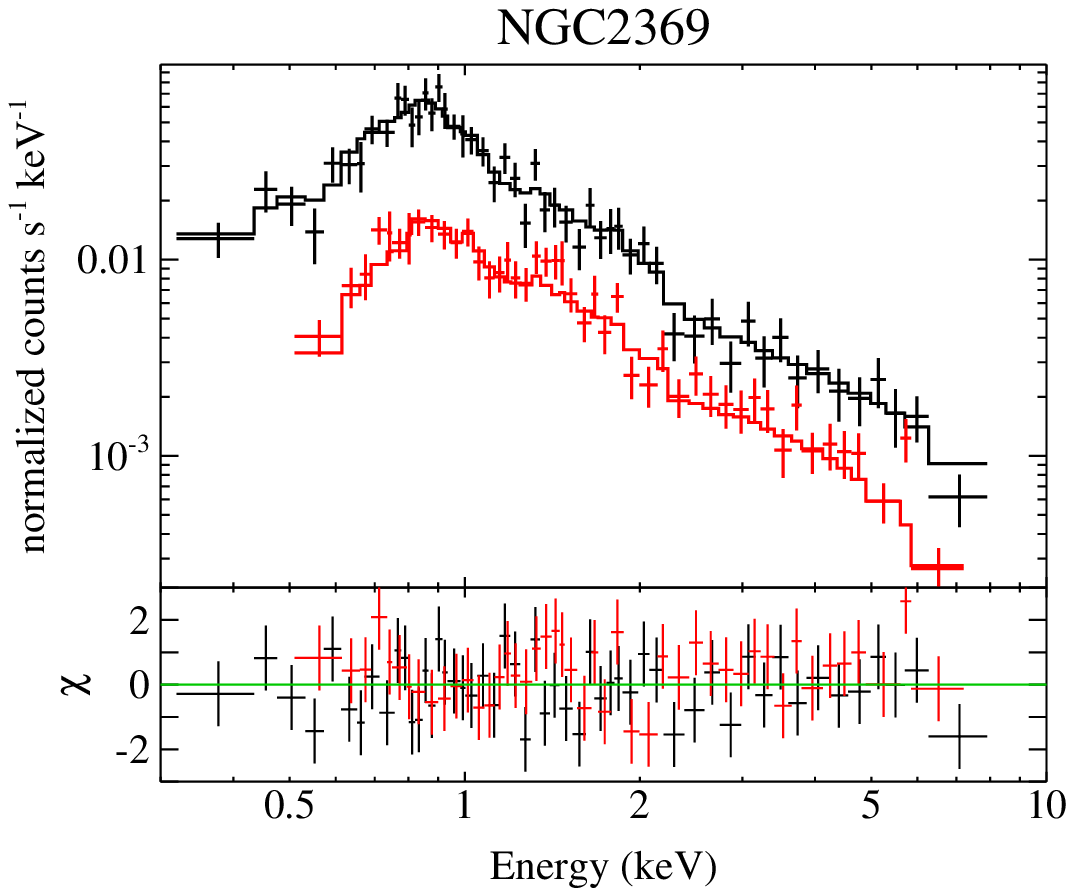} 
\includegraphics[width=0.3\hsize]{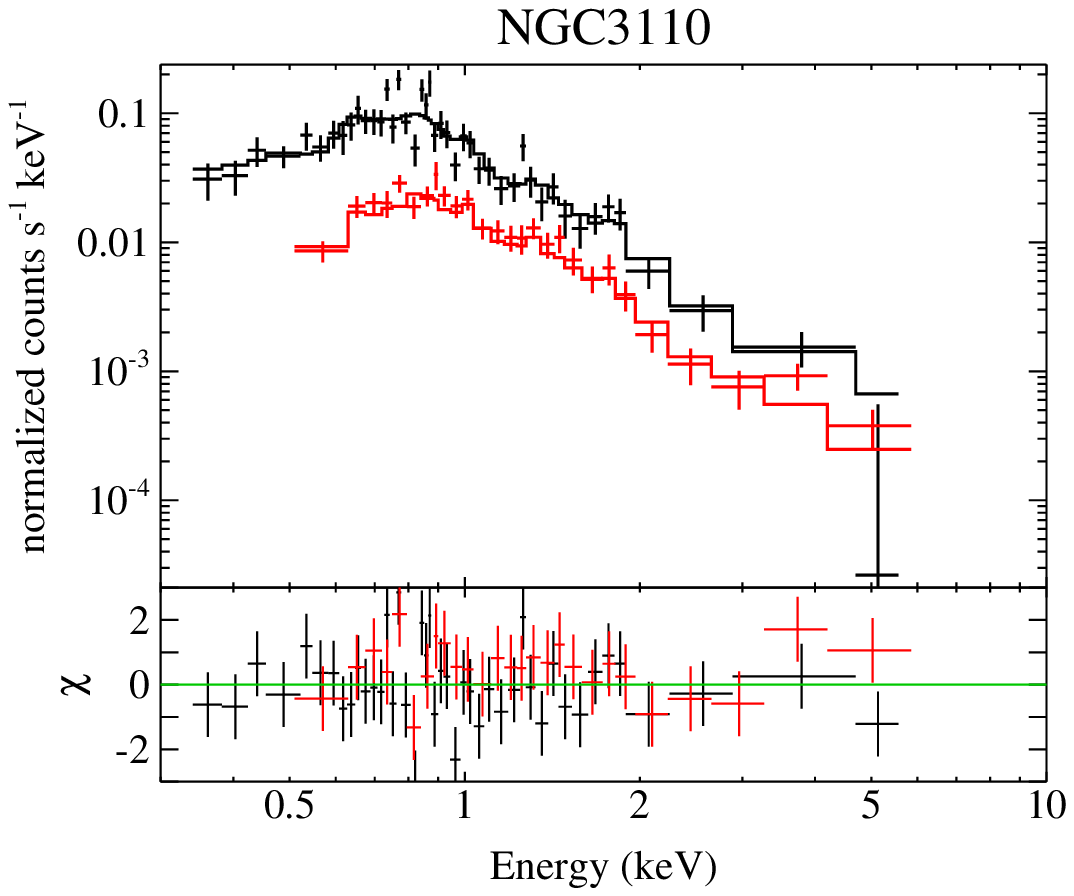}
\includegraphics[width=0.3\hsize]{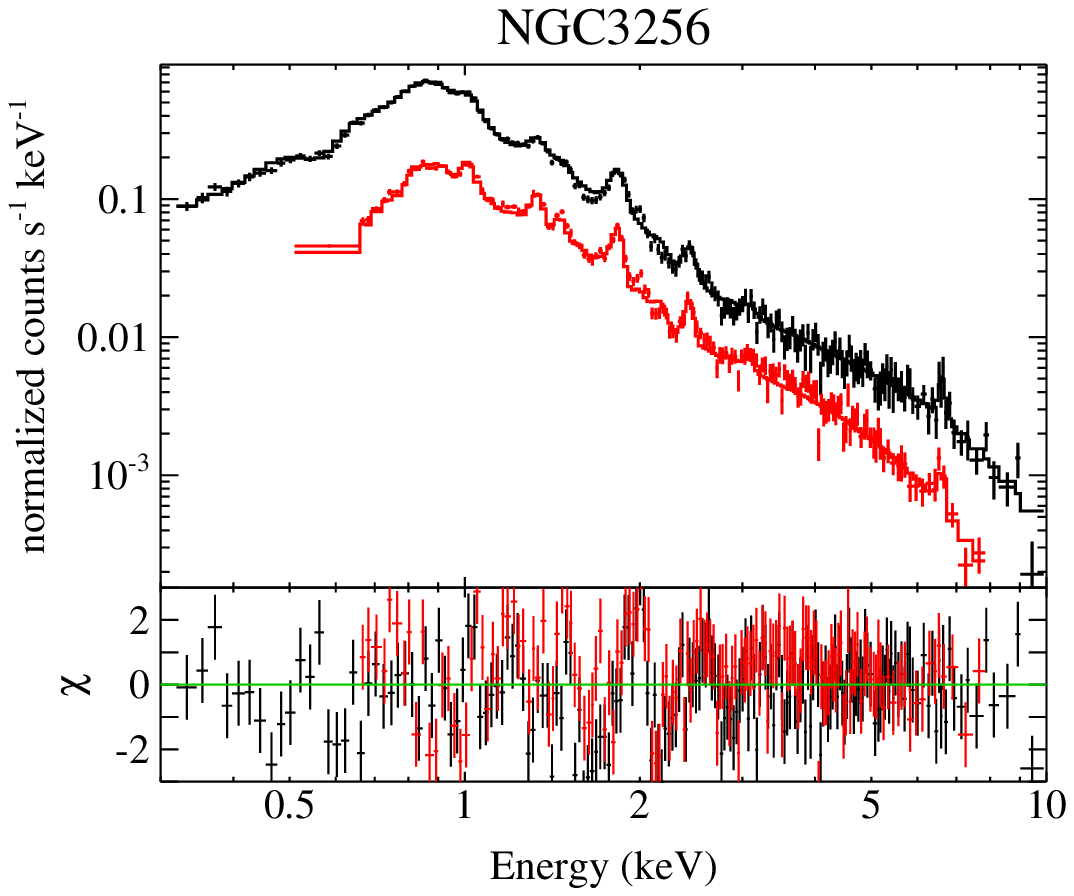}
\includegraphics[width=0.3\hsize]{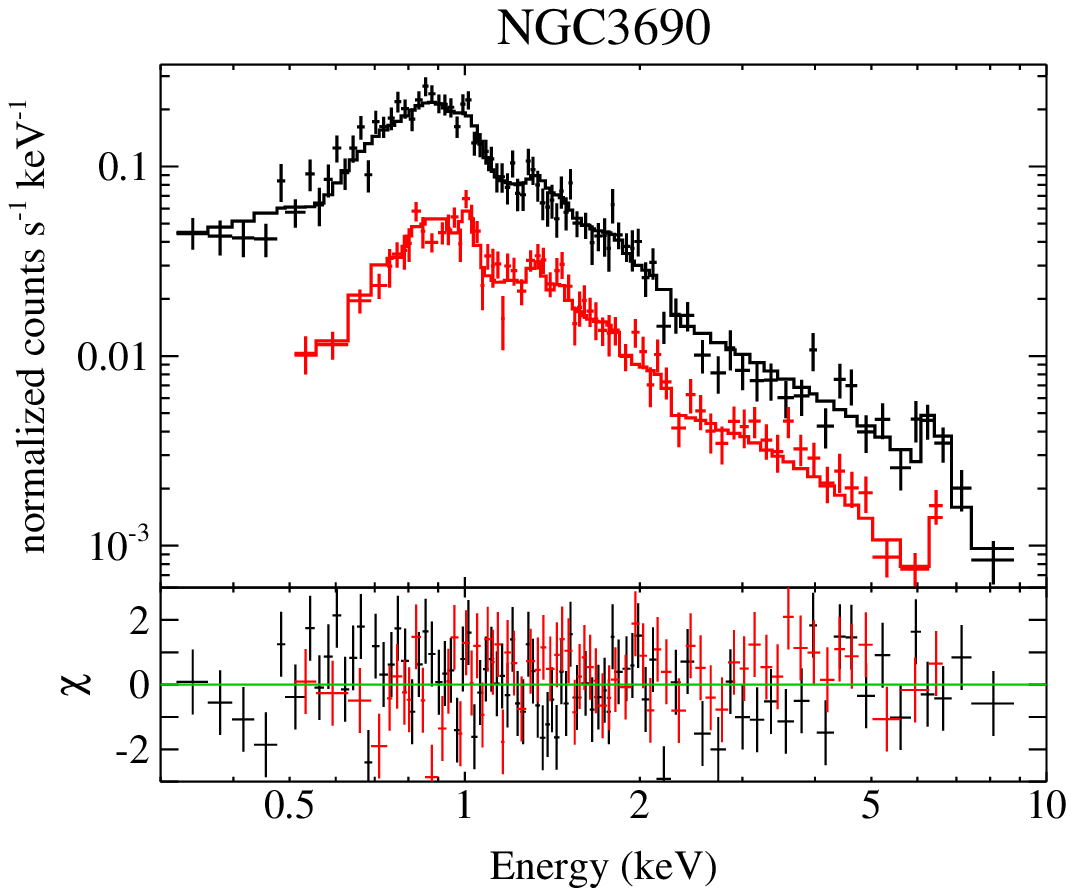} 
\includegraphics[width=0.3\hsize]{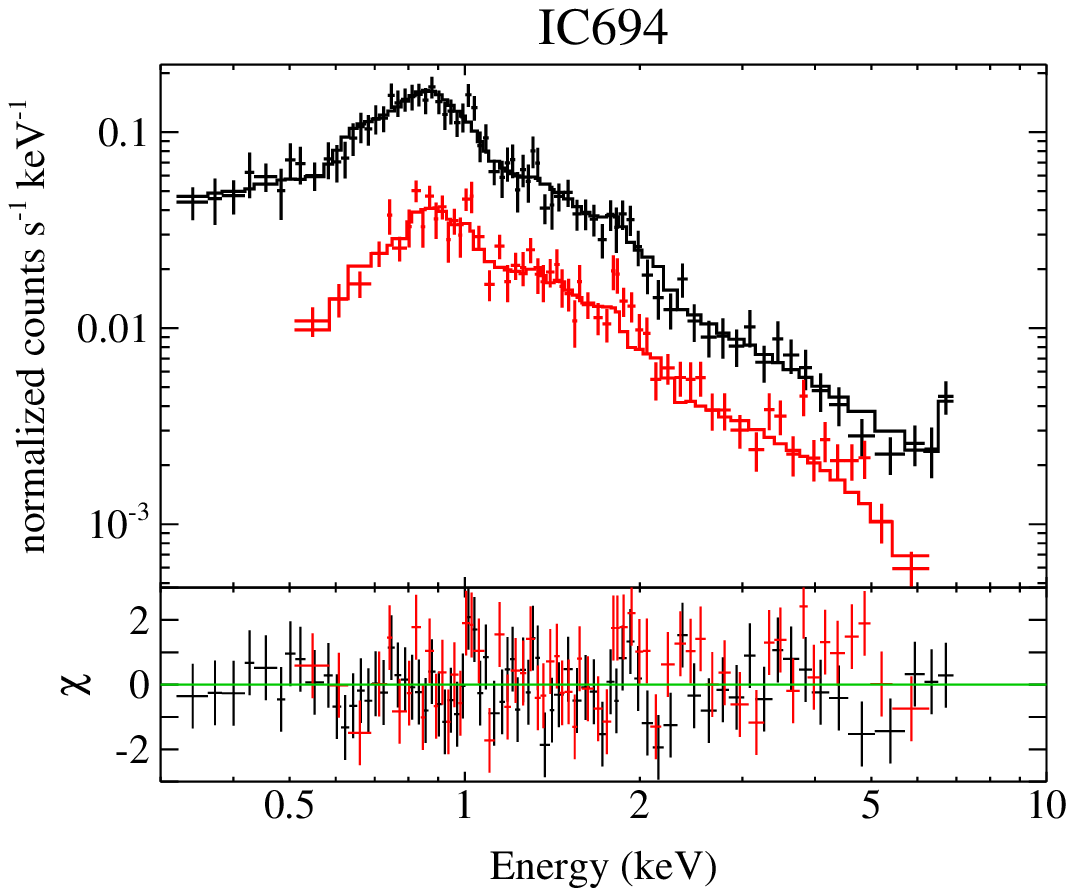}
\includegraphics[width=0.3\hsize]{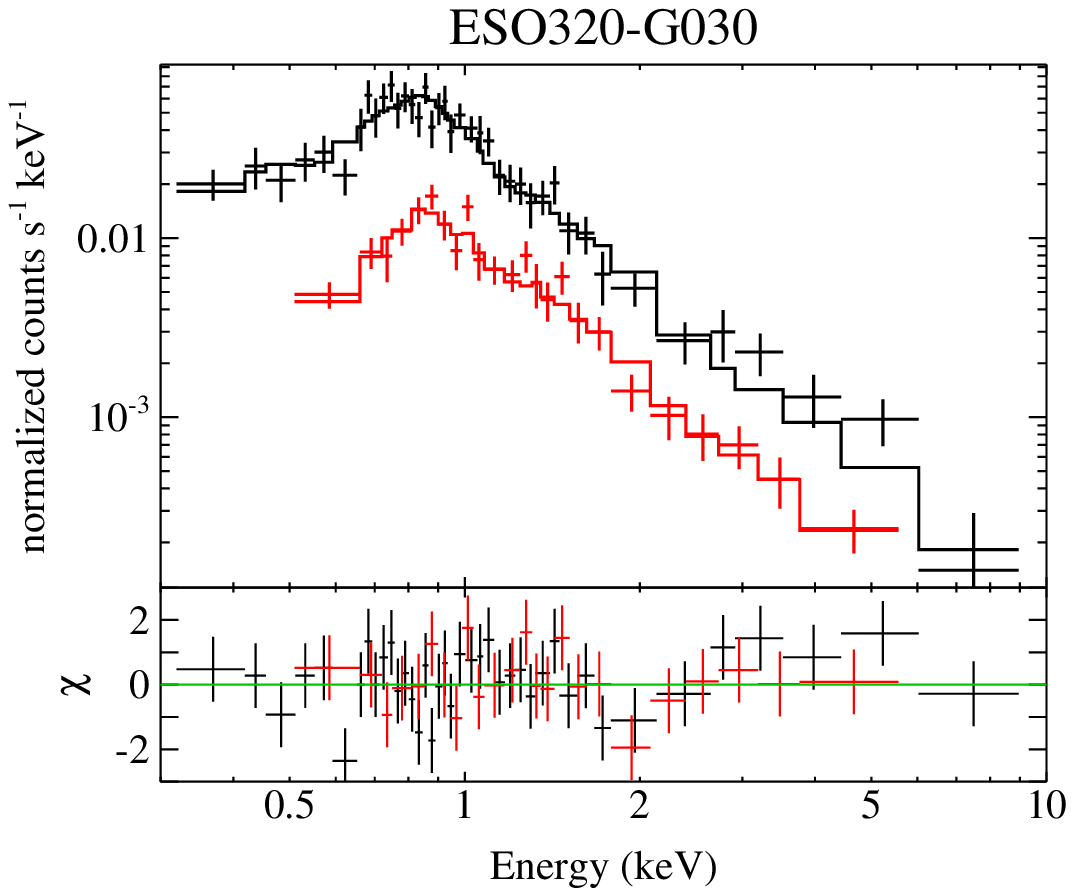} 
\includegraphics[width=0.3\hsize]{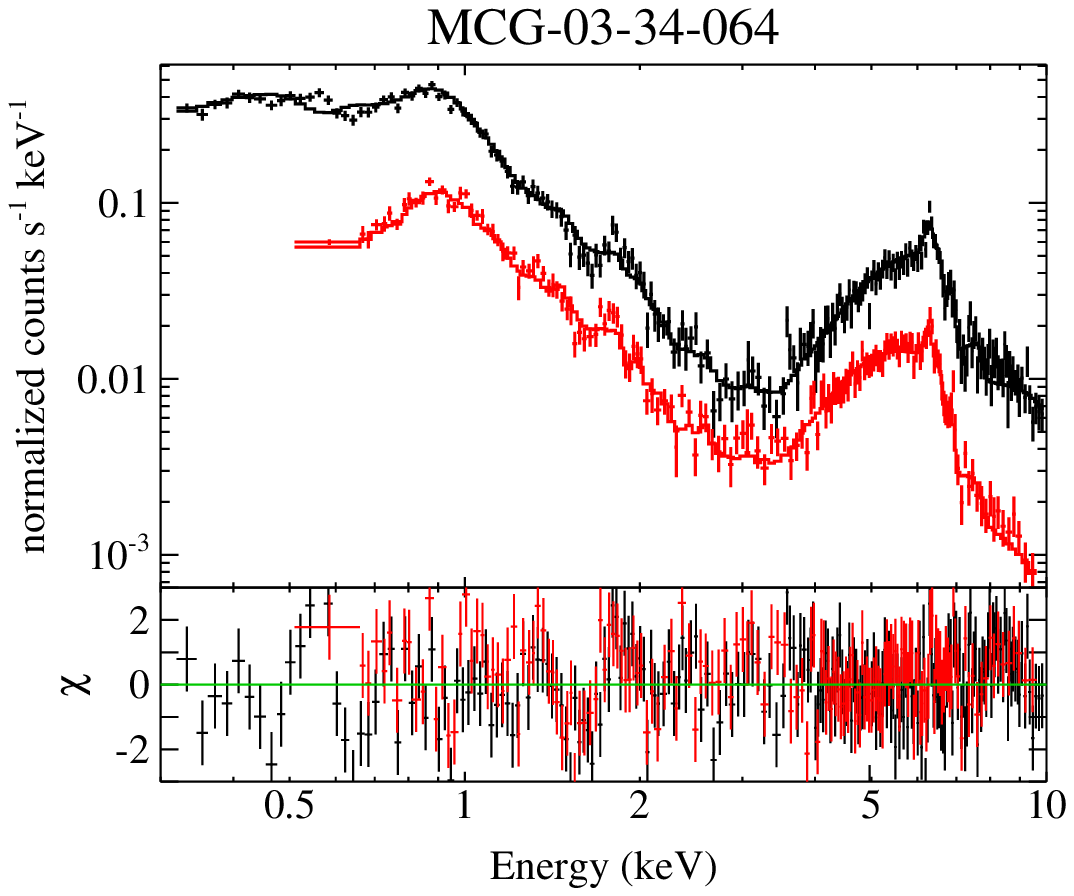}
\includegraphics[width=0.3\hsize]{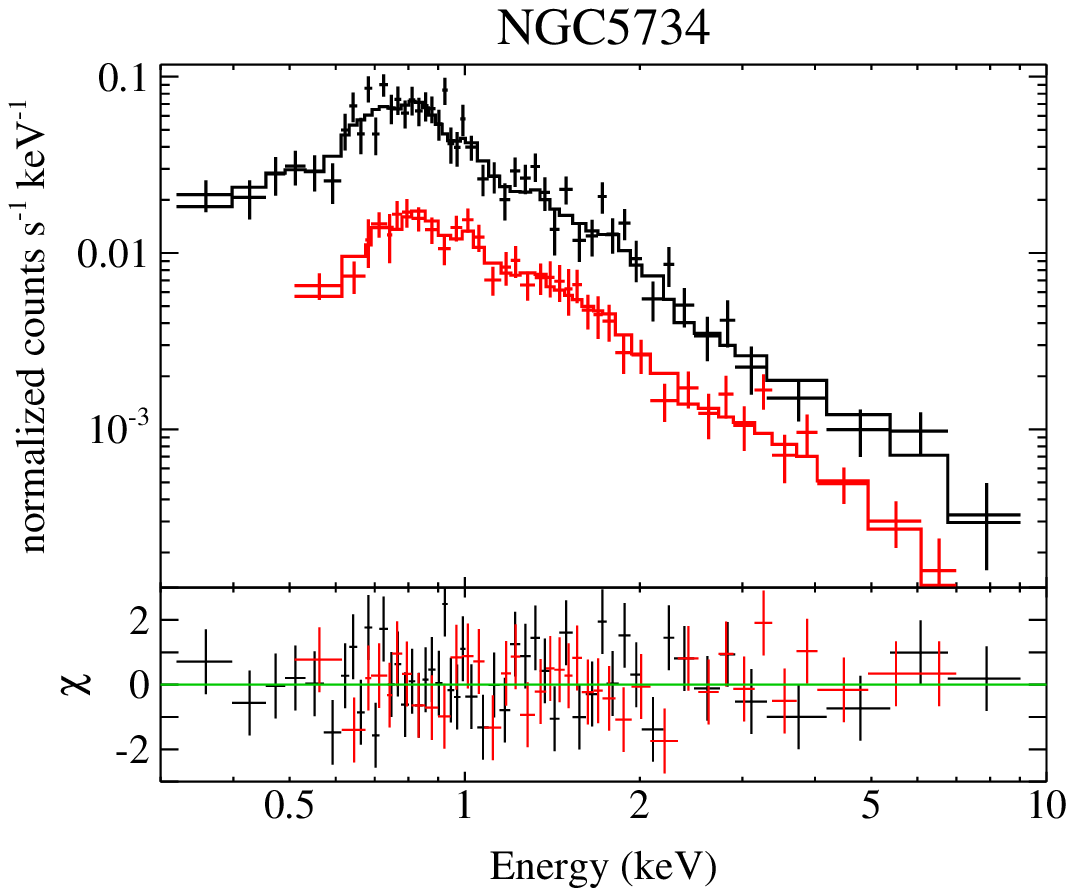}
\includegraphics[width=0.3\hsize]{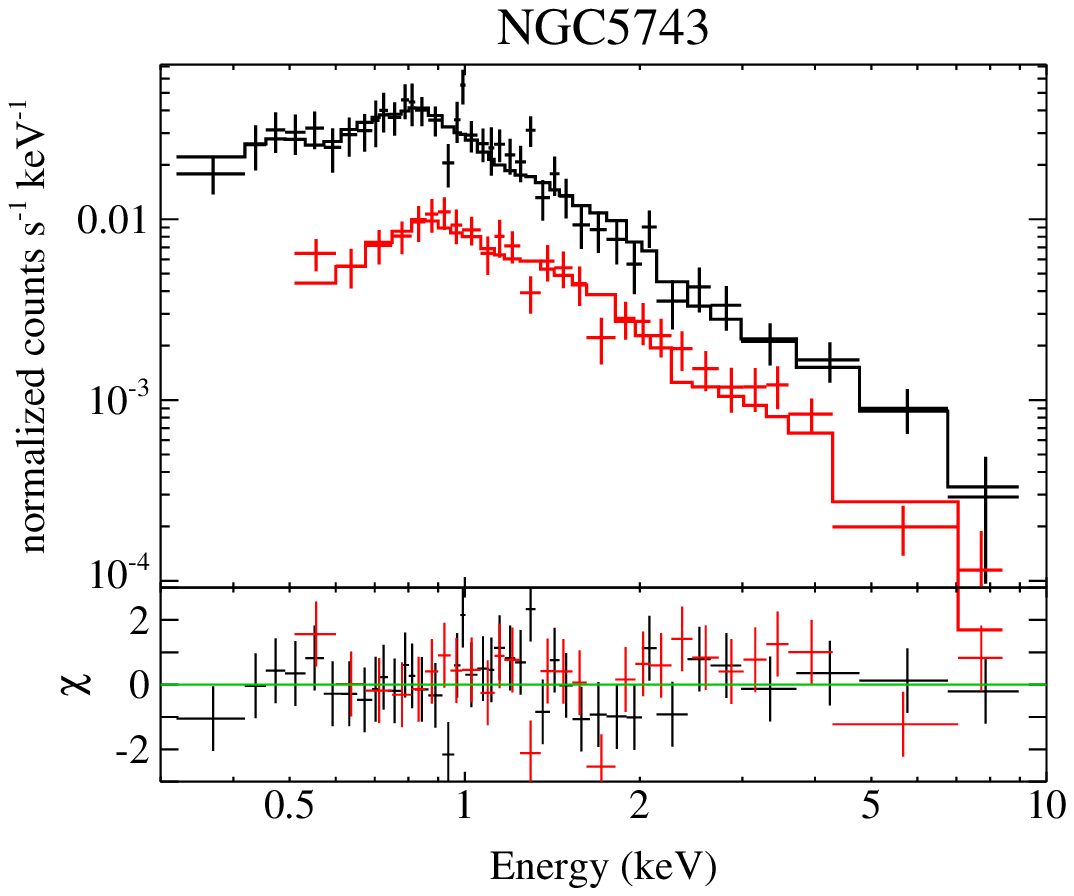}
\includegraphics[width=0.3\hsize]{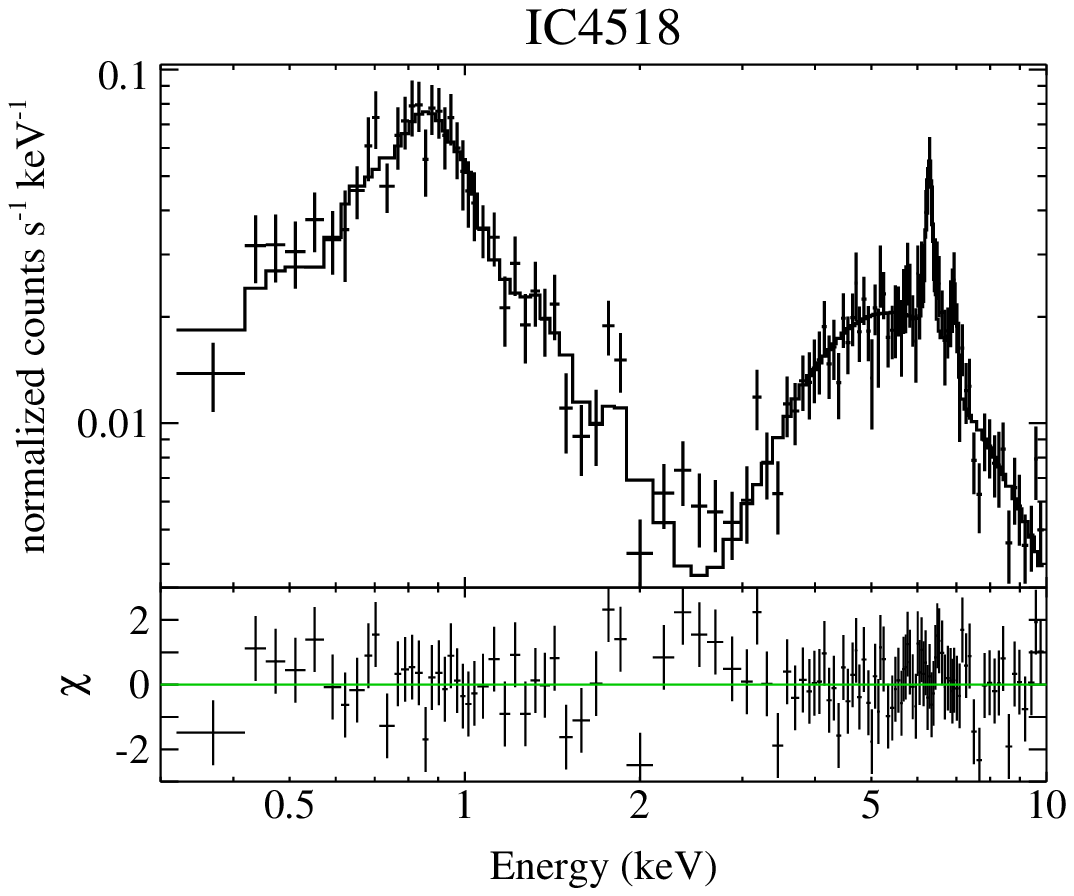}
\includegraphics[width=0.3\hsize]{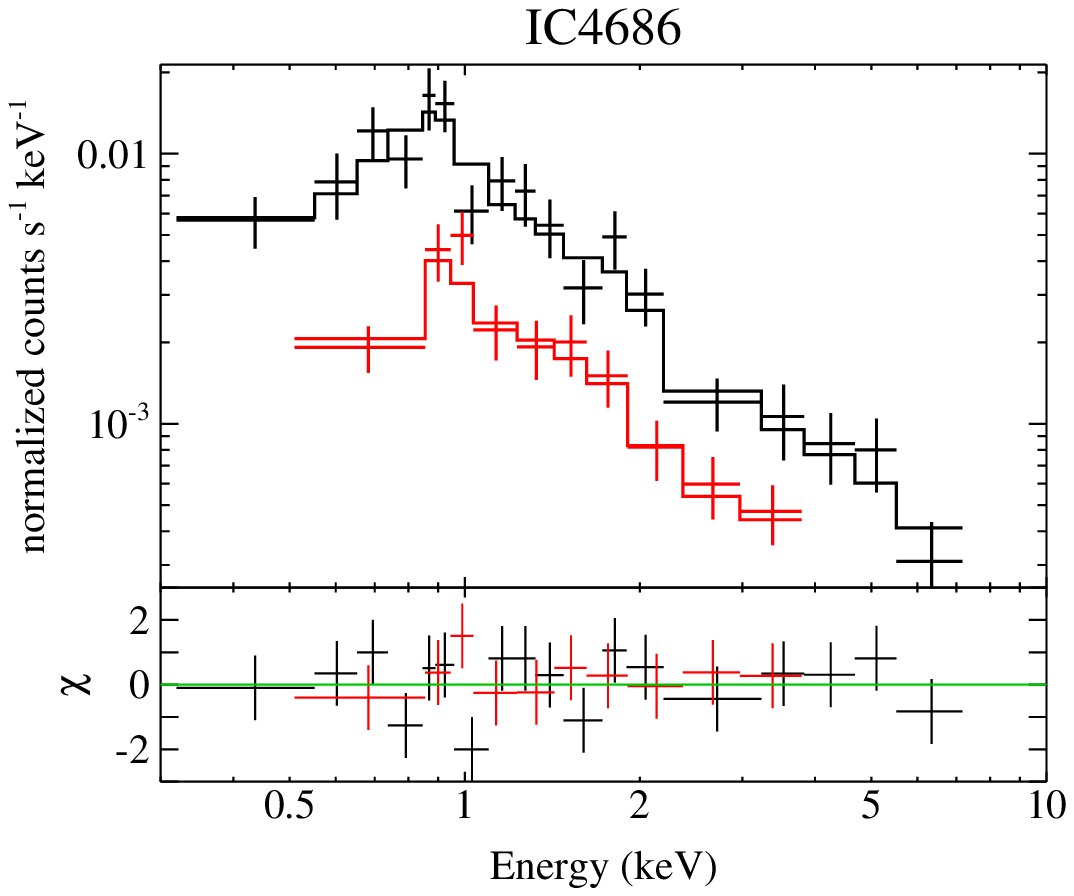}
\caption{Observed EPIC pn (black) and combined EPIC MOS (red) 0.3-10\,keV spectra, best-fitting model and residuals of the LIRGs observed by \xmm. }
\label{fig_xspectra}
\end{figure*}

\begin{figure*}
\addtocounter{figure}{-1}
\center
\includegraphics[width=0.3\hsize]{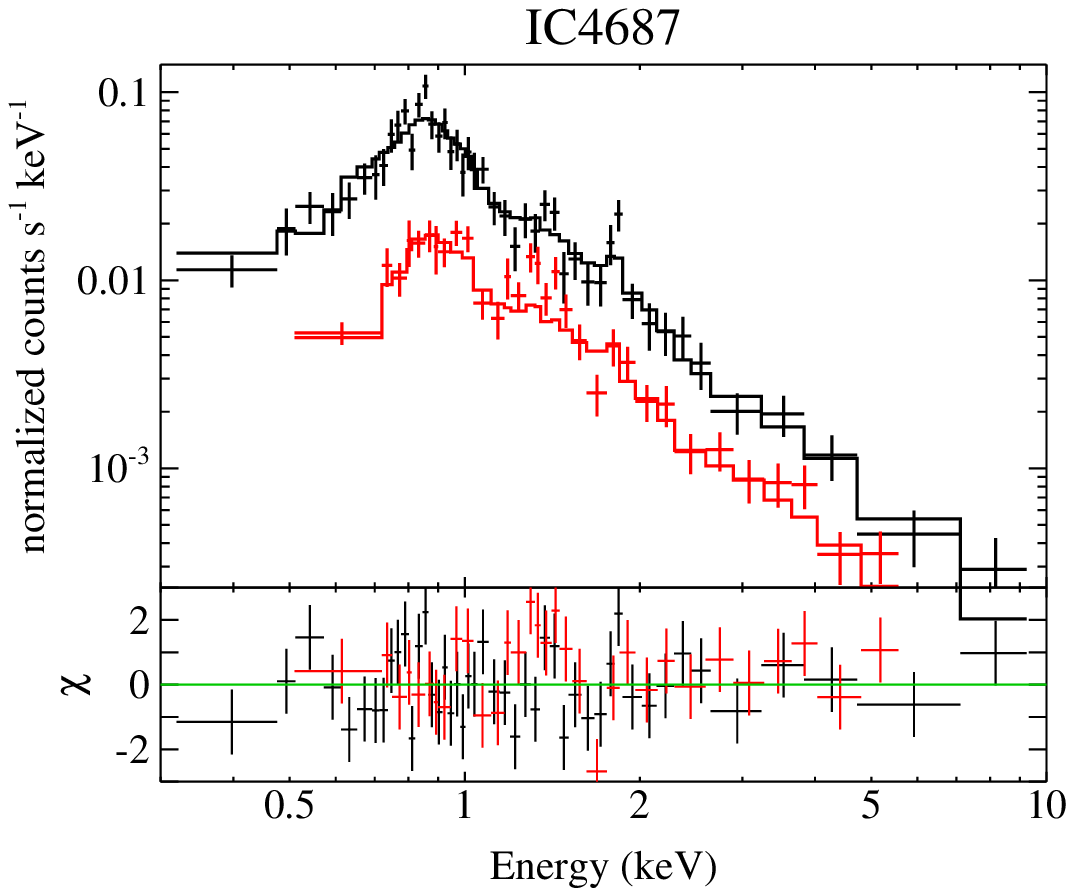}
\includegraphics[width=0.3\hsize]{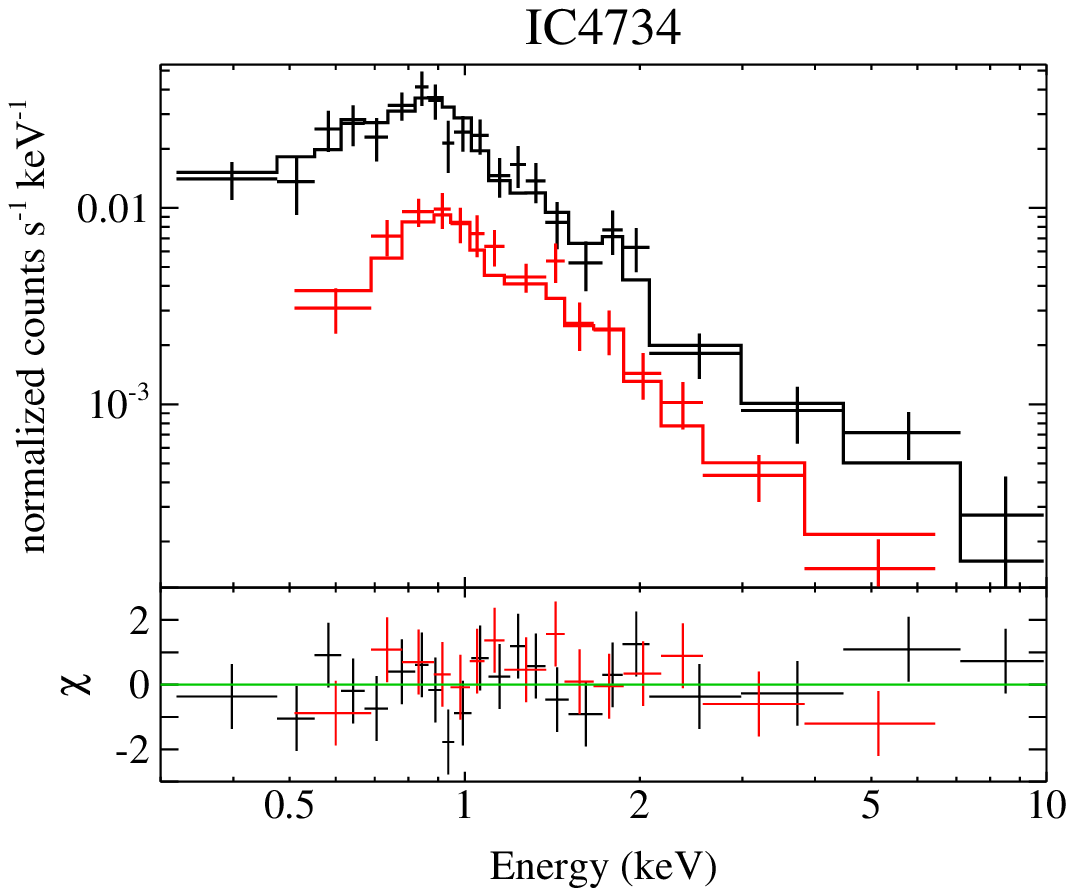}
\includegraphics[width=0.3\hsize]{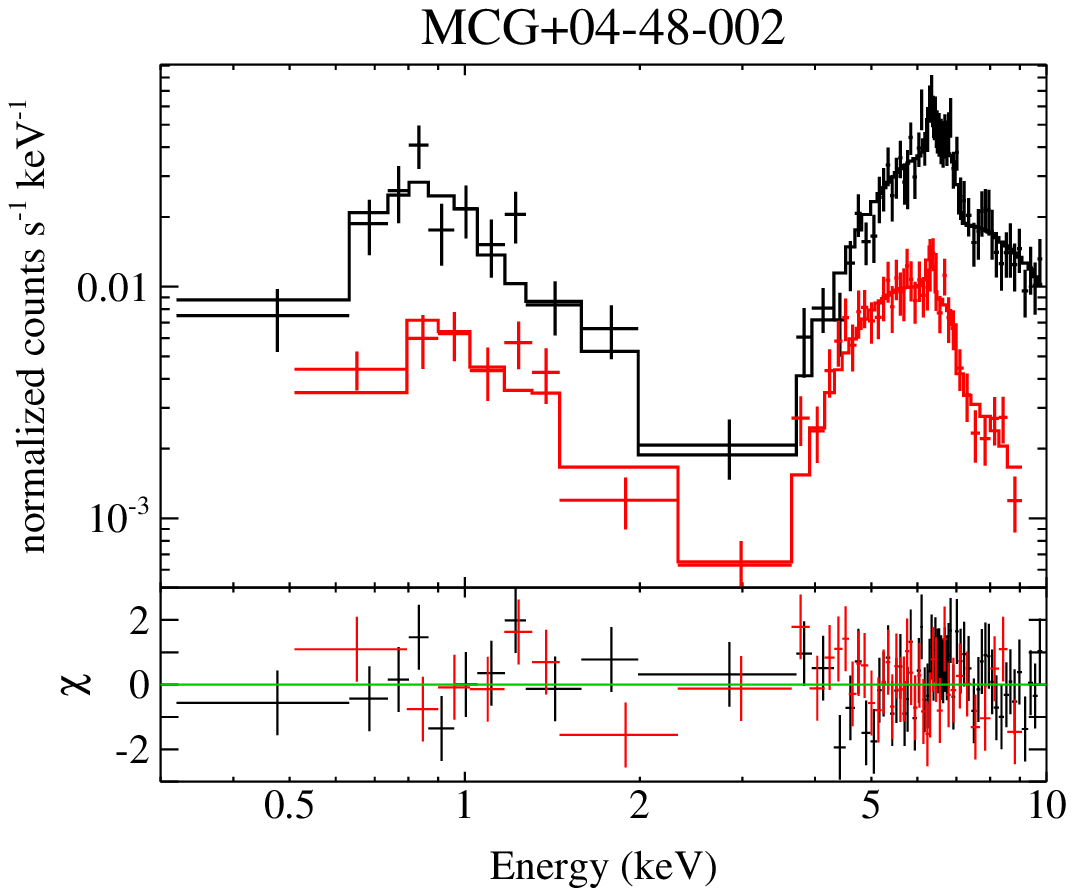}
\includegraphics[width=0.3\hsize]{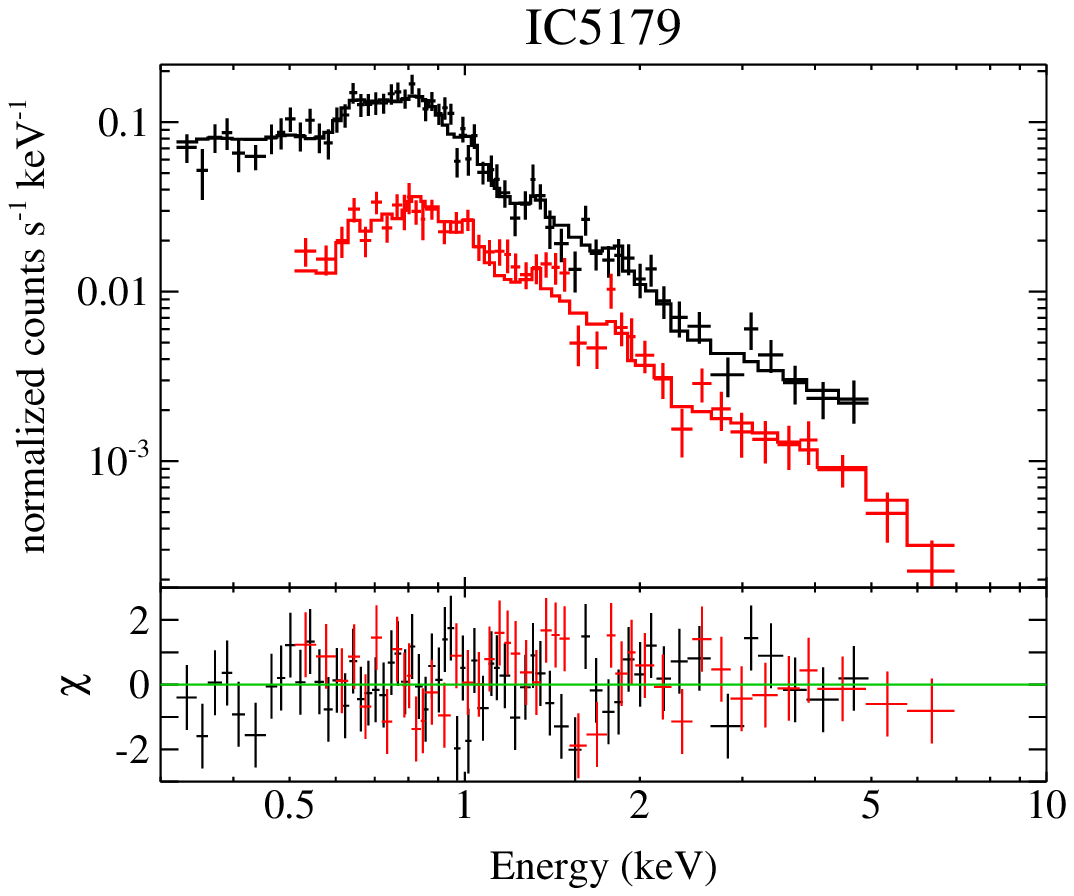}
\includegraphics[width=0.3\hsize]{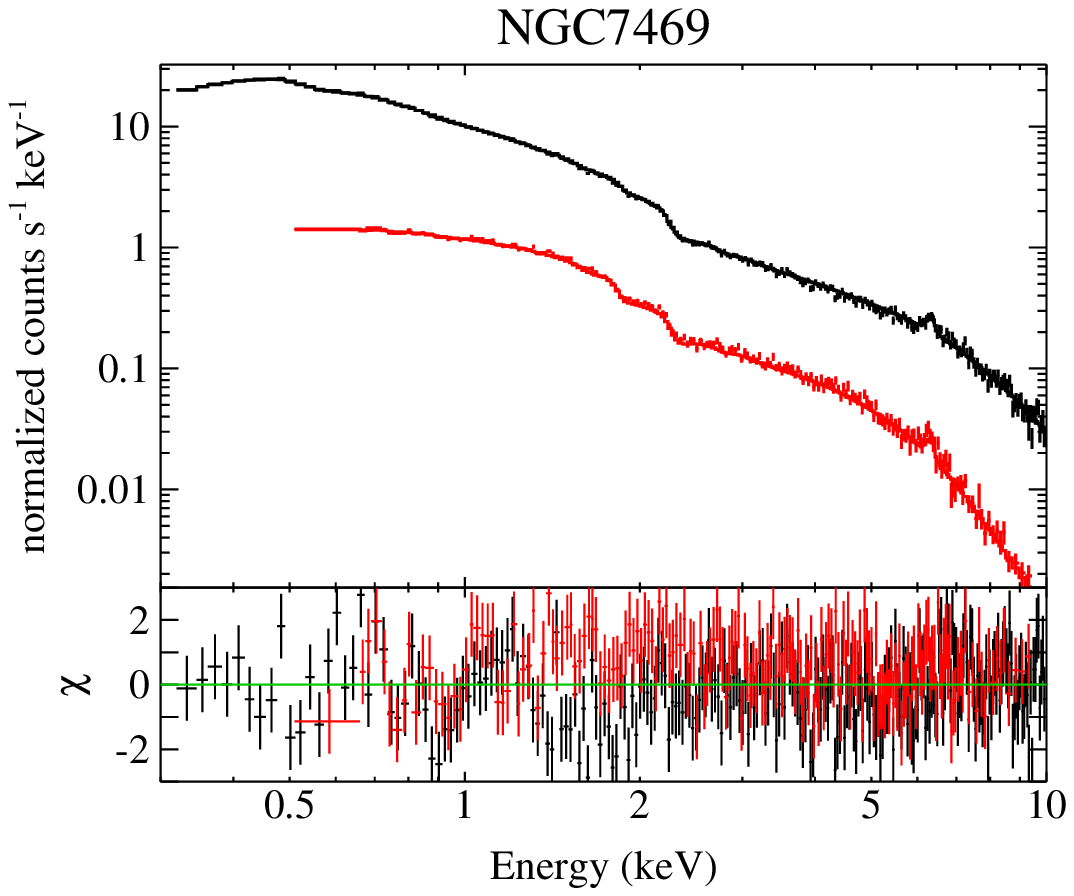}
\includegraphics[width=0.3\hsize]{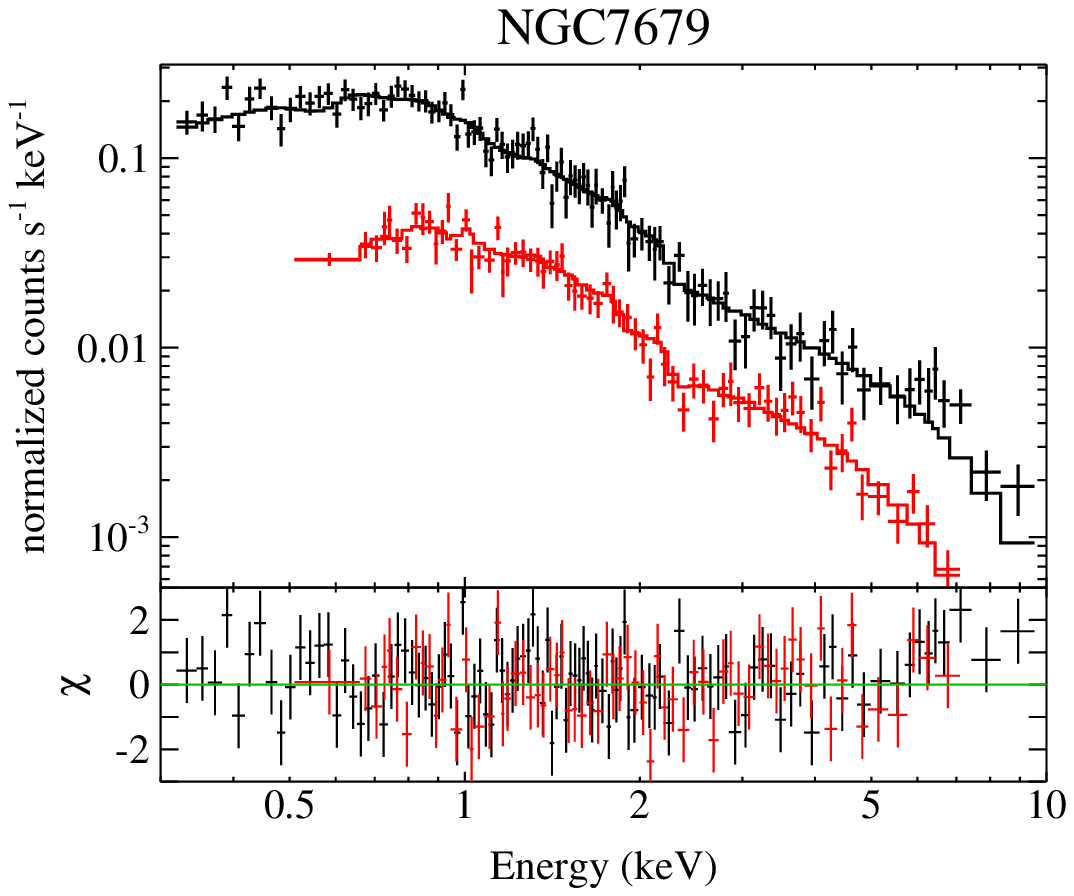}
\includegraphics[width=0.3\hsize]{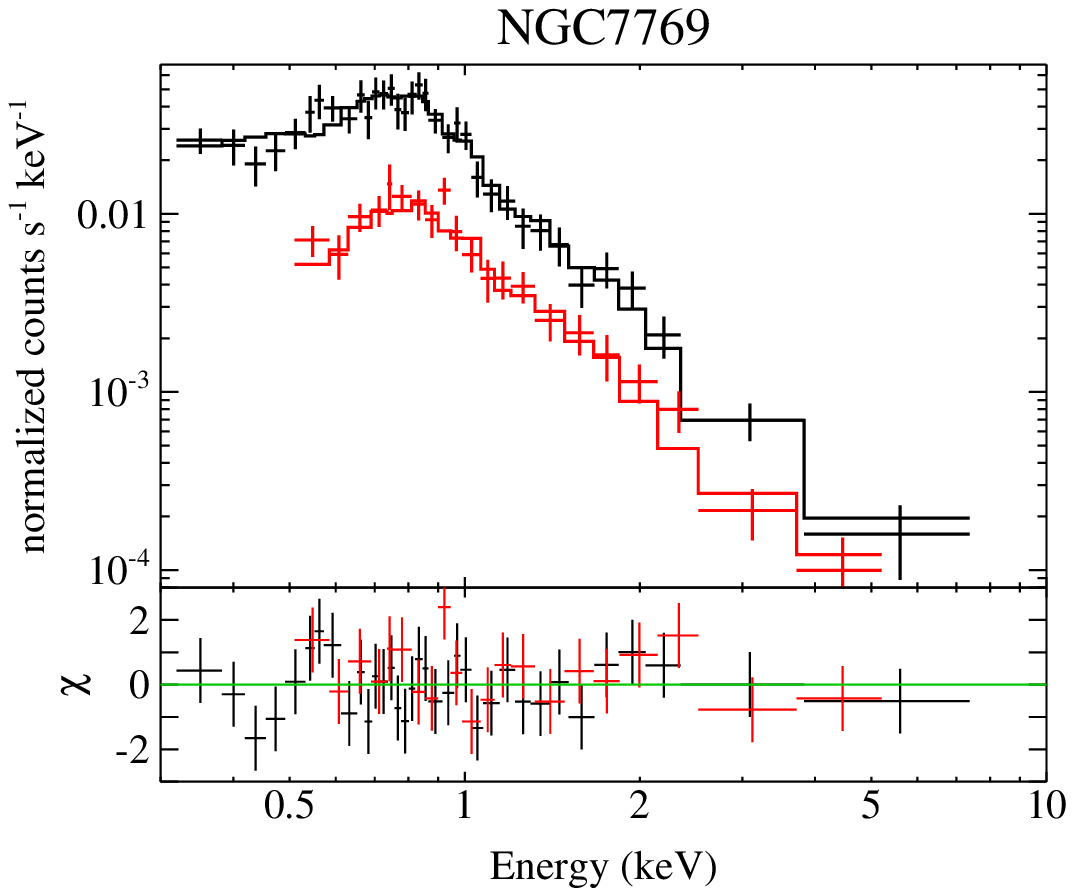}
\includegraphics[width=0.3\hsize]{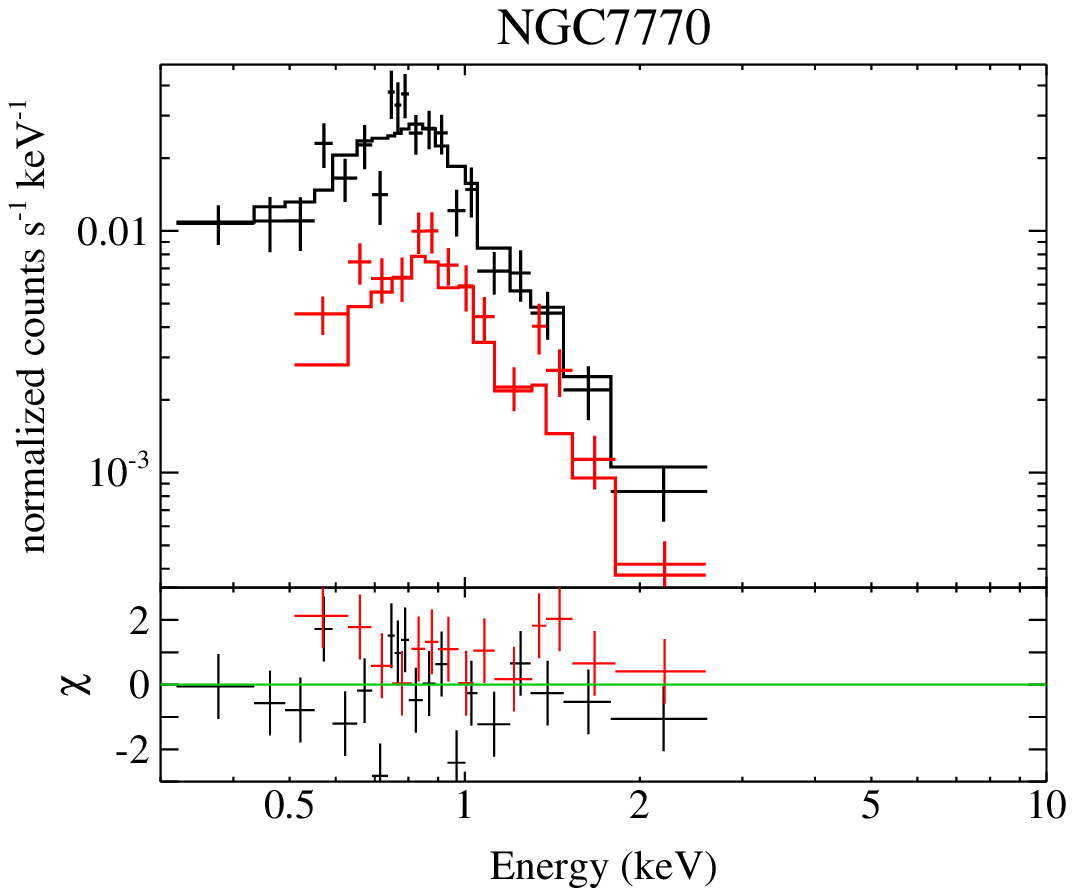}
\includegraphics[width=0.3\hsize]{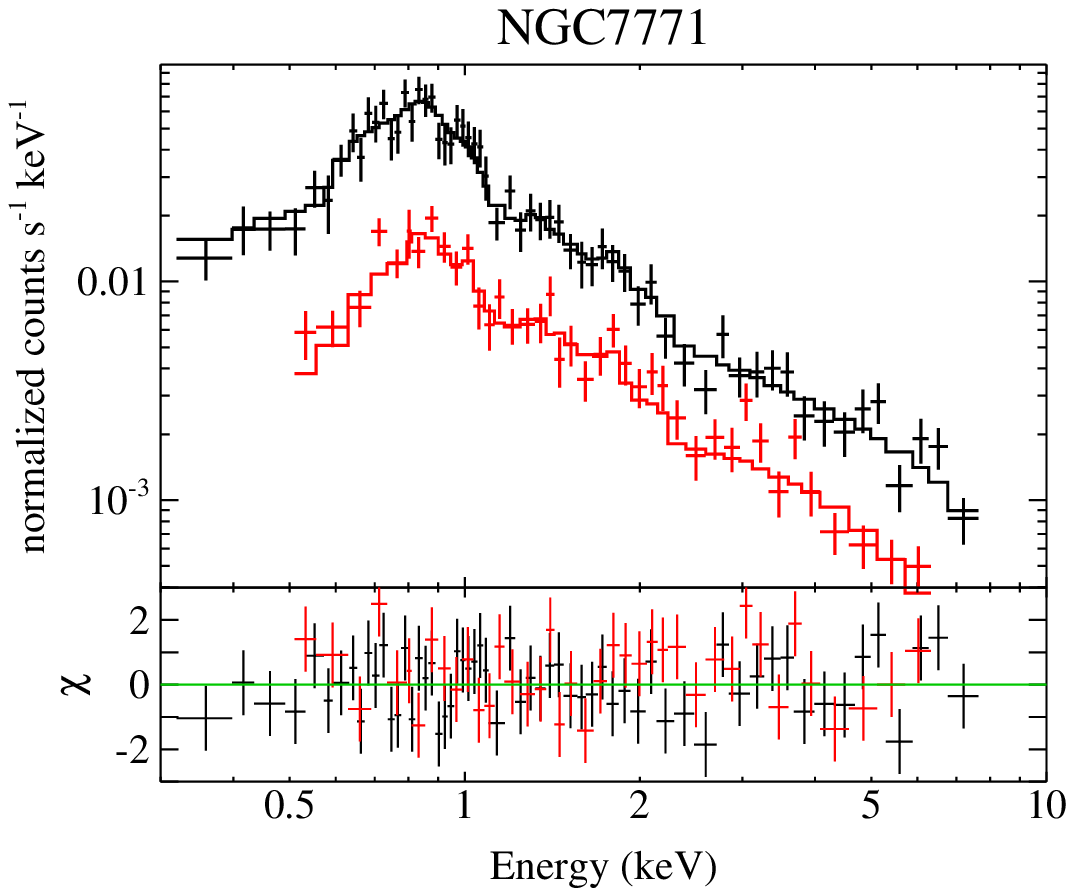}
\caption{Cont.}
\end{figure*}

\afterpage{
\begin{landscape}
\begin{table}
\centering
\caption{X-ray model fits. Starburst model.}
\label{tbl_models}
\begin{scriptsize}
\begin{tabular}{lccccccccccccccccc}
\hline\hline
Galaxy Name & $N_{\rm H,Gal}$\tablefootmark{a} & $N_{\rm H,1}$ & $kT$
& Fe\slash O\tablefootmark{b} &  $N_{\rm H,2}$ & $\Gamma$ &
$\chi^2\slash$dof & $F_{\rm 0.5-2\,keV}$\tablefootmark{c} & $F_{\rm 2-10\,keV}$\tablefootmark{c} & 
$L_{\rm 0.5-2\,keV}$\tablefootmark{d} & $L_{\rm 2-10\,keV}$\tablefootmark{d} & $L_{\rm plasma}$\tablefootmark{e} \\
& (10$^{22}$\,cm$^{-2}$) & (10$^{22}$\,cm$^{-2}$) & (keV) & & (10$^{22}$\,cm$^{-2}$) & & &
\multicolumn{2}{c}{$(\times 10^{-14}$\,erg\,cm$^{-2}$\,s$^{-1}$)} &
\multicolumn{2}{c}{$(\times 10^{40}$\,erg\,s$^{-1}$)} & (\%) \\
\hline 
NGC1614 & 0.06 & $<$0.26 & 0.73$^{+0.05}_{-0.06}$ & 0.44$^{+0.17}_{-0.09}$ & 0.38$^{+0.20}_{-0.14}$ & 1.97$^{+0.18}_{-0.13}$ & 199\slash152 & 17.7$^{+1.7}_{-2.4}$ & 27.6$^{+1.7}_{-4.2}$ & 19.4 & 16.4 & 2.1 \\
NGC2369 & 0.10 & $<$0.13 & 0.60$^{+0.05}_{-0.06}$ & 0.51$^{+0.25}_{-0.17}$ & 0.13$^{+0.28}_{-0.07}$ & 1.49$^{+0.11}_{-0.12}$ & 87\slash96 & 7.5$^{+1.0}_{-1.4}$ & 16.4$^{+1.6}_{-1.8}$ & 3.0 & 4.4 & 0.9 \\
NGC2369\tablefootmark{\star} & \nodata & \nodata & \nodata & \nodata & \nodata & \nodata & \nodata & \nodata & \nodata & 4.5 & 6.4 & \nodata \\
NGC3110 & 0.03 & $<$0.35 & 0.53$^{+0.08}_{-0.09}$ & 0.29$^{+0.17}_{-0.09}$ & 0.27$^{+0.48}_{-0.13}$ & 2.53$^{+0.48}_{-0.36}$ & 92\slash75 & 9.7$^{+1.3}_{-1.6}$ & 5.3$^{+0.4}_{-2.3}$ & 9.6 & 3.3 & 3.0 \\
NGC3256 & 0.09 & 0.18$^{+0.01}_{-0.02}$ & 0.62$^{+0.01}_{-0.01}$ & 0.37$^{+0.03}_{-0.03}$ & 0.14$^{+0.03}_{-0.03}$ & 2.02$^{+0.07}_{-0.12}$ & 561\slash335 & 71.1$^{+1.7}_{-2.9}$ & 49.2$^{+2.4}_{-2.9}$ & 29.5 & 9.7 & 12.3 \\
NGC3690 & 0.01 & 0.11$^{+0.16}_{-0.08}$ & 0.63$^{+0.06}_{-0.05}$ & 0.48$^{+0.33}_{-0.14}$ & 0.29$^{+0.22}_{-0.12}$ & 1.76$^{+0.17}_{-0.18}$ & 168\slash142 & 30.0$^{+2.6}_{-5.9}$ & 45.8$^{+3.9}_{-7.7}$ & 11.2 & 10.8 & 1.9 \\
IC694 & 0.01 & $<$0.18 & 0.66$^{+0.03}_{-0.03}$ & 0.46$^{+0.21}_{-0.08}$ & 0.37$^{+0.22}_{-0.19}$ & 1.86$^{+0.18}_{-0.14}$ & 131\slash128 & 25.4$^{+2.6}_{-4.0}$ & 42.5$^{+3.6}_{-7.3}$ & 10.0 & 10.5 & 1.3 \\
ESO320-G030 & 0.09 & $<$0.21 & 0.61$^{+0.06}_{-0.06}$ & 0.70$^{+0.53}_{-0.32}$ & 0.07$^{+0.07}_{-0.06}$ & 2.13$^{+0.44}_{-0.28}$ & 62\slash67 & 5.8$^{+0.5}_{-1.6}$ & 4.7$^{+0.6}_{-0.6}$ & 2.0 & 1.1 & 2.0 \\
NGC5734 & 0.07 & $<$0.28 & 0.49$^{+0.07}_{-0.06}$ & 0.56$^{+0.20}_{-0.18}$ & 0.13$^{+0.17}_{-0.06}$ & 1.92$^{+0.18}_{-0.17}$ & 78\slash88 & 7.3$^{+0.6}_{-1.3}$ & 9.0$^{+0.6}_{-1.1}$ & 4.7 & 3.8 & 0.7 \\
NGC5743 & 0.07 & \nodata & 0.61$^{+0.10}_{-0.14}$ & 0.63$^{+0.37}_{-0.37}$ & \nodata & 1.68$^{+0.15}_{-0.17}$ & 61\slash71 & 5.4$^{+0.2}_{-0.7}$ & 9.6$^{+1.1}_{-1.1}$ & 2.9 & 4.1 & 0.5 \\
IC4686 & 0.07 & \nodata & 0.72$^{+0.15}_{-0.10}$ & \nodata & \nodata & 1.27$^{+0.12}_{-0.16}$ & 18\slash25 & 2.4$^{+0.3}_{-0.2}$ & 7.1$^{+0.9}_{-1.1}$ & 1.7 & 4.2 & 0.4 \\
IC4687 & 0.07 & $<$0.19 & 0.67$^{+0.04}_{-0.04}$ & 0.54$^{+0.16}_{-0.13}$ & 0.48$^{+0.36}_{-0.23}$ & 2.22$^{+0.32}_{-0.28}$ & 100\slash78 & 7.7$^{+0.9}_{-2.1}$ & 7.7$^{+1.0}_{-2.8}$ & 10.7 & 5.6 & 3.0 \\
IC4734 & 0.07 & \nodata & 0.69$^{+0.09}_{-0.08}$ & 0.31$^{+0.29}_{-0.13}$ & \nodata & 1.46$^{+0.29}_{-0.20}$ & 35\slash42 & 4.0$^{+0.3}_{-0.6}$ & 5.4$^{+0.9}_{-2.8}$ & 2.7 & 3.0 & 3.0 \\
IC5179 & 0.01 & $<$0.06 & 0.55$^{+0.04}_{-0.04}$ & 0.35$^{+0.11}_{-0.08}$ & $<$0.10 & 1.44$^{+0.12}_{-0.12}$ & 91\slash100 & 12.7$^{+0.6}_{-0.8}$ & 15.2$^{+1.2}_{-2.0}$ & 3.9 & 4.4 & 1.4 \\
NGC7679 & 0.05 & $<$0.10 & 0.48$^{+0.09}_{-0.10}$ & 0.29$^{+0.20}_{-0.14}$ & $<$0.01 & 1.64$^{+0.07}_{-0.07}$ & 172\slash172 & 35.6$^{+0.7}_{-3.3}$ & 66.3$^{+3.7}_{-3.7}$ & 27.7 & 44.5 & 0.2 \\
NGC7769 & 0.04 & \nodata & 0.49$^{+0.04}_{-0.07}$ & 0.52$^{+0.21}_{-0.18}$ & \nodata & 2.17$^{+0.18}_{-0.18}$ & 42\slash53 & 5.4$^{+0.3}_{-0.6}$ & 2.7$^{+0.3}_{-0.5}$ & 2.7 & 1.2 & 2.0 \\
NGC7770 & 0.04 & \nodata & 0.58$^{+0.06}_{-0.06}$ & 0.35$^{+0.28}_{-0.13}$ & \nodata & 1.84$^{+0.45}_{-0.63}$ & 50\slash29 & 3.0$^{+0.2}_{-0.5}$ & 1.3$^{+0.3}_{-0.9}$ & 1.5 & 0.5 & 6.9 \\
NGC7771 & 0.04 & $<$0.11 & 0.60$^{+0.04}_{-0.04}$ & 0.44$^{+0.10}_{-0.08}$ & 0.51$^{+0.31}_{-0.29}$ & 1.46$^{+0.20}_{-0.18}$ & 101\slash95 & 7.9$^{+0.6}_{-1.3}$ & 18.6$^{+1.8}_{-3.4}$ & 6.1 & 8.9 & 1.0 \\
NGC7771\tablefootmark{\star} & \nodata & \nodata & \nodata & \nodata & \nodata & \nodata & \nodata & \nodata & \nodata & 7.5 & 11.3 & \nodata\\
\hline
\end{tabular}
\end{scriptsize}
\tablefoot{The model used for the fits was Abs($N_{\rm H,Gal}$)\{Abs($N_{\rm H,1}$)(v)mekal($kT$, Fe\slash O) $+$ Abs($N_{\rm H,2}$)[power-law($\Gamma$)]\}, where Abs is a photo-electric absorption model and (v)mekal is a thermal plasma (with variable metal abundances). 
\tablefoottext{a}{Galactic absorbing hydrogen column density from \citet{Kalberla2005}.}
\tablefoottext{b}{Fe\slash O abundances ratio with respect to the solar values of \citet{Anders1989}.}
\tablefoottext{c}{0.5--2\,keV and 2--10\,keV X-ray observed flux.}
\tablefoottext{d}{0.5--2\,keV and 2--10\,keV X-ray luminosity corrected for Galactic extinction ($N_{\rm H,Gal}$) and intrinsic absorption ($N_{\rm H}$).}
\tablefoottext{e}{Fraction of the 2--10\,keV luminosity from the thermal plasma component.}
\tablefoottext{\star}{Integrated luminosity including extranuclear ULX sources.}
}
\end{table}

\addtocounter{table}{+1}
\begin{table}
\centering
\caption{X-ray model fits. AGN model.}
\label{tbl_models2}
\begin{scriptsize}
\begin{tabular}{lccccccccccccccccc}
\hline\hline
Galaxy Name & $N_{\rm H,Gal}$\tablefootmark{a} & $kT$ & Fe\slash O\tablefootmark{b}
 &  $N_{\rm H,SB}$ & $\Gamma_{SB}$ &  $N_{\rm H,AGN}$ & $\Gamma_{AGN}$ &
$\chi^2\slash$dof & $F_{\rm 0.5-2\,keV}$\tablefootmark{c} & $F_{\rm 2-10\,keV}$\tablefootmark{c} & 
$L_{\rm 0.5-2\,keV}$\tablefootmark{d} & $L_{\rm 2-10\,keV}$\tablefootmark{e} \\
& (10$^{22}$\,cm$^{-2}$) & (keV) & & (10$^{22}$\,cm$^{-2}$) & & (10$^{22}$\,cm$^{-2}$) & & &
\multicolumn{2}{c}{$(\times 10^{-14}$\,erg\,cm$^{-2}$\,s$^{-1}$)} &
\multicolumn{2}{c}{$(\times 10^{40}$\,erg\,s$^{-1}$)} \\
\hline
IC4518W & 0.09 & 0.66$^{+0.05}_{-0.06}$ & 0.49$^{+0.41}_{-0.23}$ & $<$0.10 & 1.85 & 24.28$^{+2.24}_{-2.05}$ & 1.59$^{+0.17}_{-0.29}$ & 106\slash105 & 9.4$^{+0.8}_{-1.8}$ & 169.8$^{+30.9}_{-27.7}$ &  7.0 & 220.2 \\
MCG+04-48-002 & 0.21 & 0.53$^{+0.12}_{-0.14}$ & \nodata & $<$1.30 & 1.85 & 63.19$^{+10.34}_{-6.51}$ & 1.75$^{+0.55}_{-0.30}$ & 89\slash109 & 3.2$^{+0.7}_{-1.8}$ & 275.3$^{+9.2}_{-116.3}$ & 2.8 & 688.0  \\
\hline
\end{tabular}
\end{scriptsize}
\tablefoot{The model used for the fits was Abs($N_{\rm H,Gal}$)\{(v)mekal($kT$, Fe/O) $+$ Abs($N_{\rm H,SB}$)[power-law($\Gamma_{SB}$)] $+$ Abs($N_{\rm H,AGN}$)[power-law($\Gamma_{AGN}$)]\}, where Abs is  a photo-electric absorption model and mekal is a thermal plasma. A Gaussian emission line was added $\sim$6.4\,keV to account for the \Feka\ emission line. An extra Gaussian emission line at 7.1\,keV is needed for IC4518W (see Table \ref{tbl_feka}).
\tablefoottext{a}{Galactic absorbing hydrogen column density from \citet{Kalberla2005}.}
\tablefoottext{b}{Fe\slash O abundances ratio with respect to the solar values of \citet{Anders1989}.}
\tablefoottext{c}{0.5--2\,keV and 2--10\,keV X-ray observed flux.}
\tablefoottext{d}{0.5--2\,keV X-ray luminosity corrected for Galactic extinction ($N_{\rm H,Gal}$).}
\tablefoottext{e}{2--10\,keV X-ray luminosity corrected for Galactic extinction ($N_{\rm H,Gal}$) and intrinsic absorption ($N_{\rm H,AGN}$).}
}
\end{table}
\end{landscape}
}

At 60\,Mpc (typical distance of these LIRGs) the \xmm\ spatial resolution (6\,\arcsec) corresponds to 1.7\,kpc. This means that we are not able to resolve individual emitting sources. Instead, the \xmm\ spectra of these LIRGs probably include the emission from X-ray binaries (low- and high-mass), SNRs and diffuse hot plasma. An AGN may be present as well. Therefore, ideally we would include in the X-ray model one component for each one. However, this is not possible because: (1) It is complicated to determine the characteristic spectrum of these objects and even more complicated to determine the characteristic integrated spectrum of these objects in a galaxy; and (2) the S/N ratio of our data is not sufficiently high to obtain statistically meaningful results with a very complex model.
We used the \texttt{XSPEC} package (version 12.5) to fit simultaneously the EPIC MOS and pn spectra.
The RGS data, of the 3 galaxies with sufficient counts (see Sect. \ref{ss:xmm_reduction}), are compatible with the fit obtained using just the EPIC data. Adding the RGS data does not improve significantly the constraints on the model parameters. Thus the RGS data are not used in the spectral analysis.
The fits of some individual sources are discussed in Appendix \ref{apx:notes}.

\subsection{The X-ray spectra of star-forming galaxies}

We fit the spectra of the star-forming galaxies using a simple model consisting of a soft thermal plasma (\textit{mekal}) plus an absorbed power-law. The absorption of the thermal plasma component is not well constrained and it is compatible with no absorption for most of the galaxies. It was only necessary in the fit of the NGC~3256 and NGC~3690 spectra.
The thermal plasma represents the soft X-ray emitting gas heated by SN shocks, whereas the power-law reproduces the observed hard X-ray continuum produced by X-ray binaries and\slash or AGN.
We also added to the model the absorption due to the Galactic hydrogen column density \citep{Kalberla2005}.
The absolute value of the plasma metallicity is not well constrained for our relatively low S/N ratio spectra. However the \hbox{[Fe\slash O]} ratio can be determined since the most prominent spectral features in the soft X-ray range are produced by these elements (the Fe L-shell and the O K-shell). Thus, the plasma abundances were fixed to the solar values except for the Fe abundance. The latter was left as a free parameter in order to calculate the [Fe\slash O] ratio.
It should be noted that a degeneracy exists between the plasma abundances and temperatures. 
This model provides a reasonable fit to the data ($\chi^2_{\rm red}$ $<$ 1.2) for most of the galaxies.
We included a Gaussian line when a \Feka\ emission line was present in the spectrum (NGC~3256, NGC~3690, and IC~694). For those galaxies with the 6.4\,keV emission line undetected we calculated the upper limits for a narrow emission line.
Fig. \ref{fig_xspectra} shows the observed X-ray spectra together with the model for all the galaxies with \xmm\ data.

The parameters of the fits are listed in Table \ref{tbl_models}.
The typical values of the model parameters are: $\Gamma$ $\sim$ 1.3--2.2, $N_{\rm H}$ $\sim$ 1$\times$10$^{21}$--5$\times$10$^{21}$\,cm$^{-2}$, $kT$ $\sim$ 0.5--0.7\,keV and [Fe\slash O] $\sim$ $-0.5$--$-0.1$.
The measured $N_{\rm H}$ corresponds to $A_{\rm V}$ $\sim$ 0.5--2.3\,mag using the  \citet{Guver2009} conversion factor. In general, the X-ray derived absorption is lower than that obtained from the near-IR colors ($\sim$3\,mag) and the Pa$\alpha$\slash H$\alpha$ ratio ($\sim$2--5\,mag) for these galaxies \citep{AAH06s}.
The temperature of the plasma and its contribution to the hard X-ray emission of these LIRGs are comparable with those of local starbursts (0.8\,keV and 3\,\%, \citealt{Persic2003}). However the power-law component is slightly steeper than in local starbursts, $\Gamma$ $=$ 1.2 \citep{Persic2003} versus $\Gamma$ $=$ 1.8 in these LIRGs.

The upper limits and fluxes of the \Feka\ emission line are listed Table \ref{tbl_feka}.

\subsection{AGN X-ray spectra}
The X-ray spectra of two galaxies, IC~4518W, MCG+04-48-002, are not well fitted by the star-formation model described above. The hard X-ray emission of these galaxies is dominated by the AGN (Fig. \ref{fig_xspectra}).
We added to the star-formation model an absorbed power-law and a Gaussian emission line at 6.4\,keV to account for the AGN emission. For IC~4518W we added another Gaussian emission line at 7.1\,keV.
For these galaxies the power-law index of the star-formation component is not well constrained due to the AGN contribution to the hard X-ray emission. Hence we fixed it to the median value obtained for the other LIRGs ($\Gamma$ = 1.85).
This model provides a good fit to the data, $\chi^2_{\rm red}$ $\leq$ 1, for the two galaxies. For this reason, we did not include an AGN reflection component (\textit{pexrav}). By doing this we may underestimate the absorbing column density towards the AGN. This model fits well the soft X-ray emission of these galaxies. However this does not imply a star-formation origin of the soft X-ray emission since the thermal plasma and the power-law continuum can be produced by an AGN. The origin of the soft X-ray emission is discussed in Section \ref{s:sfr_xrayir}. The model parameters for these two galaxies are given in Table \ref{tbl_models2}.

\subsection{Literature X-ray data}

The X-ray luminosities taken from the literature are listed in Table \ref{tbl_model_lit}. This table includes the 6 galaxies observed with \textit{Chandra} plus 2 galaxies observed with \xmm. For these two objects, MCG$-$03-34-064 and NGC~7469, we repeated the fits of the \xmm\ data using the models given by \citet{Miniutti2007} and \citet{Blustin2003}, respectively, to take advantage of the latest calibration.

\addtocounter{table}{-2}
\begin{table}
\centering
\caption{Fe\,K line}
\label{tbl_feka}
\begin{tabular}{lcccccccccc}
\hline\hline
Galaxy Name & $E$\tablefootmark{a} & $EW$ & $F_{\rm Fe\,K}$ \\
& (keV) & (keV) & (10$^{-15}$ erg cm$^{-2}$ s$^{-1}$) \\
\hline
NGC23 & \nodata & \nodata & \nodata \\
NGC1614 & 6.4 & $<$0.62 & $<$16 \\
NGC2369 & 6.4 & $<$0.34 & $<$5.7 \\
NGC3110 & 6.4 & $<$0.76 & $<$3.0 \\
NGC3256 & 6.4 & $<$0.07 & $<$3.5 \\
  & 6.60$^{+0.10}_{-0.04}$ & 0.2 & 8.2 \\
NGC3690 & 6.6$^{+0.2}_{-0.3}$ & 0.93 & 67 \\
IC694 & 6.4 & $<$0.2 & $<$9.4 \\
  & 6.67$^{+0.10}_{-0.11}$ & 0.85 & 60 \\
ESO320-G030 & 6.4 & $<$1.8 & $<$6.5 \\
IC860 & \nodata & \nodata & \nodata \\
MCG$-$03-34-064 & 6.39$^{+0.02}_{-0.02}$ & 0.11 & 92 \\
NGC5135\tablefootmark{b} & 6.39$^{+0.03}_{-0.04}$ & 2.4 & $\sim$50 \\
NGC5653 & \nodata & \nodata & \nodata \\
NGC5734 & 6.4 & $<$0.8 & $<$6.6 \\
NGC5743 & 6.4 & $<$1.4 & $<$13 \\
IC4518W & 6.39$^{+0.03}_{-0.03}$ & 0.46 & 120 \\
 & 7.1$^{+0.1}_{-0.2}$ & 0.21 & 52 \\
Zw049.057 & \nodata & \nodata & \nodata \\
IC4686 & 6.4 & $<$0.8 & $<$5.1 \\
IC4687 & 6.4 & $<$1.4 & $<$8.7 \\
IC4734 &6.4 & $<$3.5 & $<$11 \\
MCG+04-48-002 & 6.47$^{+0.05}_{-0.06}$ & 0.12 & 66 \\
NGC7130\tablefootmark{c} & 6.40$^{+0.05}_{-0.05}$ & 1.8 & $\sim$35 \\
IC5179 & 6.4 & $<$0.41 & $<$6.3 \\
NGC7469 & 6.42$^{+0.03}_{-0.03}$ & 0.070 & 203 \\
NGC7679 & 6.4 & $<$0.42 & $<$30 \\
NGC7769 & 6.4 & $<$1.8 & $<$3.5 \\
NGC7770 & 6.4 & $<$8.5 & $<$9.6 \\
NGC7771 & 6.4 & $<$0.47 & $<$10 \\
\hline
\end{tabular}
\tablefoot{Observed fluxes and EW of the Fe\,K emission lines. Upper limits are calculated assuming an unresolved Gaussian emission line at 6.4\,keV.
\tablefoottext{a}{Rest frame energy of the emission line. When no uncertainties are quoted the value was fixed. }
\tablefoottext{b,c}{Data from \citet{Levenson2004} and \citet{Levenson2005} respectively.}
}
\end{table}
\addtocounter{table}{1}

\section{X-ray emission from star-formation activity}\label{s:sfr_xrayir}

\subsection{Soft X-ray emission versus SFR}

\begin{figure*}
\center
\includegraphics[width=0.48\textwidth]{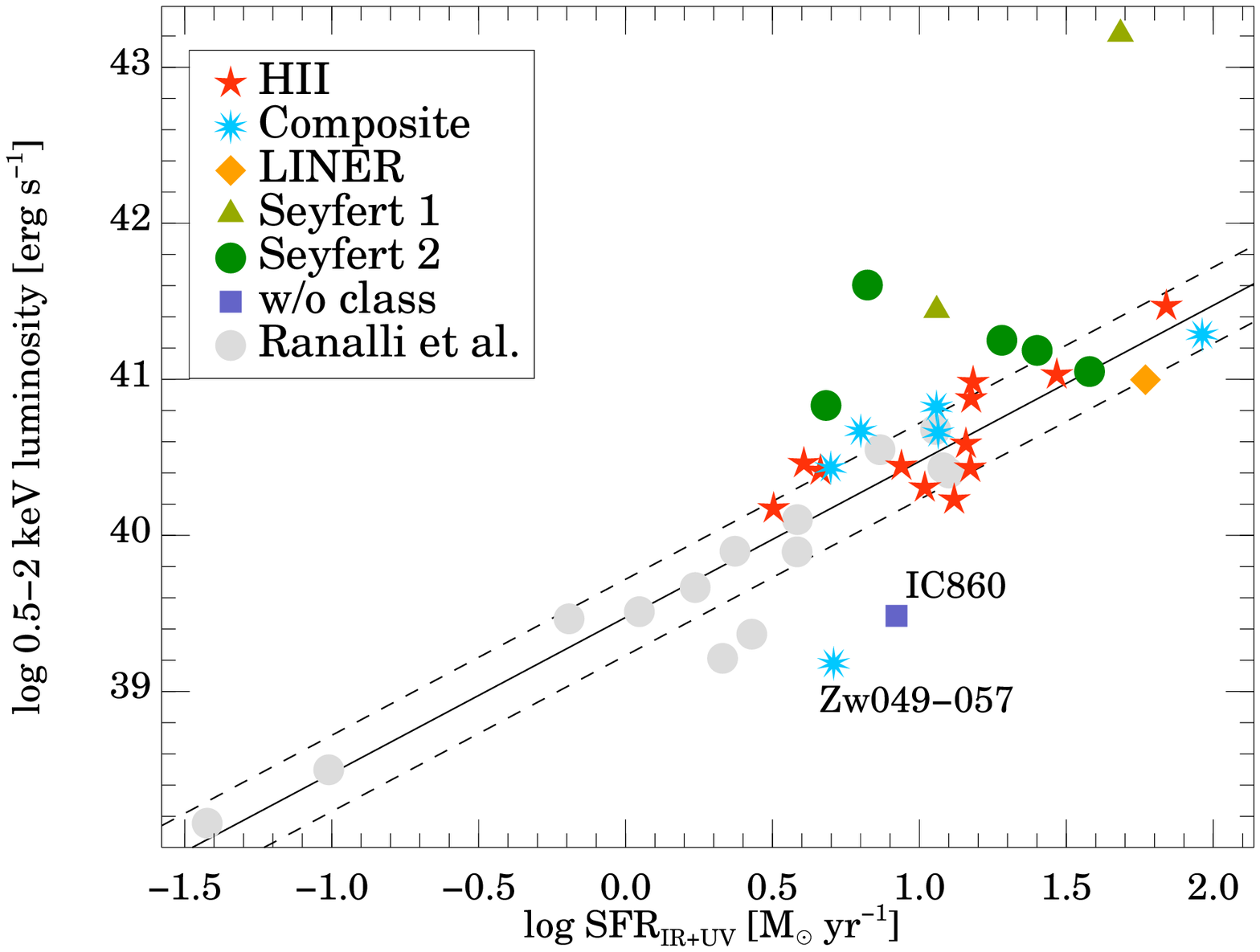}
\includegraphics[width=0.48\textwidth]{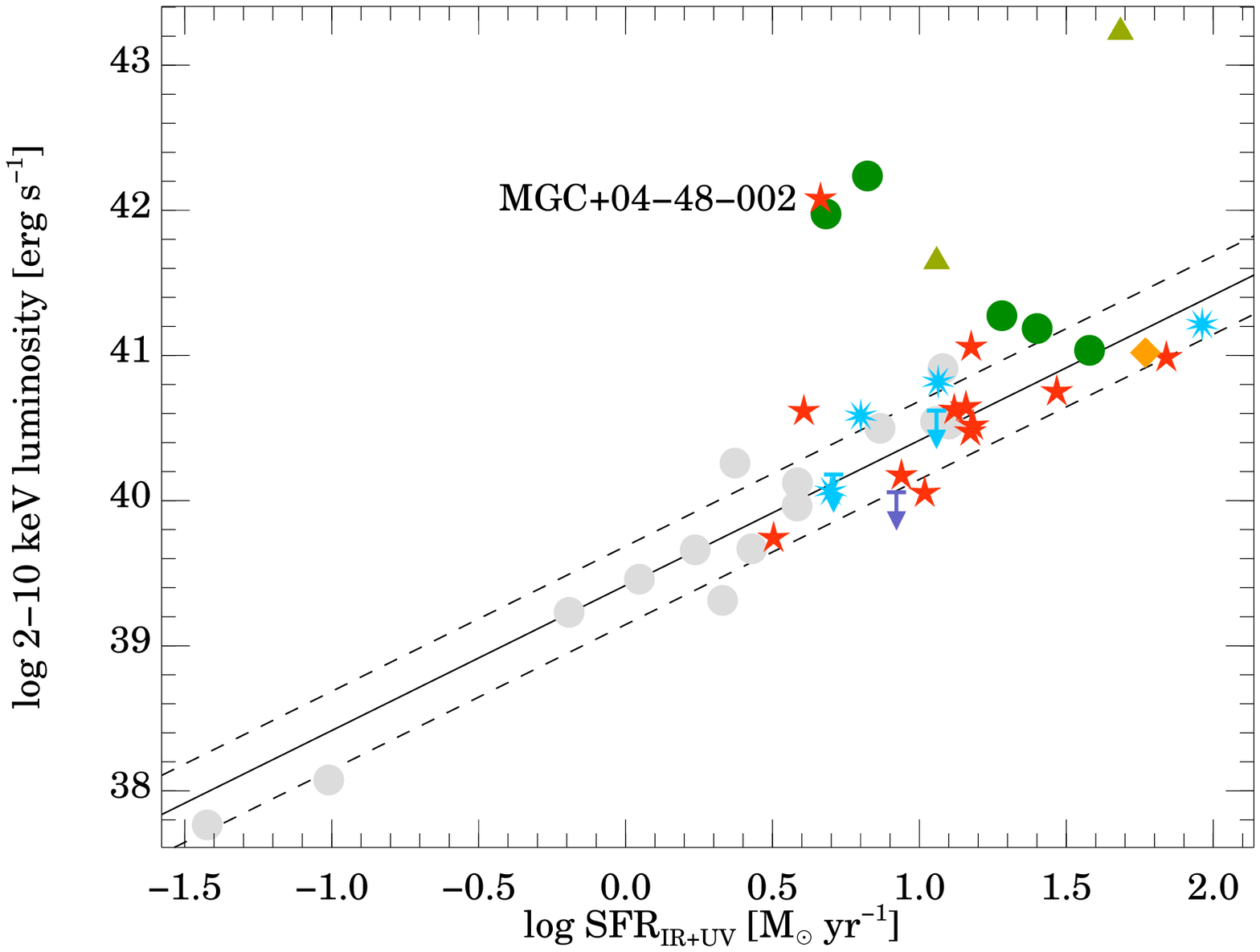}
\caption{Soft (0.5--2\,keV; left) and hard (2--10\,keV; right) X-ray luminosity corrected for absorption vs. SFR calculated combining the UV and IR luminosities (Sect. \ref{s:observations}). Red stars are \HII\ galaxies. Light green triangles and dark green circles are Type 1 and 2 Seyfert galaxies, respectively. Blue stars are composite galaxies, orange diamonds are LINERs and blue squares are galaxies without classification from optical spectroscopy. Galaxies from \citet{Ranalli2003} are plotted as gray circles. The solid line is our best linear fit to those LIRGs classified as \HII\ and the \citet{Ranalli2003} galaxies. The dashed lines indicate the $\pm$1$\sigma$ dispersion in this relation.}
\label{fig_sfrsofthard}
\end{figure*}

At soft energies (0.5--2-0\,keV), the X-ray emission is produced by both diffuse hot gas heated by supernova explosions and X-ray binaries. Therefore the soft X-ray emission is expected to be correlated with the SFR \citep{Ranalli2003, Rosa2009}.

The correlation found by \citet{Ranalli2003} between the soft X-ray and the far-infrared (FIR, 40--500\micron) luminosities is not linear. The galaxies in their study cover a large range in FIR luminosities and it is possible that, for those galaxies with the lowest SFR, the total SFR is not dominated by the obscured SFR traced by the FIR luminosity (see \citealt{PerezGonzalez2006, Kennicutt2009}). To account for the unobscured star-formation we used the near-UV (2267\,\AA) fluxes from \citet{GildePaz2007} that we translated into SFR (see Sect. \ref{s:observations}). We found near-UV fluxes for 65\,\% of the \citet{Ranalli2003} sample.
To calculate the obscured SFR we used the \textit{IRAS} fluxes to obtain the total IR luminosity (8--1000\micron). Then we used the calibration of \citet{Kennicutt1998} correcting for our adopted Kroupa IMF. We added both IR and UV SFR to obtain the total SFR. The SFR traced by the UV light contributes to the total SFR between 5\,\% and 60\,\% with a median contribution of 20\% for the galaxies of the \citet{Ranalli2003} sample.
In the case of the LIRGs we neglected the unobscured star-formation as it contributes less than 10\,\% for most of the galaxies (see Table \ref{tbl_sample_data} and \citealt{Howell2010,RodriguezZaurin2011}).
In the left panel of Fig. \ref{fig_sfrsofthard} we compare the SFR$_{\rm IR + UV}$ with the soft X-ray luminosity for our sample of LIRGs together with the nearby galaxies of \citet{Ranalli2003}. The best fit\footnote{We only used for the fit the galaxies classified as \HII\ and the galaxies of \citet{Ranalli2003}} slope in log-log space is 1.1 $\pm$ 0.1, which is compatible with a linear relation. Assuming a constant SFR$_{\rm IR + UV}$\slash $L_{\rm 0.5-2\,keV}$ ratio we found:

\begin{equation}\label{eqn:sfr_soft}
{\rm SFR_{\rm IR + UV}}\,(M_{\rm \odot}\;{\rm yr^{-1})} = 3.4\times 10^{-40} L_{\rm 0.5-2\,keV}\,{\rm (erg\;s^{-1})}
\end{equation}

with a 0.24\,dex scatter.
\citet{Mas2008} modeled the soft X-ray luminosity expected from a starburst. They assumed that the mechanical energy from the starburst (SN and stellar winds) heats the interstellar diffuse gas with an efficiency 1--5\,\%. After correcting for the different IMF normalization, their calibration for a young extended burst is consistent with Equation \ref{eqn:sfr_soft} within the scatter.

The Seyfert 2 galaxies in our sample of LIRGs lie on the correlation (NGC~3690 and IC~4518W), or have a small (less than a factor of 3) soft X-ray emission excess (MCG$-$03-34-064, NGC~5135, and NGC~7130). In Type 2 Seyferts the absorbing hydrogen column density towards the AGN is high and thus most of the soft X-ray emission coming from the AGN is absorbed. Therefore we conclude that when a sufficiently powerful starburst is present it may contribute significantly to the observed soft X-ray emission.
The two Seyfert 1s (NGC~7469 and NGC~7679) in our sample have a soft X-ray emission excess relative to their SFR due to the AGN emission.
Two objects (IC~860 and Zw~049.057) lie below the correlation. The low number of counts of these galaxies does not allow us to correct properly the soft X-ray fluxes for their internal absorption. In addition, the large 9.7\micron\ silicate absorption of these galaxies \citep{AAH2011} suggests that they are highly obscured, thus this correction is likely to be large \citep{Shi2006}.

\subsection{Hard X-ray emission versus SFR}\label{ss:hard_sfr}

HMXB dominate the hard X-ray (2--10\,keV) emission of a starburst galaxy when an AGN is not present. Thus the hard X-ray emission is also a tracer of the SFR \citep{Ranalli2003, Grimm2003, Persic2004, Lehmer2010}.

The right panel of Fig. \ref{fig_sfrsofthard} shows that there is a good correlation between the hard X-ray emission and the SFR when there is no AGN. The best fit slope is 1.1$\pm$0.1 (in log-log space). Thus assuming a directly proportional relation between the SFR$_{\rm IR+UV}$ and the $L_{\rm 2-10\,keV}$ we obtained:

\begin{equation}\label{eqn:sfr_hard}
{\rm SFR_{\rm IR + UV}}\,(M_{\rm \odot}\;{\rm yr^{-1})} = 3.9\times 10^{-40} L_{\rm 2-10\,keV}\,{\rm (erg\;s^{-1})}
\end{equation}

with a 0.27\,dex scatter.
In the fit we used all the \HII\ galaxies (excluding MGC+04-48-002 whose X-ray spectra resembles that of a Seyfert 2 galaxy, see Appendix \ref{apx:notes}) and the galaxies of the \citet{Ranalli2003} sample. This calibration agrees, within the uncertainties, with that of \citet{Ranalli2003}. However, \citet{Lehmer2010} found a highly non-linear relation (slope $=$ 0.76) between the SFR and the $L_{\rm 2-10\,keV}$. In their fit they included high-luminosity LIRGs and ULIRGs. These galaxies are underluminous in the 2--10\,keV range (see \citealt{Iwasawa2009,  Lehmer2010}), thus this may affect the relation slope.

Due to the low number of galaxies with SFR of less than $\sim$4\,$M_{\rm \odot}$\,yr$^{-1}$ in our sample, it is uncertain whether the correlation is still valid in the low SFR range or not. Actually a change in the slope of the $L_{\rm 2-10\,keV}$ versus SFR relation
is expected for this range \citep{Grimm2003}. For these galaxies with low SFR if a bright HMXB is present it can dominate the galaxy integrated hard X-ray luminosity.

As can be seen in the right panel of Fig. \ref{fig_sfrsofthard}, the hard X-ray emission of 3 Seyfert galaxies (NGC~3690, NGC~5135, and NGC~7130) is compatible (within 2$\sigma$) with that expected from star-formation. These galaxies are known to host powerful starbursts that might dominate their energy output \citep{GonzalezDelgado1998, AAH09Arp299, Bedregal2009, AAH2011}. For these three galaxies high angular resolution \textit{Chandra} images were used to isolate and quantify the AGN emission, which was found to be approximately 70\,\% of the total hard X-ray emission \citep{Zezas2003, Levenson2004, Levenson2005}.

In the previous fit we neglected the contribution of the LMXB to the hard X-ray luminosity. The emission of the LMXB is proportional to the stellar mass of the galaxy \citep{Gilfanov2004}, and LMXBs may be important for galaxies with the lowest SFR/M${\rm \star}$ ratios.
Assuming that there is a linear correlation between the 2--10\,keV galaxy integrated emission of LMXB and HMXB with the stellar mass and the SFR respectively, \citet{Lehmer2010} constrained the relation:
\begin{equation}\label{eqn_lxsfrm}
L_{\rm 2-10\,keV} = \alpha M_{\star} + \beta {\rm SFR}
\label{eqn_lxsfr}
\end{equation}
for a sample of nearby normal galaxies, LIRGs, and ULIRGs. They found $\alpha = (9.05 \pm 0.37)\times 10^{28}$ erg s$^{-1}$ $M_{\odot}^{-1}$ and $\beta = (1.62 \pm 0.22)\times 10^{39} $ erg s$^{-1}$ ($M_{\odot}$\,yr$^{-1}$)$^{-1}$.
From this equation we estimate that the contribution of LMXB to the integrated hard X-ray luminosity is less than 15\,\% for our sample of LIRGs. This is much lower than the scatter of the SFR versus hard X-ray luminosity correlation, thus it would not be the main cause of the observed scatter.

\begin{figure}
\center
\resizebox{\hsize}{!}{\includegraphics{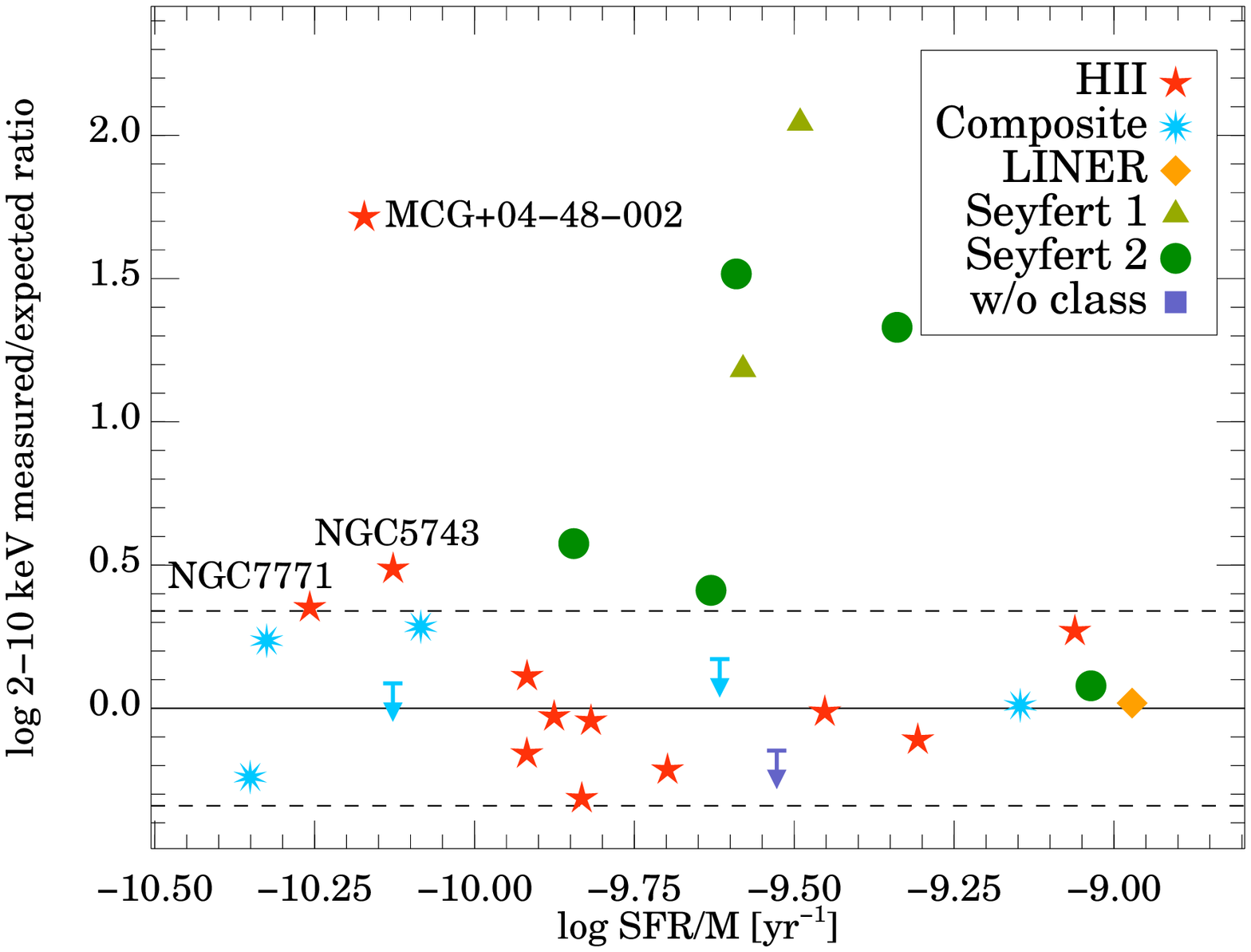}}
\caption{Observed\slash expected hard X-ray (2--10\,keV) luminosity ratio vs. SFR/$M_{\rm \star}$ ratio.
The expected X-ray luminosity only includes the X-ray emission from LMXB and HMXB and it is based on the relation $L_{\rm X} = \alpha M_{\star} + \beta$SFR from \citet{Lehmer2010}. The dashed lines indicate the scatter in this relation. Galaxy symbols are as in Fig. \ref{fig_sfrsofthard}.}
\label{fig_sfrm}
\end{figure}

Fig. \ref{fig_sfrm} shows that the predicted X-ray luminosity using Equation \ref{eqn_lxsfr} agrees with that observed for most of the \HII\ galaxies in our sample of LIRGs. Likewise, most of the Seyfert galaxies have hard X-ray luminosities 10 times larger than that expected from star-formation. The 3 Seyferts (NGC~3690, NGC~5135, and NGC~7130) that lie within 2$\sigma$ of the expected relation for star-formation are those with powerful starbursts.

\subsection{\Feka\ line from star-formation}\label{ss:feka_sf}

\begin{figure}
\center
\resizebox{\hsize}{!}{\includegraphics{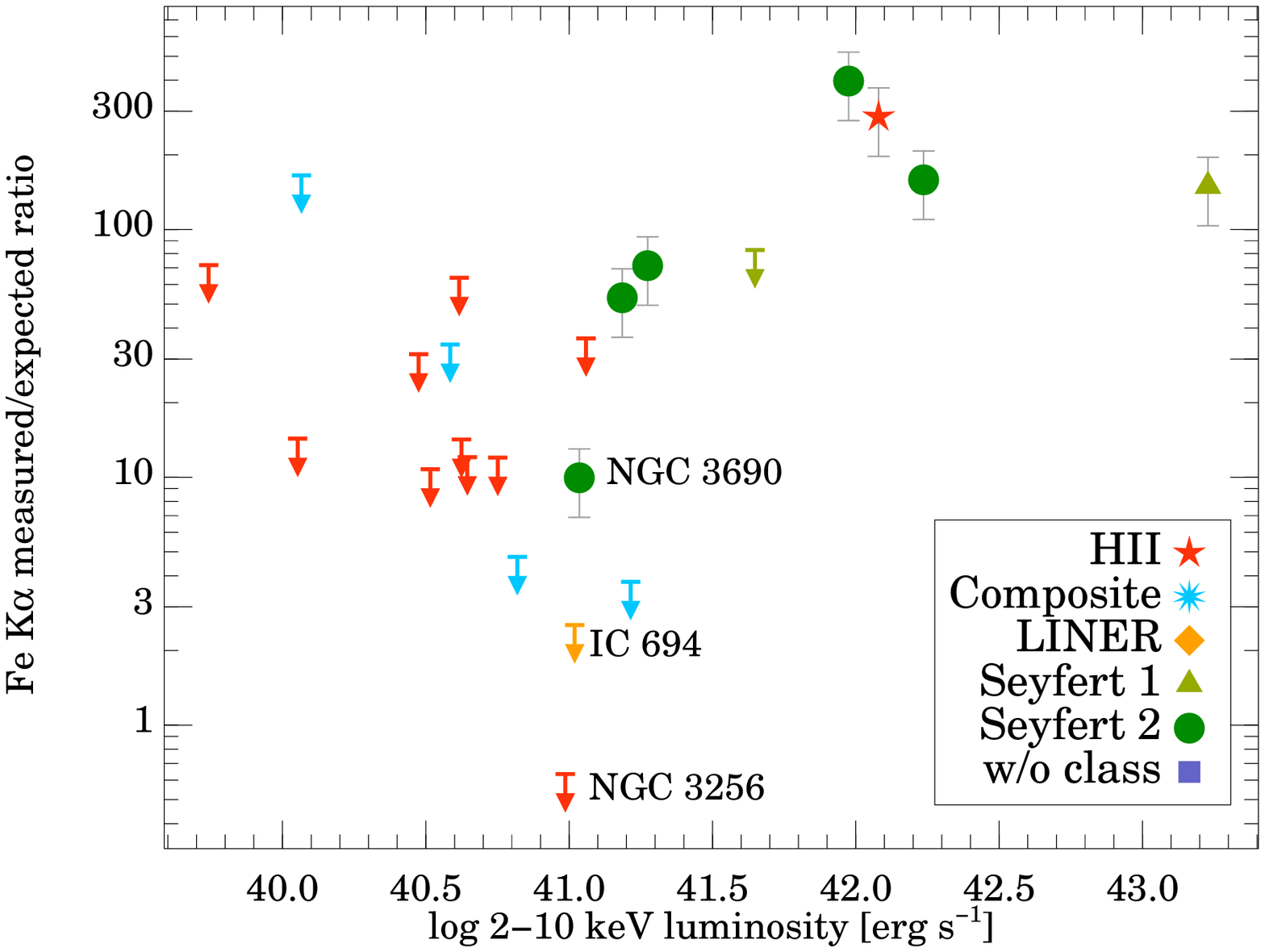}}
\caption{Observed\slash expected 6.4\,keV \Feka\ luminosity from star-formation (see Sect. \ref{ss:feka_sf}) vs. 2--10\,keV luminosity.
Galaxy symbols are as in Fig. \ref{fig_sfrsofthard}.}
\label{fig_feka}
\end{figure}

Although the \Feka\ emission line is detected mainly in active galaxies, it has been found in starbursts (e.g., M~82 and NGC~253, \citealt{Cappi1999}). In the latter its origin is associated with X-ray binaries and SNRs. The 6.4\,keV \Feka\ line is observed in Galactic X-ray binaries \citep{White1983,Torrejon2010}. Using high-resolution X-ray spectroscopy \citet{Torrejon2010} studied a sample of 41 X-ray binaries. They detected the \Feka\ line in all of the HMXB (10), but only in 10\,\% of the LMXB. For this reason and since the X-ray binary population in starbursts is dominated by HMXB we only consider HMXB in the following.

We estimated the median EW of the \Feka\ line produced in X-ray binaries using the data of \citet{Torrejon2010}, $EW$ $=$ $0.07\pm0.04$\,keV.
Assuming that the spectra of the HMXB in the 2--10\,keV energy range can be represented as a power-law with photon index $\Gamma=1.2$ \citep{Persic2002} we obtain log $L_{\rm Fe K\alpha}$\slash$L_{\rm 2-10\,keV}$ $=$ $-2.1\pm0.6$. The variation of this ratio with $\Gamma$ is small, being less than 0.1\,dex for $\Gamma$ between 1 and 2.

The integrated hard X-ray luminosity, $L_{\rm 2-10\,keV}$, from HMXB is directly related with the SFR (see \citealt{Lehmer2010} and Equation \ref{eqn_lxsfrm}). Combining these relations we obtain the following:

\begin{equation}\label{eqn_lfehmxb}
L_{\rm Fe K\alpha}^{\rm HMXB}\;({\rm erg\,s^{-1}}) = (1.3\pm0.4)\times 10^{37}\,{\rm SFR}\;(M_{\rm \odot}\,{\rm yr^{-1}})
\end{equation}

Using this equation we estimated the expected flux of the 6.4\,keV \Feka\ line from HMXB. In Fig. \ref{fig_feka} we show the expected\slash observed \Feka\ emission line ratio.
For all the galaxies without line detection, the upper limits are compatible with our flux estimation for star-formation. As this line is correlated with the luminosity of the AGN when present, these upper limits also put an upper limit on the AGN contribution to the total energy output of the galaxy (see Sect. \ref{s:low_lum}).

The detection of the ionized iron line at 6.7\,keV in IC~694 and NCG~3256 (Table \ref{tbl_feka}) indicates the presence of hot gas ($kT$ $>$ 3\,keV) in these galaxies. Therefore their hard X-ray emission may be dominated by hot gas as is the case in high luminosity LIRGs and ULIRGs \citep{Iwasawa2009,Iwasawa2011,Colina2011}. This is in contrast with local starbursts where the hard X-ray emission is mostly due HMXB. Fig. \ref{fig_feka} shows that the upper limit for the 6.4\,keV line in NGC~3256 is smaller than the expected value for HMXB. This provides further support for a noticeable contribution from hot gas to the hard X-ray emission, at least, in some LIRGs.

For Seyfert galaxies the observed to expected from star-formation \Feka\ emission ratio is larger than $\sim$50, reaching $\sim$300 in some cases. It is clear that the \Feka\ emission is dominated by the AGN in these galaxies.

\subsection{Metal abundances of the thermal plasma}\label{s:metal_abun}

The soft X-ray spectrum of starburst galaxies is dominated by the emission of a diffuse thermal plasma with temperatures in the range 0.1 to 1\,keV. It is believed that it is heated by shock-fronts generated by SN explosions and stellar winds \citep{Persic2004}.
Table \ref{tbl_models} shows the Fe\slash O ratio respect to the solar values\footnote{We used the solar abundances of \citet{Anders1989}.} of our sample. Due to the limited S/N ratio of the spectra we cannot obtain the absolute abundances. The average Fe\slash O ratio respect to the solar abundance is 0.5 $\pm$ 0.2. The underabundance of Fe relative to $\alpha$ elements has been observed in nearby starbursts \citep{Strickland2004, Grimes2005} and local (U)LIRGs \citep{Iwasawa2011}.

Various processes have been proposed to explain these results. In the dwarf starburst galaxy NGC~1559 the $\alpha$ elements abundance with respect to Fe is consistent with the enhanced production of $\alpha$ elements in Type II SN \citep{Martin2002}. Indeed the Fe\slash O ratio measured in these LIRGs is consistent with the IMF-averaged Fe relative to $\alpha$ elements ratio expected from Type II SN \citep{Gibson1997}.
Alternatively \citet{Strickland2000} suggested that the X-ray emission is produced in the boundary layer between the cold interstellar medium and the SN winds, thus the underabundance of Fe could be due to the Fe depletion into dust grains. To distinguish between Fe depletion and enhanced $\alpha$ elements production in Type II SN it is needed to determine the abundances relatives to hydrogen and compare them with the galaxy metallicity \citep{Strickland2004}.

\section{AGN activity of LIRGs from X-ray emission}\label{s:agn_activity}

\subsection{Low luminosity AGN}\label{s:low_lum}

The \Feka\ emission line is one of the most prominent signatures of obscured AGN in the hard X-ray energy range.
In Compton thick AGN ($N_{\rm H} > 10^{24}$ cm$^{-2}$)  the \Feka\ line at 6.4\,keV has a large EW ($>$ 1\,keV; \citealt{Levenson2006}). This is due to the different gas column densities affecting the AGN continuum and the emission line. The X-ray continuum in a Compton thick AGN is highly absorbed below 10\,keV by the gas around the AGN. The 6.4\,keV \Feka\ emission line, on the other hand, is produced in regions far from the AGN with an absorbing column density in our line of sight significantly lower than that of the continuum (e.g., \citealt{JimenezBailon2005}).
\Feka\ emission lines with large EW have been detected in ULIRGs, indicating that some ULIRGs ($\sim$20--30\,\%) contain buried AGNs, although their X-ray emission is dominated by star-formation \citep{Ptak2003,Franceschini2003}.

Table \ref{tbl_feka} shows the \Feka\ EW measured for our sample of LIRGs. The \Feka\ EW upper limits for 40\,\% of the galaxies are compatible with EW larger than 1\,keV. These galaxies could host a Compton thick AGN. However, if they do, it would not be a bright AGN. Their bolometric luminosity would be lower than 10$^{43}$\,erg\,s$^{-1}$ (estimated from the upper limits of the \Feka\ emission line flux, see Sect. \ref{s:agn_compton}). That is, the AGN contribution in these galaxies would be less than 10\,\% of the total luminosity (IR luminosity).
By comparison, for the Seyfert galaxies in the sample the AGN contribution ranges from less than 1\,\% to 35\,\% (see Sect. \ref{s:agn_lirgs}).

Apart from the \Feka\ emission line, the AGN continuum, absorbed or reflected (depending on the absorbing column density), might be detected in the hard X-ray (2--10\,keV) range.
Three of the non-Seyfert galaxies (NGC~5743, MCG+04-48-002, and NGC~7771) have hard X-ray luminosities between 5 and 100 times larger than those expected from star-formation (Fig. \ref{fig_sfrm}). NGC~5743 is classified as \HII\ from optical spectroscopy, although the detection of the [\ion{Ne}{V}] lines at 14.3 and 24.3\micron\ \citep{Pereira2010c} indicates that an AGN is present in this galaxy.
These [\ion{Ne}{V}] lines are also detected in the mid-infrared spectrum of MCG+04-48-002 \citep{Pereira2010c}. The presence of an AGN in this object is further confirmed by its 20--100\,keV emission \citep{Bassani2006}.
In the case of NGC~7771 the nuclear activity type is \HII\ \citep{AAH09PMAS}. The only evidence of an AGN in this galaxy is the detection of the \Feka\ line\footnote{In our spectral analysis of NGC~7771 we do not detect the \Feka\ line, although the upper limit is compatible with the line flux measured by \citet{Jenkins2005}.} and the excess hard X-ray emission \citep{Jenkins2005} .

\subsection{Obscured AGN}\label{s:agn_compton}

Five of the LIRGs (NGC~3690, MCG$-$03-34-064, NGC~5135, IC~4518W, and NGC~7130) are Seyfert 2 galaxies. In addition, the mid-IR and X-ray properties of MCG+04-48-002 suggest the presence of a bright AGN, although it is not detected in the optical spectrum \citep{Masetti2006}. These galaxies are likely to host an obscured AGN.

We can constrain the hydrogen column density towards the AGN from their X-ray spectra for three galaxies: MCG$-$03-34-064, IC~4518W, and MCG+04-48-002. It is $<$ 10$^{24}$\,cm$^{-2}$ for all of them, therefore they are not Compton thick AGN. After correcting the observed luminosity for this absorption we obtain the intrinsic AGN X-ray luminosity (Table \ref{tbl_agnlum}). These 3 galaxies are also detected in the \textit{Swift}\slash BAT 14--195\,keV all-sky survey \citep{Tueller2010}. The 14--195\,keV emission is less affected by absorption than the 2--10\,keV emission. Therefore it is a direct indicator of the intrinsic AGN luminosity except for the most obscured Compton thick AGN\footnote{However due to the angular resolution of the \textit{Swift}-BAT survey (19.5\,arcmin) the 14--195\,keV fluxes can be contaminated by other nearby Seyfert galaxies (MCG$-$03-34-064 and MCG$-$03-34-063, MCG+04-48-002 and NGC~6921).} ($N_{\rm H}>10^{26}$\,cm$^{-2}$; e.g., \citealt{Matt1997}).

Using the \citet{Marconi2004} AGN template \citet{Rigby2009} calculated that the $L_{\rm 2-10\,keV}$\slash$L_{\rm 14-195\,keV}$ ratio is 0.37. This ratio is between 0.77 and 0.12 for these 3 LIRGs and is comparable to that observed in a sample of local \textit{Swift}\slash BAT selected AGNs (Figure 6 of \citealt{Winter2009}).
MCG$-$03-34-064 has the highest ratio and also the steepest continuum ($\Gamma\sim2.7$). Likewise, IC~4518W has the lowest ratio and the lowest photon index ($\Gamma\sim1.6$). Thus the continuum slope might affect the $L_{\rm 2-10\,keV}$\slash$L_{\rm 14-195\,keV}$ ratio. However, the  \textit{Swift}\slash BAT flux contamination by nearby sources, the uncertainty in the contribution of the AGN reflected continuum to the 14--195\,keV luminosity (which represents about 40\,\% of the total AGN emission at 30\,keV, \citealt{Ueda2003}), and the AGN variability may be important factors  affecting the observed $L_{\rm 2-10\,keV}$\slash$L_{\rm 14-195\,keV}$ ratio.

The other 3 Seyfert 2s in our LIRGs sample (NGC~3690, NGC~5135, and NGC~7130) might be Compton thick AGN. In fact, NGC~5135 and NGC~7130 have been classified as Compton thick based on their large \Feka\ EW (see \citealt{Levenson2004, Levenson2005}). The \Feka\ EW of NGC~3690 is 0.93\,keV (see Table \ref{tbl_feka}). It is slightly less than the typical values of Compton thick AGN ($>$1\,keV). However the star-formation contribution to the hard X-ray continuum is $\sim$30\,\% in NGC~3690 (see Sect. \ref{ss:hard_sfr}) and thus it decreases the observed EW of the \Feka\ emission line.
To estimate the AGN X-ray luminosity of these objects we used the flux of the \Feka\ emission line since it seems to be a good indicator of the intrinsic AGN luminosity \citep{Ptak2003, Levenson2006, LaMassa2009}. We assumed $L_{\rm Fe K\alpha}$\slash$L_{\rm 2-10\,keV}^{int}$ = $2 \times 10^{-3}$ \citep{Levenson2006}. However we note that this ratio depends on both the geometry of the AGN obscuring material and the column density in our line of sight \citep{Liu2010, Yaqoob2010, Murphy2009}. Thus a large uncertainty, a factor of $\sim$5, is expected in the intrinsic AGN luminosities of these galaxies

None of these 3 Compton thick candidates are detected in the \textit{Swift}-BAT 14--195\,keV survey. The 14--195\,keV luminosity upper limits are slightly lower than the expected luminosity for their 2--10\,keV emission.
The large scatter (a factor of 6) in the $L_{\rm 14-195\,keV}$\slash$L_{\rm 2-10\,keV}$ ratio for the detections and the uncertainties discussed above might explain this.

The 2--10\,keV and 14--195\,keV luminosities are listed in Table \ref{tbl_agnlum}. For completeness, the two Seyfert 1 galaxies in our sample are also included in the table.

\subsection{AGN contribution to the LIRGs luminosity}\label{s:agn_lirgs}

We calculated the fraction of the bolometric luminosity produced by AGN in our sample of LIRGs. We used the $L_{\rm IR}$(8--1000\micron) as the total luminosity of the LIRGs. The AGN luminosity was estimated from the X-ray data.

There are 8 active galaxies in our sample. This represents 30\,\% of the sample, although the AGN does not dominate the luminosity of any of them. For these galaxies we estimated the AGN luminosity from their X-ray spectral model or from their \Feka\ line luminosity (Sect. \ref{s:agn_compton}). To obtain the AGN bolometric luminosity we applied the bolometric correction of \citet{Marconi2004}. Comparing the bolometric AGN luminosity with the IR luminosity (Table \ref{tbl_agnlum}) we find that the median AGN contribution is 25\,\% and it ranges from less than 1\,\% to 35\,\% for the Seyfert LIRGs
For the rest of the sample we used the upper limit of the \Feka\ line luminosity to obtain the upper limit for the AGN luminosity (Sect. \ref{s:low_lum}).

The AGN luminosity of the active LIRGs contributes 7\,\% of the total luminosity of the sample. If we also consider the upper limits of the AGN luminosity of the star-forming galaxies, the AGN contribution is $<$ 10\,\%. That is, AGN contribute between 7\,\% and 10\,\% to the total energy output of our sample of local LIRGs. This is in agreement with the value obtained for local LIRGs by \citet{Petric2011}, 12\,\%, and \citet{AAH2011} using mid-IR diagnostics.

\begin{table*}[ht]
\caption{AGN luminosity}
\label{tbl_agnlum}
\centering
\begin{tabular}{lcccccccc}
\hline\hline
Galaxy Name & Type & $L_{\rm 2-10\,keV}^{int}$ & $L_{\rm 14-195\,keV}$\tablefootmark{a} & $L_{\rm bol}^{\rm AGN}$ & $L_{\rm IR}/L_{\rm bol}^{\rm AGN}$ \\
& & \multicolumn{3}{c}{(10$^{42}$ erg s$^{-1}$)} \\
\hline
NGC3690 & Sy2 & 3.9\tablefootmark{b}& $<$10 & 55 & 19\\
MCG$-$03-34-064 & Sy2 & 15 & 20 & 290 & 1.7 \\
NGC5135 & Sy2 & 10\tablefootmark{b} & $<$16 & 180 & 3.9 \\
IC4518W & Sy2 & 2.2 & 18 & 26 & 21 \\
MCG+04-48-002 & \nodata & 6.9 & 38 & 120 & 3.0 \\
NGC7130 & Sy2 & 10\tablefootmark{b} & $<$23 & 180 & 5.1 \\
NGC7469 & Sy1 & 17 & 39 & 340 & 5.0 \\
NGC7679 & Sy1 & 0.4 & 15\tablefootmark{c} & 4.1 & 120 \\
\hline
\end{tabular}
\tablefoot{Intrinsic AGN 2--10\,keV, 14--195\,keV and bolometric luminosities. We used the bolometric corrections of \citet{Marconi2004}.
\tablefoottext{a}{Observed \textit{Swift}\slash BAT 14--195\,keV luminosity from \citet{Tueller2010}. For non-detections we assumed a flux $<$3.9$\times$10$^{-11}$\,erg\,cm$^{-2}$\,s$^{-1}$, which is the 4.8\,$\sigma$ sensitivity achieved for 95\,\% of the sky in this survey. }
\tablefoottext{b}{Estimated from the 6.4\,keV \Feka\ emission line luminosity using the relation $L_{\rm Fe K\alpha}$\slash$L_{\rm 2-10\,keV}$ = $2 \times 10^{-3}$ from \citet{Levenson2006}.}
\tablefoottext{c}{The $L_{\rm 14-195\,keV}$ of NGC~7679 is likely to be contaminated by the nearby Seyfert 2 NGC~7682.}
}
\end{table*}

\section{Conclusions}\label{s:conclusions}

We have analyzed the X-ray properties of a representative sample of 27 local LIRGs. The median $\log L_{\rm IR}/L_\odot$ is 11.2, thus the low-luminosity end of the LIRG class is well represented. The main results are as follows:

\begin{enumerate}
\item For most of the galaxies the soft X-ray emission (0.5--2\,keV) can be associated to the star-formation activity. This is true even for some Seyfert 2s that host powerful starbursts and highly obscured AGN. We find a proportional correlation between the SFR (unobscured plus obscured) and the $L_{\rm 0.5-2\,keV}$ (Equation \ref{eqn:sfr_soft}). This relation is compatible with that obtained from synthesis models \citep{Mas2008}. Only LIRGs hosting Seyfert 1 deviate significantly from this correlation.

\item We find that the hard X-ray (2--10\,keV) emission of those LIRGs classified as \HII\ like is also proportional to the SFR (Equation \ref{eqn:sfr_hard}).  This correlation is compatible with that found for nearby starbursts \citep{Ranalli2003, Persic2004}. 
In this relationship LIRGs hosting Seyfert nuclei (type 1 and type 2) show in general an excess of 2--10\,keV emission clearly attributed to the AGN. However, some LIRGs hosting a Seyfert 2 nucleus and with powerful starbursts relative to their obscured AGN also lie on the correlation.

\item The soft X-ray emission can be modeled with a thermal plasma. The plasma abundance has subsolar Fe\slash O ratios. This can be explained by the $\alpha$ elements enrichment due to Type II SNe or by the Fe depletion into dust grains. The data analyzed in here does not allow us to reject any of these possibilities.

\item We do not detect the \Feka\ emission line at 6.4\,keV in most ($>$90\,\%) of the \HII\ LIRGs. Only in one \HII\ LIRG (MCG+04-48-002) is the presence of an obscured AGN evident from the X-ray data. Thus we can rule out the presence of luminous obscured (or Compton thick) AGN in these \HII\ LIRGs. If present, the AGN contribution to the bolometric luminosity would be less than 10\,\%.

\item Three Seyfert LIRGs (10\,\%) in our sample are Compton thick AGN candidates based on their large \Feka\ EW. The rest are Seyfert 2s (2, 7\,\%) with $N_{\rm H}$ $<$ 10$^{24}$\,cm$^{-2}$ or Seyfert 1s (2, 7\,\%). The median AGN contribution to the bolometric luminosity of those LIRGs hosting a Seyfert nucleus is 25\,\%, ranging from 1\,\% to 35\,\%.

\item The AGN emission represents about 7\,\% of the total energy output of the sample. Taking into account the upper limits for the AGN contribution in the \HII\ LIRGs, the AGN contribution is between 7\,\% and 10\,\%. This is in agreement with the values estimated from mid-IR data \citep{AAH2011, Petric2011}.

\end{enumerate}

\begin{acknowledgements}

We thank the anonymous referee for useful comments and suggestions. The authors thank C. Done for helpful discussion. MP-S thanks the Durham University for their hospitality during his stay while part of this work was done. MP-S also acknowledges support from the CSIC under grant JAE-Predoc-2007. AA-H and MP-S acknowledge support from the Spanish Plan Nacional del Espacio under grant ESP2007-65475-C02-01 and Plan Nacional de Astronom\'ia y Astrof\'isica AYA2009-05705-E.

This work is based on observations obtained with \xmm, an ESA science mission with instruments and contributions directly funded by ESA Member States and the USA (NASA).
This research has made use of the NASA\slash IPAC Extragalactic Database (NED) which is operated by the Jet Propulsion Laboratory, California Institute of Technology, under contract with the National Aeronautics and Space Administration.
\end{acknowledgements}

\appendix

\section{Notes on individual sources}\label{apx:notes}

In this Appendix we discuss the X-ray spectral analysis of some galaxies with \xmm\ data.

\paragraph{NGC~3256}
It is the most luminous nearby (z$<$0.01) merger system. Its energy output is dominated by a powerful starburst. Previous \textit{ASCA}, \textit{Chandra} and \xmm\ X-ray observations of this galaxy have been analyzed in detail by \citet{Moran1999}, \citet{Lira2002} and \citet{Jenkins2004}, respectively.
We used a simple model (absorbed \textit{vmekal} $+$ power-law) to fit the \xmm\ spectrum. It provides an acceptable fit ($\chi^2_{\rm red}\sim 1.7$) for our analysis.
\citet{Jenkins2004} tentatively detected a \Feka\ emission line at $\sim$6--7\,keV. The higher S/N ratio data analyzed here shows clearly an emission line at 6.60$^{+0.10}_{-0.04}$\,keV (Table \ref{tbl_feka}). The energy of the line suggests that it is produced by ionized Fe, possibly related to supernovae activity. The upper limit for the EW of a neutral \Feka\ line at 6.4\,keV is $<$70\,eV (Table \ref{tbl_feka}). This low EW is not compatible with that expected from a luminous Compton thick AGN.

\paragraph{Arp~299 (NGC~3690 and IC~694)}
It is a luminous infrared ($L_{\rm IR}$ $=$ 6 $\times$ 10$^{11}$ $L_{\rm \odot}$) merger system. It hosts one of the most powerful starbursts in local galaxies \citep{AAH09Arp299}. The X-ray emission below 10\,keV is dominated by star-formation, however a Compton thick AGN is found in the system \citep{DellaCeca2002}. The hard X-ray spectrum of the nucleus of NGC~3690 indicates that the obscured AGN is probably located there \citep{Zezas2003}.
\citet{Ballo2004} detected the \Feka\ emission feature in both system components NGC~3690 and IC~694.
They found that the energy of the emission line in NGC~3690 is consistent with neutral iron, however in our fit the energy of the emission line is not well constrained (Table \ref{tbl_feka}). 
The measured 6.4\,keV \Feka\ flux is $\sim$10 times larger than that expected from star-formation suggesting the presence of an AGN in NGC~3690.
The energy of the emission line in IC~694 is consistent with the \Feka\ from ionized iron that may be produced in highly ionized gas around the AGN or SN explosions \citep{Ballo2004}. For the fit we used a model consisting of an absorbed thermal plasma plus a power-law. We included a Gaussian profile to account for the \Feka\ emission lines.

\paragraph{MCG$-$03-34-064} This galaxy is classified as Seyfert 1.8. The soft X-ray spectrum is dominated by a thermal component likely produced by star-formation and gas photoionized by the AGN. The AGN absorbed component dominates the spectrum in the hard X-ray range. An \Feka\ emission line is detected at 6.4\,keV, consistent with neutral iron. A detailed analysis of the \xmm\ data of this galaxy was presented by \citet{Miniutti2007}.

\paragraph{IC~4518W}
It is a Seyfert 2 galaxy. Its \xmm\ and \textit{INTEGRAL} observations are described by \citet{deRosa2008}.
This is the only galaxy in the sample in which we detect two prominent emission lines in the hard X-ray spectrum, one at 6.39$\pm$0.03\,keV and a weaker emission line at 7.1$^{+0.1}_{-0.2}$ (Table \ref{tbl_feka}). The former is compatible with \Feka\ emission from neutral Fe. 
The latter may be \Feka\ produced by highly ionized iron, Fe\,K$\beta$ or these lines blended. \citet{Comastri2010} found the Fe\,K$\beta$ line in some obscured AGN. However the Fe\,K$\beta$\slash Fe\,K$\alpha$ ratio from neutral iron is 0.12--0.17 \citep{Palmeri2003} and the ratio between the 6.39 and the 7.1\,keV emission lines in IC~4518W is larger (0.4). Thus there might be a contribution from the \ion{Fe}{XXVI} K$\alpha$ emission line.

\paragraph{MCG+04-48-002}
This galaxy is classified as \HII\ from optical spectroscopy, however its radio, hard X-ray emission
and mid-IR spectrum suggest the presence of an obscured AGN \citep{Masetti2006, Pereira2010c}. 
\textit{Suzaku} observations of this galaxy were analyzed by \citep{Winter2009}.
We detect an emission line at 6.47$^{+0.05}_{-0.06}$ that is compatible with neutral \Feka. The high hydrogen column density ($N_{\rm H}$ $=$ 63.2$^{+10.0}_{-5.5}\times 10^{22}$\,cm$^{-2}$) towards the AGN and the powerful star-formation might explain why no AGN signatures are found in its optical spectrum.

\paragraph{NGC~7679}
Is a composite Seyfert 1\slash starburst galaxy. It is sometimes misclassified as Seyfert 2 (see \citealt{Shi2010}). The hard X-ray spectrum is well reproduced by a power-law model, but we had to add a soft thermal plasma component to account for the soft X-ray excess.
\citet{DellaCeca2001} reported X-ray fluxes $\sim$7 times larger in the soft and hard bands from the analysis of contemporaneous (1998) \textit{ASCA} and \textit{BeppoSax} observations of this galaxy.
Previous X-ray observations of NGC~7679 are available with \textit{Einstein} (1981) and \textit{ROSAT} (1990). The fluxes in the 0.2--4\,keV (\textit{Einstein}) and 0.1--2.4\,keV (\textit{ROSAT}) bands are factor of $\sim$2 larger than those measured in the \xmm\ data \citep{DellaCeca2001}. These flux variations reflect the long term variability of the X-ray emission of this galaxy.

\section{Optical classification}\label{apx:optical_class}

The optical spectra of 7 galaxies in the parent sample of LIRGs \citep{AAH06s,AAH2011} without a previous activity classification were obtained as part of the six-degree Field (6dF) Galaxy Survey (6dFGS DR3; \citealt{Jones2004, Jones2009}). Only 4 of these 7 LIRGs are members of the subsample studied in this paper. However we present the optical spectra of all of them since in Section \ref{s:sample} we compare the nuclear activity of both samples.

\begin{figure*}
\center
\includegraphics[width=0.3\hsize]{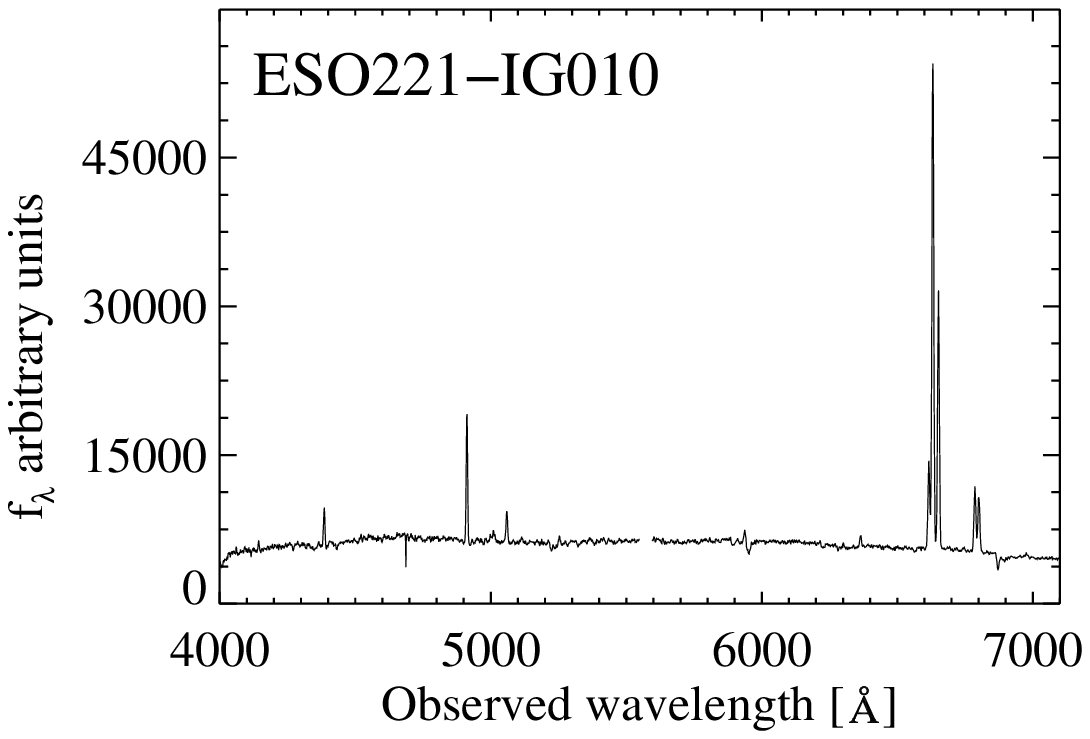}
\includegraphics[width=0.3\hsize]{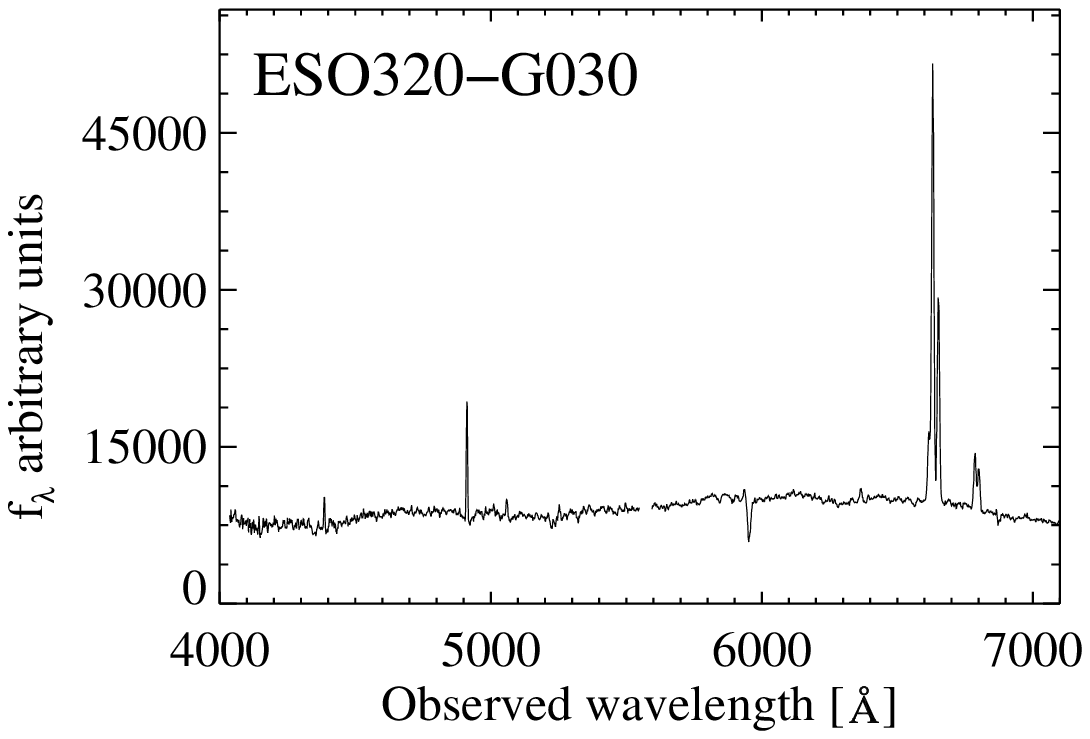}
\includegraphics[width=0.3\hsize]{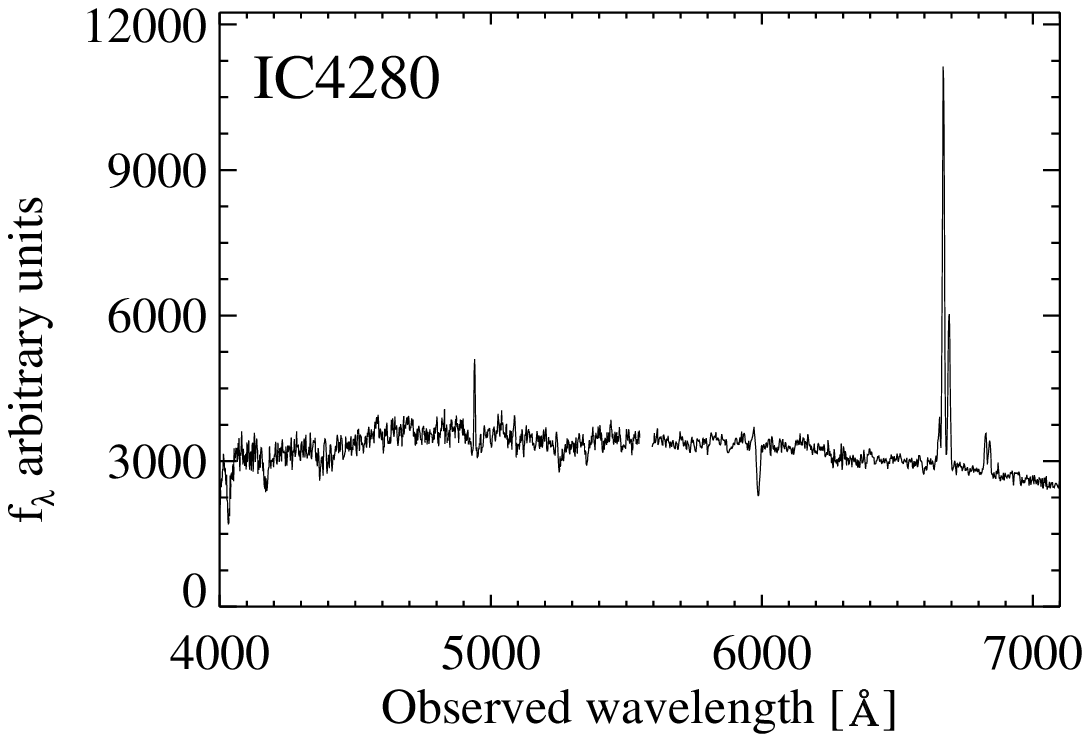}
\includegraphics[width=0.3\hsize]{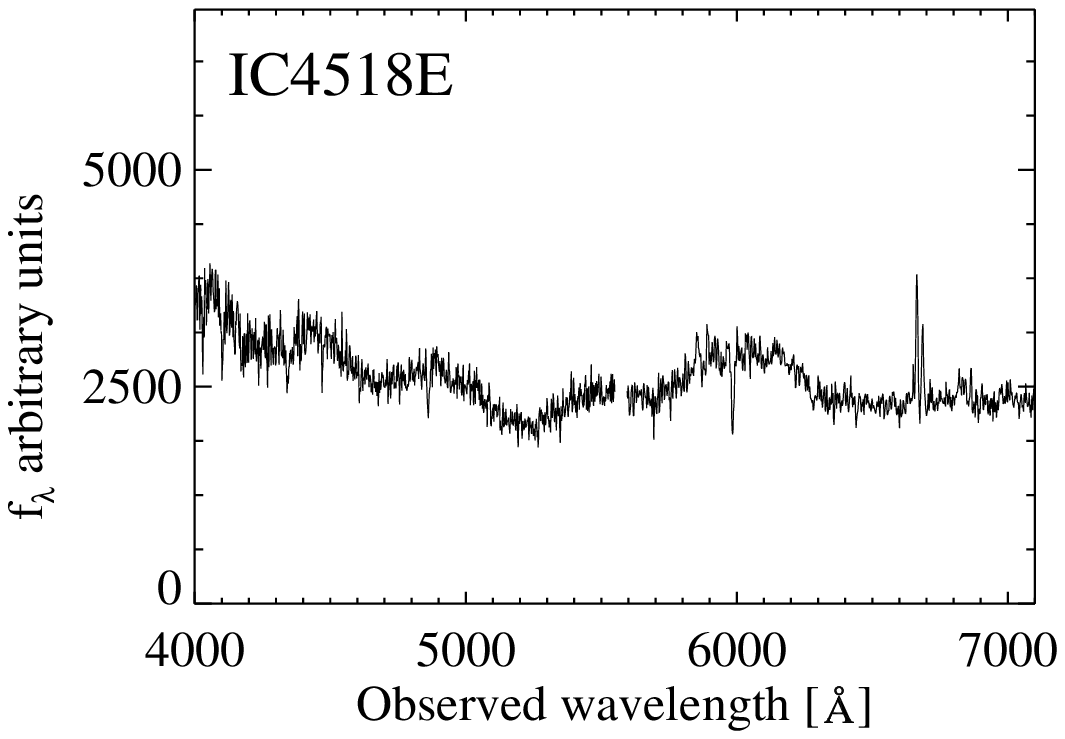}
\includegraphics[width=0.3\hsize]{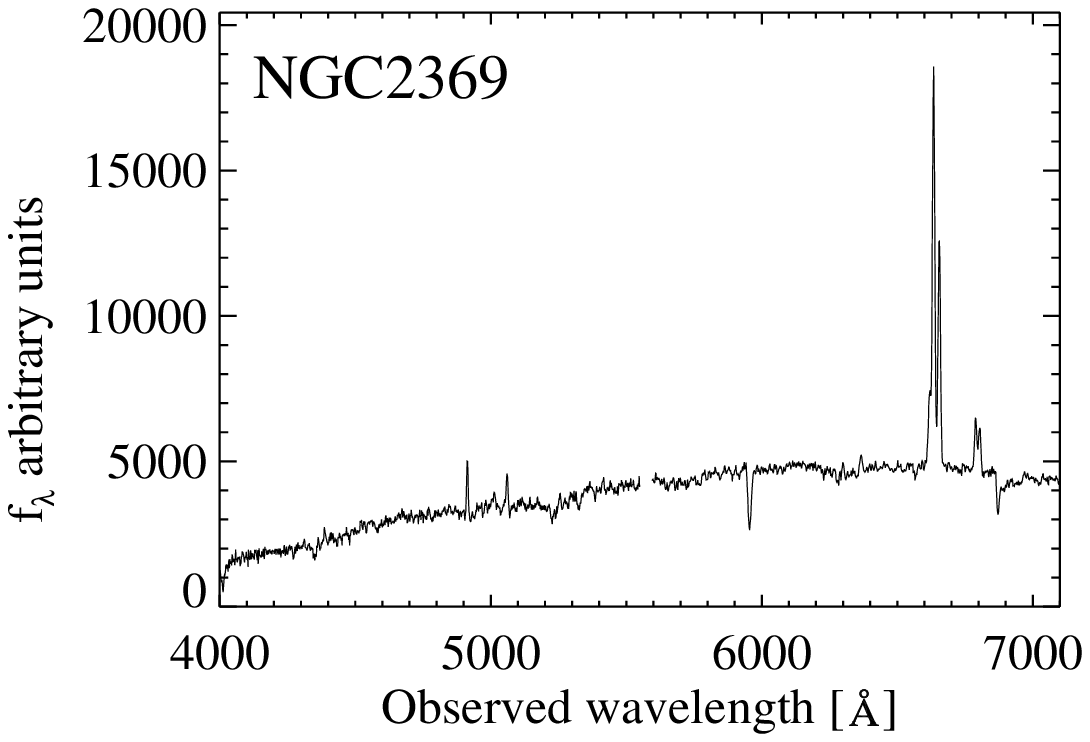}
\includegraphics[width=0.3\hsize]{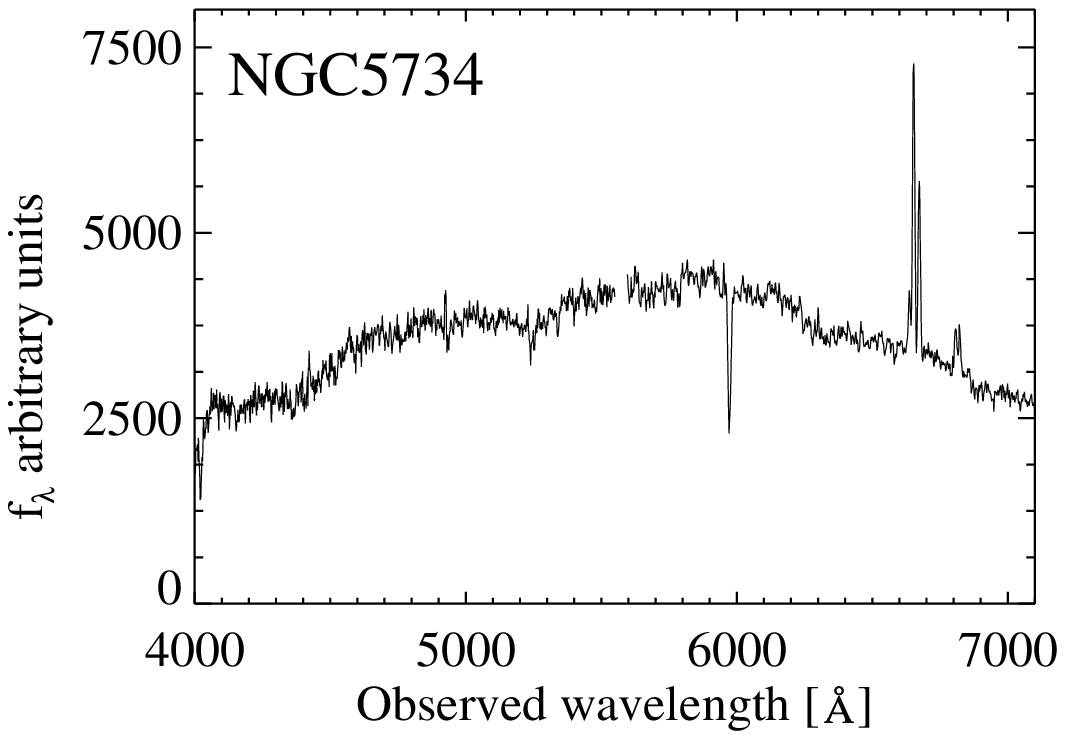}
\includegraphics[width=0.3\hsize]{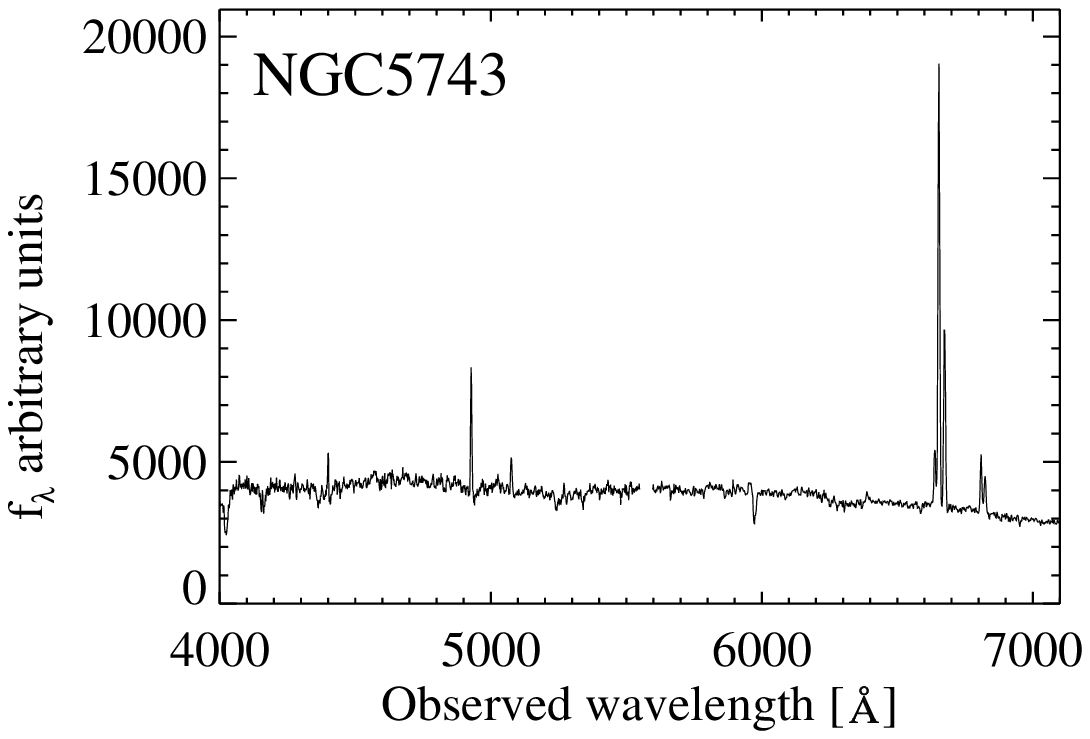}
\hspace{0.3\hsize}
\hspace{0.3\hsize}
\caption{Observed optical spectra of 7 LIRGs from the 6dFGS database.}
\label{fig_opticalspectra}
\end{figure*}

The optical spectra were obtained with the 6dF multi-object fibre spectrograph on the United Kingdom Schmidt Telescope (UKST) over 2001 to 2006. The fiber angular diameter is 6\farcs7, thus at the distance of these LIRGs it covers the central $\sim$2\,kpc of the galaxies (i.e., similar to the physical regions covered by the \citealt{Veilleux1995} spectra of local LIRGs). Each object was observed with two gratings in the V (3900--5600\,\AA) and R (5400--7500\,\AA) bands for, at least, 1\,hr and 0.5\,hr, respectively. These two spectra were later spliced to obtain the final object spectrum. The spectral resolution is 5--6\,\AA\ in V and 9--12\,\AA\ in R. The spectra are not accurately flux calibrated, however they can be used to calculate ratios of emission lines near in wavelength (see \citealt{Lee2011}). The spectra of the 7 LIRGs are shown in Fig. \ref{fig_opticalspectra}.

We measured the H$\beta$, [\ion{O}{III}]$\lambda$5007\,\AA, H$\alpha$, and [\ion{N}{II}]$\lambda$6584\,\AA\ emission lines in the spectra by fitting a Gaussian to each emission line. The measured [\ion{N}{II}]\slash H$\alpha$ and [\ion{O}{III}]\slash H$\beta$ line ratios are listed in Table \ref{tbl_opticlass}. We do not correct H$\beta$ for stellar absorption, so the calculated [\ion{O}{III}]\slash H$\beta$ ratio should be considered as an upper limit.

\begin{figure}
\center
\includegraphics[width=\hsize]{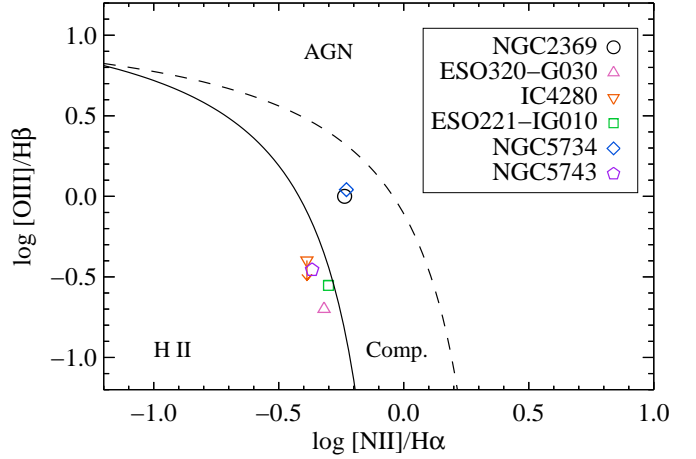}
\caption{[\ion{N}{II}]$\lambda$6584\slash H$\alpha$ versus [\ion{O}{III}]$\lambda$5007\slash H$\beta$ diagnostic diagram for the nuclear spectra of 6 LIRGs. The black lines show the empirical separation between \HII, AGN, and composite galaxies of \citet{Kewley2006}.}
\label{fig_bpt}
\end{figure}

To determine the nuclear activity of these galaxies we used the standard optical diagnostic diagram [\ion{N}{II}]$\lambda$6584\,\AA\slash H$\alpha$ vs. [\ion{O}{III}]$\lambda$5007\,\AA\slash H$\beta$ \citep{Baldwin1981}. We adopted the boundaries between \HII, AGN, and composite galaxies of \citet{Kewley2006}.
The diagram for these LIRGs is shown in Fig. \ref{fig_bpt} and the adopted classifications are listed in Table \ref{tbl_opticlass}. Neither the H$\beta$ nor the [\ion{O}{III}]$\lambda$5007 emission lines are detected in IC~4518E, thus we do not plot this galaxy in Fig. \ref{fig_bpt}. We do not detect the [\ion{O}{III}]$\lambda$5007\,\AA\ emission line in the spectrum of IC~4280, so we plotted the upper limit of the [\ion{O}{III}]\slash H$\beta$ ratio in Fig. \ref{fig_bpt}. This does not affect the \HII\ activity classification of this galaxy (see Fig. \ref{fig_bpt}).

\begin{table}[h]
\centering
\caption{Observed optical emission line ratios and classification}
\label{tbl_opticlass}
\begin{tabular}{lccccccc}
\hline\hline
Galaxy Name & [\ion{N}{II}]\slash H$\alpha$ & [\ion{O}{III}]\slash H$\beta$ & Class. \\
\hline
NGC2369 & 0.58 & 1.00 & composite \\
ESO320-G030 & 0.48 & 0.20 &  \HII \\
IC4280 & 0.41 & $<$0.40 & \HII \\
ESO221-IG010 & 0.50 & 0.28 & \HII \\
NGC5734 & 0.59 & 1.10 & composite \\
NGC5743 & 0.43 & 0.35 &  \HII \\
IC4518E & 0.57 & \nodata & \nodata \\
\hline
\end{tabular}
\end{table}

\end{document}